\documentclass{aastex61}
\pdfoutput=1 
\usepackage{amsmath,amstext}
\usepackage[T1]{fontenc}
\usepackage{apjfonts} 
\usepackage[figure,figure*]{hypcap}

\usepackage{color}  

\newcommand{\ha}{H$\alpha$}

\def\la{\mathrel{\mathpalette\fun <}}
\def\ga{\mathrel{\mathpalette\fun >}}
\def\fun#1#2{\lower3.6pt\vbox{\baselineskip0pt\lineskip.9pt
        \ialign{$\mathsurround=0pt#1\hfill##\hfil$\crcr#2\crcr\sim\crcr}}}

\shorttitle{ATLAS Probe}
\shortauthors{Wang et al.}

\begin{document}

\title{ATLAS Probe: Breakthrough Science of Galaxy Evolution, Cosmology,\\ Milky Way, and the Solar System}

\author{Yun Wang$^*$}\footnote{*wang@ipac.caltech.edu}
\affiliation{IPAC, California Institute of Technology, Mail Code 314-6, 1200 E. California Blvd., Pasadena, CA 91125}
\author{Massimo Robberto}
\affiliation{Space Telescope Science Institute, 3700 San Martin Drive, Baltimore, MD 21218}
\affiliation{Dept. of Physics \& Astronomy, Johns Hopkins University, 3400 N. Charles Street,
Baltimore, MD 21218}
\author{Mark Dickinson}
\affiliation{NOAO, 950 North Cherry Ave., Tucson, AZ 85719}
\author{Lynne A. Hillenbrand}
\affiliation{Dept. of Astronomy, California Institute of Technology, MC 249-17,
1200 East California Blvd, Pasadena CA 91125}
\author{Wesley Fraser}
\affiliation{School of Mathematics and Physics, Queen's University Belfast, University Road,
BT7 1NN, Belfast, United Kingdom}
\author{Peter Behroozi}
\affiliation{Steward Observatory, University of Arizona, 933 N Cherry Ave, Tucson, AZ 85719}
\author{Jarle Brinchmann}
\affiliation{Leiden Observatory, Leiden Univ., P.O. Box 9513, NL-2300 RA Leiden,The Netherlands}
\affiliation{Instituto de Astrof{\'\i}sica e Ci{\^e}ncias do Espaço, Universidade do Porto, CAUP, Rua das Estrelas, PT4150-762 Porto, Portugal}
\author{Chia-Hsun Chuang}
\affiliation{Kavli Institute for Particle Astrophysics and Cosmology and Department of Physics, Stanford University, Stanford, CA 94305, USA}
\author{Andrea Cimatti}
\affiliation{Department of Physics and Astronomy, Alma Mater Studiorum University of Bologna, via Gobetti 93/2, I-40129 Bologna, Italy}
\affiliation{INAF - Osservatorio Astrofisico di Arcetri, Largo E. Fermi 5, I-50125, Firenze, Italy}
\author{Robert Content}
\affiliation{Australian Astronomical Optics, Macquarie University, 105 Delhi Road, North Ryde, NSW 2113
Australia}
\author{Emanuele Daddi}
\affiliation{CEA, IRFU, DAp, AIM, Universit\'e Paris-Saclay, Universit\'e Paris Diderot,  Sorbonne Paris Cit\'e, CNRS, F-91191 Gif-sur-Yvette, France}
\author{Henry C. Ferguson}
\affiliation{Space Telescope Science Institute, 3700 San Martin Drive, Baltimore, MD 21218}
\author{Christopher Hirata}
\affiliation{Center for Cosmology and Astroparticle Physics, The Ohio State Univ., 191 West Woodruff Avenue, Columbus, OH 43210}
\author{Michael J. Hudson}
\affiliation{Dept. of Physics \& Astronomy, University of Waterloo, 200 University Avenue West,
Waterloo, ON, Canada  N2L 3G1}
\author{J. Davy Kirkpatrick}
\affiliation{IPAC, California Institute of Technology, Mail Code 314-6, 1200 E. California Blvd., Pasadena, CA 91125}
\author{Alvaro Orsi}
\affiliation{Centro de Estudios de F\'isica del Cosmos de Arag\'on,  Plaza de San Juan 1, Teruel, 44001, Spain}
\author{Russell Ryan}
\affiliation{Space Telescope Science Institute, 3700 San Martin Drive, Baltimore, MD 21218}
\author{Alice Shapley}
\affiliation{Dept. of Physics \& Astronomy, UCLA, 430 Portola Plaza, Box 951547, Los Angeles, CA 90095-1547}
\author{Mario Ballardini}
\affiliation{Department of Physics \& Astronomy, University of the Western Cape, Cape Town 7535,
South Africa}
\affiliation{INAF - Osservatorio di Astrofisica e Scienza dello Spazio di Bologna, via Gobetti 93/3, I-40129 Bologna, Italy}
\author{Robert Barkhouser }
\affiliation{Dept. of Physics \& Astronomy, Johns Hopkins University, 3400 N. Charles Street,
Baltimore, MD 21218}
\author{James Bartlett}
\affiliation{Jet Propulsion Laboratory, California Institute of Technology, 4800 Oak Grove Drive, Pasadena, CA 91109}
\author{Robert Benjamin}
\affiliation{Dept. of Physics, University of Wisconsin - Whitewater, 800 W. Main Street
Whitewater, WI 53190-1790}
\author{Ranga Chary}
\affiliation{IPAC, California Institute of Technology, Mail Code 314-6, 1200 E. California Blvd., Pasadena, CA 91125}
\author{Charlie Conroy}
\affiliation{Harvard Smithsonian Center for Astrophysics, 60 Garden Street,
Cambridge, MA 02138}
\author{Megan Donahue}
\affiliation{Physics and Astronomy Dept., Michigan State University,
567 Wilson Rd., East Lansing, MI 48824}
\author{Olivier Dor\'{e}}
\affiliation{Jet Propulsion Laboratory, California Institute of Technology, 4800 Oak Grove Drive, Pasadena, CA 91109}
\author{Peter Eisenhardt}
\affiliation{Jet Propulsion Laboratory, California Institute of Technology, 4800 Oak Grove Drive, Pasadena, CA 91109}
\author{Karl Glazebrook}
\affiliation{Centre for Astrophysics \& Supercomputing, Mail number H29, Swinburne University of Technology, PO Box 218, Hawthorn, Victoria 3122, Australia}
\author{George Helou}
\affiliation{IPAC, California Institute of Technology, Mail Code 314-6, 1200 E. California Blvd., Pasadena, CA 91125}
\author{Sangeeta Malhotra}
\affiliation{Goddard Space Flight Center, 8800 Greenbelt Rd, Greenbelt, MD 20771}
\author{Lauro Moscardini}
\affiliation{Department of Physics and Astronomy, Alma Mater Studiorum University of Bologna, via Gobetti 93/2, I-40129 Bologna, Italy}
\affiliation{INAF - Osservatorio di Astrofisica e Scienza dello Spazio di Bologna, via Gobetti 93/3, I-40129 Bologna, Italy}
\affiliation{INFN - Sezione di Bologna, viale Berti Pichat 6/2, I-40127 Bologna, Italy}
\author{Jeffrey A. Newman}
\affiliation{University of Pittsburgh and PITT PACC, 3941 O'Hara St., Pittsburgh, PA 15260}
\author{Zoran Ninkov}
\affiliation{Center for Imaging Science, Rochester Institute of Technology,
54 Lomb Memorial Drive, Rochester, NY 14623}
\author{Michael Ressler}
\affiliation{Jet Propulsion Laboratory, California Institute of Technology, 4800 Oak Grove Drive, Pasadena, CA 91109}
\author{James Rhoads}
\affiliation{Goddard Space Flight Center, 8800 Greenbelt Rd, Greenbelt, MD 20771}
\author{Jason Rhodes}
\affiliation{Jet Propulsion Laboratory, California Institute of Technology, 4800 Oak Grove Drive, Pasadena, CA 91109}
\author{Daniel Scolnic}
\affiliation{Department of Physics, Duke University, Durham, NC 27708}
\author{Stephen Smee}
\affiliation{Dept. of Physics \& Astronomy, Johns Hopkins University, 3400 N. Charles Street,
Baltimore, MD 21218}
\author{Francesco Valentino}
\affiliation{Cosmic Dawn Center (DAWN), Niels Bohr Institute, University of Copenhagen, Juliane Maries Vej 30, DK-2100 Copenhagen \O; DTU-Space, Technical University of Denmark, Elektrovej 327, DK-2800 Kgs.\ Lyngby}
\author{Risa H. Wechsler}
\affiliation{Kavli Institute for Particle Astrophysics and Cosmology and Department of Physics, Stanford University, Stanford, CA 94305, USA}
\affiliation{Particle Physics \& Astrophysics Department, SLAC National Accelerator Laboratory, Menlo Park, CA 94025}

\begin{abstract}
ATLAS (Astrophysics Telescope for Large Area Spectroscopy) Probe is a concept for a NASA probe-class space mission that will achieve  groundbreaking science in the fields of galaxy evolution, cosmology, Milky Way, and the Solar System. It is the follow-up space mission to WFIRST, boosting its scientific return by obtaining deep 1 to 4$\,\mu$m slit spectroscopy for $\sim$ 70\% of all galaxies imaged by the $\sim$2000 deg$^{2}$ WFIRST High Latitude Survey at $z>0.5$. ATLAS spectroscopy will measure accurate and precise redshifts for $\sim$ 200M galaxies out to $z < 7$, and deliver spectra that enable a wide range of diagnostic studies of the physical properties of galaxies over most of cosmic history.
ATLAS Probe and WFIRST together will produce a 3D map of the Universe with $\sim$Mpc resolution in redshift space over 2,000 deg$^2$, the definitive data sets for studying galaxy evolution, probing dark matter, dark energy and modifications of General Relativity, and quantifying the 3D structure and stellar content of the Milky Way. 
ATLAS Probe science spans four broad categories:
(1) Revolutionizing galaxy evolution studies by tracing the relation between galaxies and dark matter from galaxy groups to cosmic voids and filaments, from the epoch of reionization through the peak era of galaxy assembly; 
(2) Opening a new window into the dark Universe by weighing the dark matter filaments using 3D weak lensing with spectroscopic redshifts, and obtaining definitive measurements of dark energy and modification of General Relativity using galaxy clustering;
(3) Probing the Milky Way's dust-enshrouded regions, reaching the far side of our Galaxy;
and (4) Exploring the formation history of the outer Solar System by characterizing Kuiper Belt Objects.
ATLAS Probe is a 1.5m telescope with a field of view (FoV) of 0.4 deg$^2$, and uses Digital Micro-mirror Devices (DMDs) as slit selectors. It has a spectroscopic resolution of R = 1000, and a wavelength range of 1-4$\,\mu$m. The lack of slit spectroscopy from space over a wide FoV is the obvious gap in current and planned future space missions; ATLAS fills this big gap with an unprecedented spectroscopic capability based on DMDs (with an estimated spectroscopic multiplex factor greater than 5,000).  
ATLAS is designed to fit within the NASA probe-class space mission cost envelope; it has a single instrument, a telescope aperture that allows for a lighter launch vehicle, and mature technology (we have identified a path for DMDs to reach Technology Readiness Level 6 within two years).  
ATLAS Probe will lead to transformative science over the entire range of astrophysics: from galaxy evolution to the dark Universe, from Solar System objects to the dusty regions of the Milky Way.
\end{abstract}

\keywords{space mission --- galaxy evolution --- cosmology}

\section{Introduction}
\label{sec:intro}

The observational data from recent years have greatly improved our understanding of the Universe. The fundamental questions that remain to be studied in the coming decades include: \\
(1) How have galaxies evolved? What is the origin of the diversity of galaxies?\\
(2) What is the dark matter that dominates the matter content of the Universe?\\
(3) What is the dark energy that is driving the accelerated expansion of the Universe?\\
(4) What is the 3D structure and stellar content in the dust-enshrouded regions of the Milky Way?\\
(5) What is the census of objects in the outer Solar System?

The upcoming space missions Euclid \citep{Laureijs11}, WFIRST \citep{Spergel15}, and JWST\footnote{https://jwst.nasa.gov/}, and ground-based projects like LSST\footnote{https://www.lsst.org/} \citep{Abell09} will help us make progress in these areas through the synergy of imaging and spectroscopic data. In particular, Euclid, WFIRST, and LSST are complementary to each other, and jointly form a strong program for probing the nature of dark energy. However, the instrumental designs of Euclid and WFIRST limit their capabilities to fully address the fundamental questions described above. 
Both Euclid and WFIRST employ slitless grism spectroscopy, which increases background noise and will limit their capability to probe galaxy evolution science. 
Euclid and WFIRST spectra cover limited wavelength ranges, severely restricting opportunities to measure multiple diagnostic emission lines, and their red wavelength cutoffs are both below 2$\mu$m, limiting the redshift range over which emission lines can be detected.
JWST has slit spectroscopic capability, but a relatively small field of view (FoV), thus will not be suitable for carrying out surveys large enough to probe the relation between galaxy evolution and galaxy environment in a statistically robust manner.
LSST has no spectroscopic capability. The lack of slit spectroscopy from space over a wide FoV is the obvious gap in current and planned future space missions \citep{Cimatti09}. 
Two previous proposals aimed to address this, SPACE \citep{Cimatti09}, and Chronos \citep{Ferreras13}.
SPACE was a 1.5m space telescope optimized for NIR spectroscopy over a FoV of 0.4 deg$^2$ at a spectral resolution of R $\sim$ 400 as well as diffraction-limited imaging with continuous coverage from 0.8$\mu$m to 1.8$\mu$m, to produce the 3D evolutionary map of the Universe over the past 10 billion years.
Chronos was a dedicated 2.5m space telescope optimized for ultra-deep NIR spectroscopy over a FoV of 0.2 deg$^2$ at a spectral resolution of R=1500 in the 0.9-1.8$\mu$m range, to study the formation of galaxies to the peak of activity.
Both SPACE and Chronos proposed the use of digital micro-mirror devices (DMDs) as light modulators to select targets for massively parallel multi-object spectroscopy.

In order to obtain definitive data sets to study galaxy evolution in diverse environments and within the cosmological context, we need slit spectroscopy with high multiplicity and low background noise from space. 
In this paper, we present the mission concept for ATLAS (Astrophysics Telescope for Large Area Spectroscopy) Probe, a 1.5m space telescope with a FoV of 0.4 deg$^2$, a spectral resolution of $R=1000$ over 1-4$\,\mu$m, and an unprecedented spectroscopic capability based on DMDs --- with spectroscopic multiplex  greater than $\sim$ 5,000 targets per observation. 
ATLAS Probe has a similar FoV as SPACE and Chronos, but has a much broader wavelength range that extends from the NIR into the Mid IR, in order to probe galaxy formation and evolution through cosmic time reaching back to cosmic dawn.

We present an overview on the ATLAS Probe mission concept in Sec.2. 
Sec.\ref{sec:galaxy}-Sec.\ref{sec:solar-system} make the science case, with Sec.\ref{sec:galaxy} focusing on the galaxy evolution science, Sec.\ref{sec:cosmology} on cosmology, Sec.\ref{sec:Milky-Way} on the Milky Way galaxy, and Sec.\ref{sec:solar-system} on the Solar System. Sec.\ref{sec:source} presents our methodology for predicting source counts. Sec.\ref{sec:multiplex} and Sec.\ref{sec:ETC} discuss spectroscopic multiplex simulations and exposure time calculations. We present a preliminary design for the ATLAS instrument in Sec.\ref{sec:instrument}, and discuss its technological readiness. Sec.\ref{sec:mission-design-cost} discusses the mission architecture and cost estimate for ATLAS Probe. 
We conclude in Sec.\ref{sec:summary}. Some technical details are presented in the appendices.

\section{The ATLAS Probe Mission}
\label{sec:overview}

ATLAS Probe is designed to provide the key data sets to address the fundamental questions listed at the beginning of Sec.\ref{sec:intro}. It is a concept for a NASA probe-class space mission for groundbreaking science in the fields of galaxy evolution, cosmology, Milky Way, and Solar System.
ATLAS Probe is the follow-up space mission to WFIRST that leverages WFIRST imaging for targeted spectroscopy, 
and boosts the scientific return of WFIRST by obtaining spectra of $\sim$ 70\% of all galaxies at $z>0.5$ imaged by the $\sim$2000 deg$^{2}$ WFIRST High Latitude Survey. 

ATLAS Probe science goals span four broad categories that address all of the fundamental questions presented in Sec.1:
(1) Revolutionize galaxy evolution studies by tracing the relation between galaxies and dark matter from  galaxy groups to cosmic voids and filaments, from the epoch of reionization through the peak era of galaxy assembly; 
(2) Open a new window into the dark Universe by weighing the dark matter filaments in the cosmic web using 3D weak lensing,
and obtaining definitive measurements of dark energy and tests of General Relativity using galaxy clustering;
(3) Probe the Milky Way's dust-enshrouded regions, reaching the far side of our Galaxy;
and (4) Characterize Kuiper Belt Objects and other planetesimals in the outer Solar System.  

ATLAS Probe is a 1.5m telescope with a FoV of 0.4 deg$^2$, and uses DMDs as slit selectors. It has a spectroscopic resolution of R = 1000, and a wavelength range of 1-4$\,\mu$m, with an estimated spectroscopic multiplex factor greater than 5,000. The pixel scale of ATLAS Probe is 0.39$^{\prime\prime}$. Each micro-mirror in the DMD provides a spectroscopic aperture (``slit'') of dimension 0.75$^{\prime\prime}\times 0.75^{\prime\prime}$ on the sky. The effective point spread function (PSF) FWHM, based on the diffraction limit of 1.22$\lambda/D$ for a 1.5m aperture telescope, is 0.17$^{\prime\prime}$ at 1$\,\mu$m, 0.34$^{\prime\prime}$ at 2$\,\mu$m, and 0.67$^{\prime\prime}$ at 4$\,\mu$m. Table \ref{table:summary} shows the key characteristics of ATLAS Probe.
\begin{table}
\begin{tabular}{|l|l|l|}\hline
Aperture & 1.5m & Probe-class, can launch in 2030\\\hline
Field of View & 0.4 deg$^2$ & \\\hline
Wavelength range & 1-4$\mu$m & Near and mid IR \\\hline
Spectral resolution & R=1000 &   Heritage of mature designs \\\hline
Spectroscopic multiplex factor & > 5000 & Uses Digital Micro-mirror Devices (DMDs) \\\hline
Spectroscopic slit size & 0.75$^{\prime\prime}\times 0.75^{\prime\prime}$ & \\\hline
Number of galaxy spectra & 200M in 4 years & 3-tiered galaxy surveys (AB mag=23.0, 25.2, 26.5) \\\hline
Primary redshift range & $0.5 < z < 7+$ & Emission line and passive galaxies 
\\\hline
\end{tabular}
\caption{Summary of key characteristics of ATLAS Probe. The AB mag of the 3 galaxy surveys are cumulative depths, which are significantly deeper than the per random target depth for ATLAS Medium and Deep Surveys. }
\label{table:summary}
\end{table}

ATLAS Probe will be capable of transforming the state of knowledge of how the Universe works by the time it launches in $\sim$ 2030.
WFIRST will provide the imaging data which ATLAS Probe will complement and enhance with spectroscopic data.
ATLAS Probe will dramatically boost the science from WFIRST by obtaining slit spectra of $\sim$ 200M galaxies imaged by WFIRST with $0.5<z<7$.
ATLAS Probe and WFIRST together will produce a 3D map of the Universe with $\sim$Mpc resolution in redshift space over 2,000 deg$^2$, the definitive data sets for studying galaxy evolution, probing dark matter, dark energy and modification of General Relativity, quantify the 3D structure and stellar content of the Milky Way, and provide a census of objects in the outer Solar System. The four science goals of ATLAS probe can be quantified into four scientific objectives: \\
\noindent
(1) Trace the relation between galaxies and dark matter with less than 10\% uncertainty on relevant scales at $1<z<7$.\\
\noindent
(2) Measure the mass of dark-matter-dominated filaments in the cosmic web on the scales of $\sim$ 5-50$\,$Mpc$\,h^{-1}$ over 2,000 deg$^2$ at $0.5<z<3$.\\
\noindent
(3) Measure the dust-enshrouded 3D structure and stellar content of the Milky Way to a distance of 25$\,$kpc. \\
\noindent
(4) Probe the formation history of the outer Solar System through the composition of $~$3,000 comets and asteroids.

The ATLAS science objectives are transformative in multiple fields of astrophysics through extremely high multiplex slit spectroscopy. This requires a modest but not small telescope aperture (1.5m) that is beyond the range of SMEX or MIDEX class missions. ATLAS is a probe-class spectroscopic survey mission with a single instrument and mature technology (TRL 5 and higher). Compared to WFIRST, ATLAS has a much smaller telescope aperture (1.5m vs. 2.4m), half the number of major instruments, and 40\% the number of detectors; all these factors lead to significant cost savings. Compared to Spitzer (3 instruments and cryogenic cooling with liquid Helium), ATLAS has a larger mirror but 1/3 the number of instruments, and passive cooling. A preliminary cost estimate by JPL places ATLAS securely within the cost envelope of a NASA probe-class space mission.

ATLAS is designed to meet its science objectives through transformative capabilities enabled by mature technology. 
The core of our system, DMD, can be brought to TRL 6 within two years (see Sec.\ref{sec:DMD}).
Our baseline detector, Teledyne H4RG-10, is the same type currently under development for WFIRST. The long-wavelength cutoff of our spectroscopic channels match the standard cutoff of WFIRST and JWST devices respectively (see Sec.\ref{sec:detector}). 
Table \ref{table:comp} shows a comparison of the capabilities of several planned/proposed future space missions. Table \ref{table:depths} shows how the ATLAS surveys trace back to science objectives.
The designs for the surveys in Table \ref{table:depths} are justified in the corresponding science sections: galaxy surveys in Sec.3, Galactic Plane survey in Sec.5, and Solar System survey in Sec.6.
\begin{figure}
\centering
\includegraphics[width=0.55\columnwidth,clip]{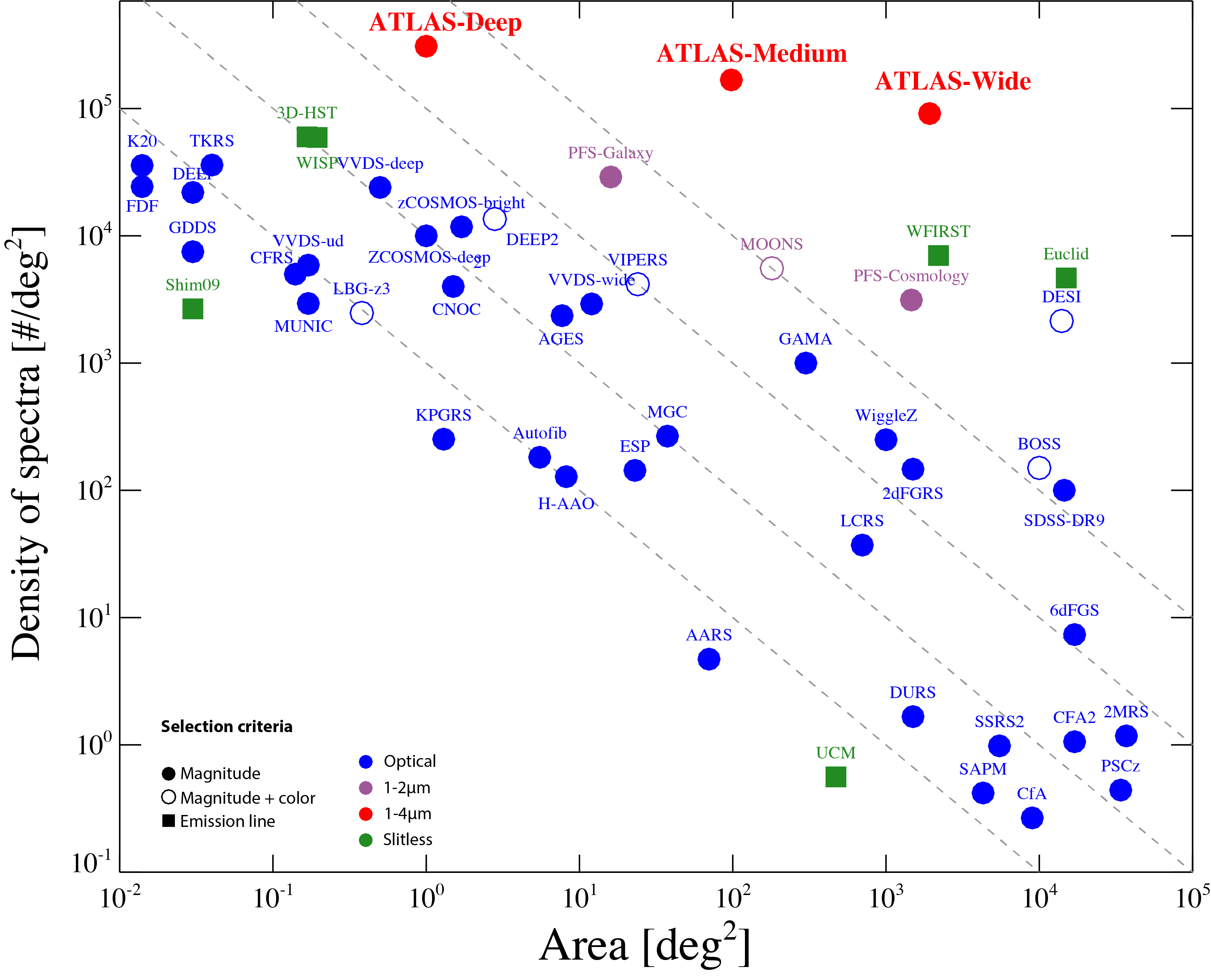}
\includegraphics[width=0.44\columnwidth,clip]{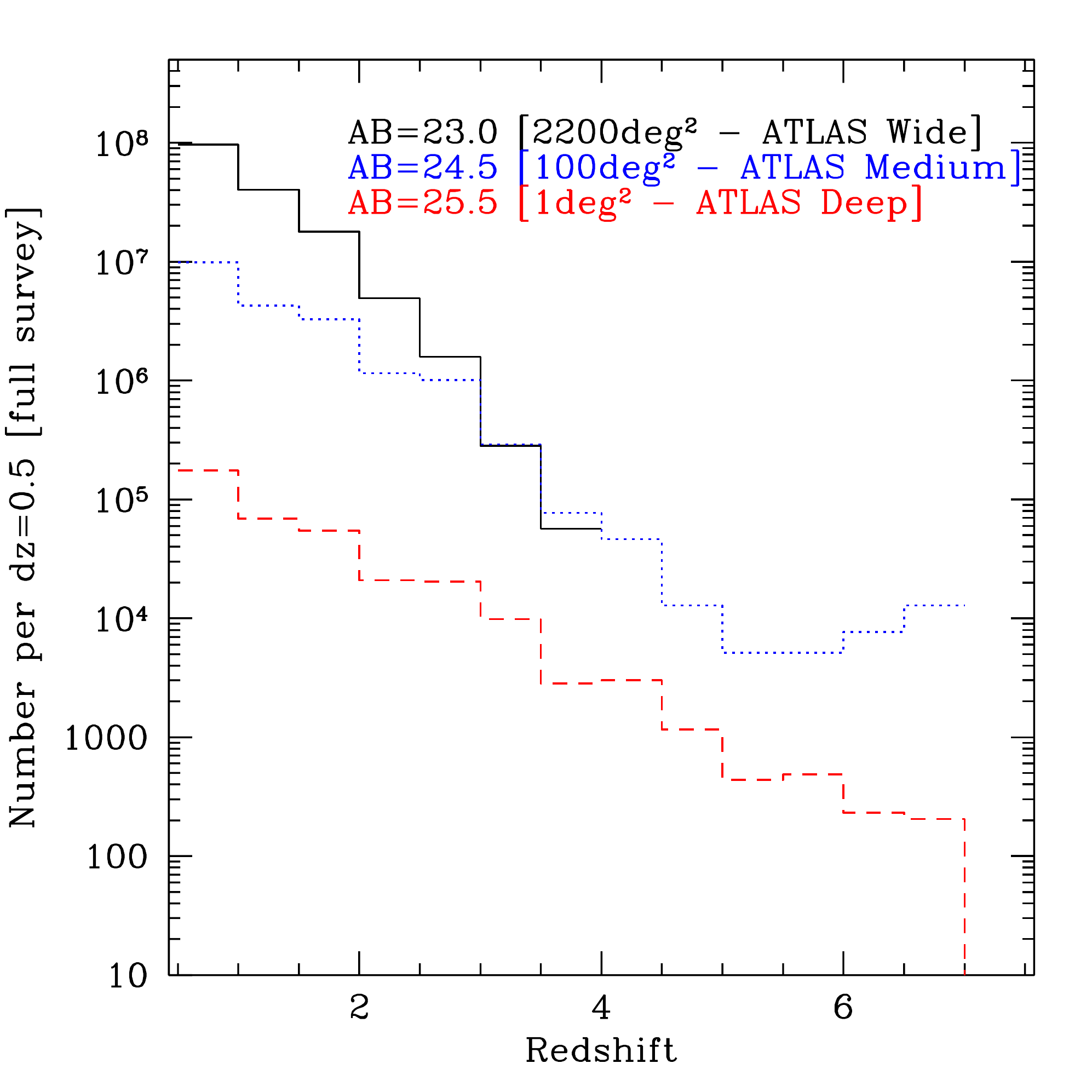}
\caption{The ATLAS galaxy surveys. Left panel: Comparison with other spectroscopic surveys (adapted from a figure originally made by Ivan Baldry). Right panel: Number of galaxies per $dz=0.5$ for the three ATLAS Probe galaxy redshift surveys (see Sec.\ref{sec:source} for a detailed discussion). Note that the forecasts are based on a single CANDELS field, CDFS (0.04 deg$^2$), hence the large uncertainty in the prediction for ATLAS Medium at $z>6$ due to cosmic variance. The depth of ATLAS Deep 
corresponds to a significantly higher number density of galaxies in CDFS, thus the effect of cosmic variance is less apparent.}
\label{fig:overview}
\end{figure}

During its 5 year prime mission, ATLAS Probe will carry out three galaxy redshift surveys at high latitude (ATLAS Wide over 2000 sq deg, ATLAS Medium over 100 sq deg, and ATLAS Deep over 1 sq deg), a Galactic plane survey over 700 sq deg, a survey of the outer Solar System over 1,200 sq deg, as well as a Guest Observer program. 
With its unprecedented capability for massively parallel slit spectroscopy at R=1000 with wavelength coverage of 1-4$\mu$m over a wide field of view, ATLAS Probe will be in great demand as a guest observatory. Approximately 5 months will be set aside during the 5 year prime mission for competitively selected GO proposals spanning a broad range of science topics. These will inform the large GO survey proposals selected for implementation during the extended mission.
The spectroscopic follow-up of the 15,000 deg$^2$ Euclid imaging survey would be an exciting legacy project; it would add a very-wide base tier to the ATLAS wedding cake of galaxy surveys, and greatly enhance the science from Euclid by mitigating systematic effects in its dark energy probes.

The galaxy evolution science objective provides the driving requirements for ATLAS Probe, which flow down to its spectral coverage of 1-4$\mu$m at the spectral resolution of R=1000; the requirements from the other science objectives are then easily met.
It is well known that this spectral range represents a major advantage for galaxy evolution studies because of: (1) Better sensitivity to stellar mass than rest-frame UV selection; (2) Smaller K-corrections (i.e. sensitivity to all galaxy types, from star-forming to passive); (3) Less affected by dust extinction;
(4) Coverage of the strongest optical rest frame lines (H$\alpha$6563 and [OIII]4959,5007) and those most useful for the physics of galaxy evolution.
In Fig.\ref{fig:overview}, the left panel compares the ATLAS galaxy redshift surveys with other spectroscopic surveys, and the right panel shows their redshift distributions.
ATLAS Wide will use the WFIRST weak lensing imaging sample as the target list.
ATLAS Medium and ATLAS Deep target lists will be magnitude selected from the lowest background regions of the WFIRST High Latitude Survey imaging, and/or from deeper WFIRST Guest Observer imaging programs.

While the spectral resolution of R=1000 meets all of the ATLAS Probe science requirements, it may be possible to increase the spectral resolution to R=1500 while maintaining the spectroscopic multiplex factor, at the cost of decreased sensitivity to continuum sources. This is a trade study that we will carry out in the future with input from the community.

In addition to definitive studies in galaxy evolution (see Sec.\ref{sec:galaxy}), ATLAS Wide Survey will enable ground-breaking studies in cosmology (see Sec.\ref{sec:cosmology}).
ATLAS Galactic Plane Survey will explore dusty regions toward the Galactic center (see Sec.\ref{sec:Milky-Way}), and ATLAS Solar System Survey will probe the formation of the outer Solar System (see Sec.\ref{sec:solar-system}).
ATLAS Probe is synergistic and complementary to ground-based spectroscopic survey instruments, e.g., 
DESI \footnote{https://www.desi.lbl.gov/},
WEAVE\footnote{http://www.ing.iac.es/Astronomy/telescopes/wht/weavepars.html}, 
4MOST\footnote{https://www.4most.eu/cms/}, and MSE\footnote{http://mse.cfht.hawaii.edu/observatory/}.

ATLAS Probe mission concept leverages the instrumentation heritage from SPACE \citep{Cimatti09}, the spectroscopic precursor to Euclid, which baselined DMDs as slit selectors in its spectrograph. SPACE motivated the space-qualification studies on DMD carried out by ESA \citep{Zamkotsian+11,Zamkotsian+17}, which were later continued by NASA \citep{Travinsky17} to advance its Technological Readiness Level (TRL).
DMD based spectrographs have been designed for ground-based telescopes \citep{Meyer04,MacKenty06,Robberto16}. 
ATLAS Probe represents the logical next step in the development of DMD-based spectroscopy. A preliminary design of the ATLAS Probe instrument is presented in Section \ref{sec:instrument}.

\begin{table}
\begin{tabular}{|l|l|l|l|l|l|l|l|l|}\hline
Mission	& $\lambda$  &	R	& FoV 	& Continuum Limit & Line Flux Limit 	& Aperture &	Cost &	Launch\\
& ($\mu$m) &&  (deg$^2$)  & (AB mag) & (erg/s/cm$^2$) & (m) && Date\\
\hline
ATLAS	& 1-4	&  1000	 &   0.4	& 23.0 (3$\sigma$) Wide &
	5$\times10^{-18}$ (5$\sigma$) Wide
	&   1.5	 &   Probe 	&  ~2030 \\
& & & & 24.5 (3$\sigma$) Medium & 1.2$\times$10$^{-18}$ (5$\sigma$) Medium &&&\\
& & & & 25.5 (3$\sigma$) Deep   &	4.6$\times 10^{-19}$ (5$\sigma$) Deep &&&\\
\hline
WFIRST	& 1.35-1.89	&  460	& 0.281 & 	20.5 (7$\sigma$)	& 10$^{-16}$ Wide (7$\sigma$) &	2.4	&  ~\$3.2B &	2025\\
\hline
SPHEREx	& 0.75-5 &	41 (0.75-2.42$\mu$m)	&   39.6	 & 19.1-19.6 (5$\sigma$)	  &   N/A	 &  0.2	& MIDEX	& 2023\\
	&  &	35 (2.42-3.82$\mu$m)	&   	 & R=41 channel	  &   	 &  	& 	& \\
	&  &	110 (3.82-4.42$\mu$m)	&   	 & 	  &   	 &  	& 	& \\
    &  &	130 (4.42-5$\mu$m)	&   	 & 	  &   	 &  	& 	& \\
\hline
Euclid &	0.92-1.85	&  380	&   0.53	&   20.0 (3.5$\sigma$)	& 2$\times 10^{-16}$ (3.5$\sigma$) Wide 
	&   1.2	& $\sim$1B Euros	&   2022\\
&&&& 21.3 (3.5$\sigma$) & 6$\times 10^{-17}$ (3.5$\sigma$) Deep &&&\\
\hline
JWST NIRSpec & 0.6-5.3 &	100, 1000, 2700	&   0.0034 &	 25.3 (10$\sigma$)	& 3.5$\times 10^{-19}$(10$\sigma$)	&   6.5	& \$9.66B	&   2021 \\
(10$^5$s)	&  &		&    &	 	&	&   & 	&   
\\
\hline
\end{tabular}
\caption{Comparison of spectroscopic capability of current \& planned/proposed future missions. The parameter values for Euclid and WFIRST are from \cite{Vavrek16} and \cite{Spergel15} respectively.
The continuum limits for WFIRST and Euclid are estimated at a wavelength of 1.5$\mu$m, assuming the same signal-to-noise ratios as for the line flux limits adopted by WFIRST and Euclid. 
The depths for JWST NIRSpec are for Band III at 3$\mu$m with $R=1000$.
The SPHEREx parameter values are from http://spherex.caltech.edu/Survey.html.
Note that for the ATLAS surveys, the flux limits correspond to the exposure time required for the faint limit of the selected galaxy sample, and not the total amount of exposure time per field.
ATLAS will visit the same fields multiple times with updated target lists.
Targets can be observed repeatedly to achieve fainter flux limits, see Table \ref{table:depths}.
}
\label{table:comp}
\end{table}

\begin{table}
\begin{tabular}{|l|l|l|l|l|l|l|l|}
\hline
      Survey	& Area 	&   Continuum Limit  &	Line Flux Limit &	Number  &	Faint limit  &	Observing 	& Traceback to  \\
      & in deg$^2$ & (AB mag) &  (erg/s/cm$^2$) & of Sources & Exposure & Time& Science Objectives\\
      \hline
ATLAS Wide	 &  2,000	&     23.0  (3$\sigma$)	& 5.0$\times 10^{-18}$ (5$\sigma$)	& 183M	&   5000s (per ELG)	&   1.6 yrs	&  Objectives 1 \& 2
\\\hline
ATLAS Medium	&  100	&   24.5 (3$\sigma$)	& 1.2$\times 10^{-18}$ (5$\sigma$) &	 17M	&  7.7$\times 10^4$s (per object)	 &  2.1 yrs	 &    Objective 1\\
	&  	&    25.2 (3$\sigma$)	&  6.2$\times 10^{-19}$ (5$\sigma$) &	 	&  2.61$\times 10^5$s (per field)	 &  	 &    \\\hline
ATLAS Deep	   &     1	 &     25.5 (3$\sigma$)&	4.6$\times 10^{-19}$ (5$\sigma$)	& 0.31M	& 4.7$\times 10^5$s (per object)	&   0.22 yr	&     Objective 1\\
	   &     	 &     26.5  (3$\sigma$)&	 1.9$\times 10^{-19}$ (5$\sigma$)	& 	& 2.79$\times 10^6$s (per field)	&  	&     \\\hline
ATLAS Galactic Plane	&     700	  &   18.2 (30$\sigma$ )& 1.6$\times 10^{-17}$ (5$\sigma$)	&	 95M	&    800s (per object) &	   0.4yr	&     Objective 3\\
& & 20.2 (30$\sigma$ )& 4.0 $\times 10^{-18}$ (5$\sigma$)	& & 7200s (per field) & &\\
\hline
ATLAS Solar System	&     1,200	  &     22.5 (3$\sigma$)  &  7.4 $\times 10^{-18}$ (5$\sigma$)   &	 3000	&    2500s &	   0.25 yr	&     Objective 4\\
\hline
\end{tabular}
\caption{The ATLAS Surveys (assuming a fiducial wavelength of 2.5$\mu$m).
Note that for ATLAS Medium, Deep, and Galactic Plane surveys, both the faint limit exposure time for the selected galaxy sample and the
total amount of exposure time per field are given; it takes multiple visits to the same field with updated target lists to obtain the spectra of all objects in a given sample. We can obtain the spectra of passive galaxies by retaining them in the target list for all visits, so that they get a much higher signal-to-noise compared to the emission line galaxies. This is feasible since passive galaxies are only a small fraction of the galaxy population at high z, especially at lower masses. }
\label{table:depths}
\end{table}

\begin{figure}
\centering
\includegraphics[width=0.49\columnwidth,clip]{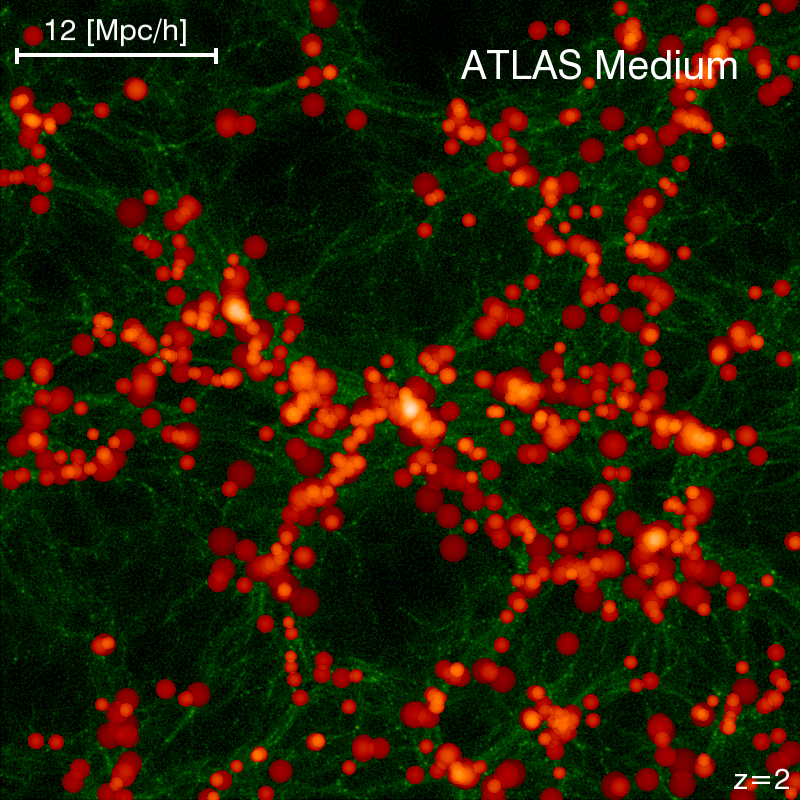}
\includegraphics[width=0.49\columnwidth,clip]{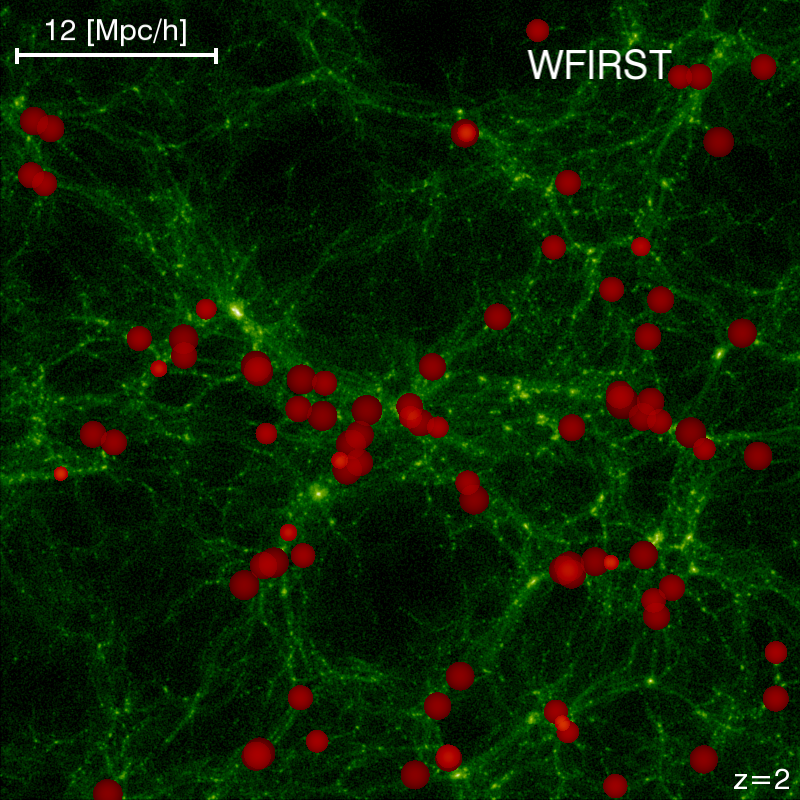}
\caption{Cosmic web of dark matter (green) at $z=2$, traced by the galaxies (red) with spectroscopy from the ATLAS Medium Survey (left) and from WFIRST slitless spectroscopy (right). See Appendix B for details on the visualization.}
\label{fig:cosmic-web1}
\end{figure}

\begin{figure}
\centering
\includegraphics[width=0.49\columnwidth,clip]{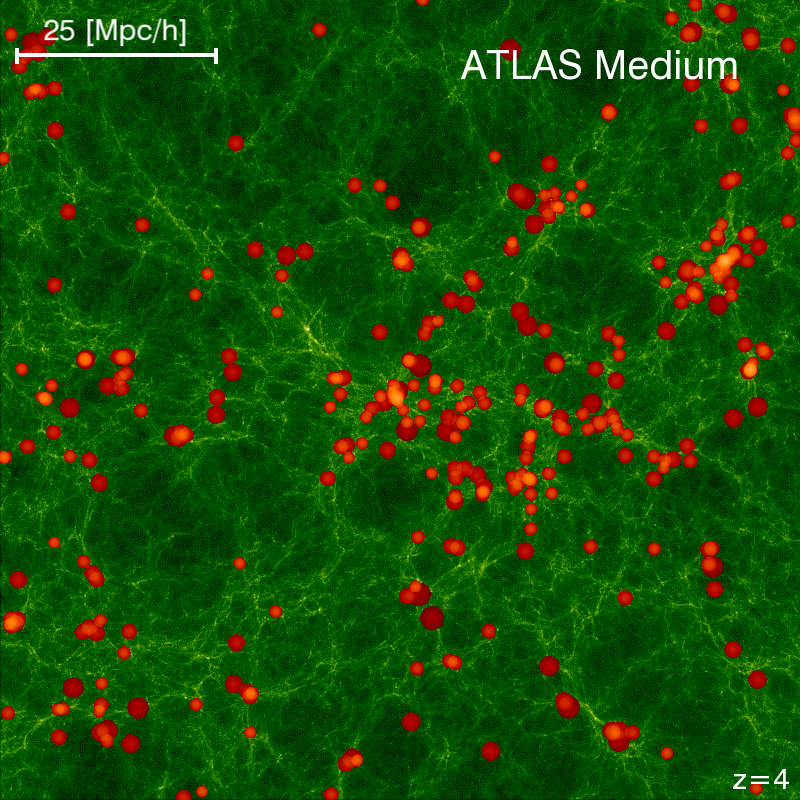}
\includegraphics[width=0.49\columnwidth,clip]{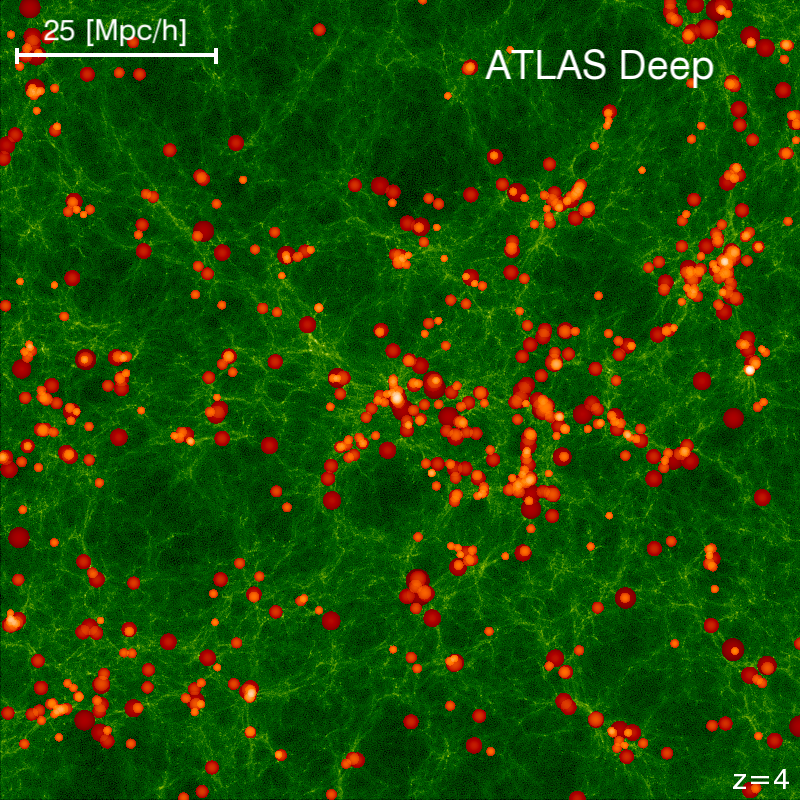}
\caption{Cosmic web of dark matter (green) at $z=4$, traced by the galaxies (red) with spectroscopy from the ATLAS Medium (left) and ATLAS Deep (right) surveys. 
ATLAS Deep (AB=25.5) is significantly deeper than ATLAS Medium (AB=24.5), and traces the cosmic web with a higher density of galaxies. See Appendix B for details on the visualization.}
\label{fig:cosmic-web2}
\end{figure}

\section{Galaxy Evolution}
\label{sec:galaxy}
ATLAS Science Objective 1, "Trace the relation between galaxies and dark matter with less than 10\% shot noise on relevant scales at $1<z<7$," flows down to three galaxy surveys nested by area and depth (Table \ref{table:depths}, see Appendix A for a quantitative justification). ATLAS Wide will observe galaxies at $z > 0.5$ in the 2000 deg$^2$ WFIRST High Latitude Imaging Survey that have robust WL shape measurements, to $H < 24.7$, with 70\% redshift completeness from emission line detections. ATLAS Medium will expose longer over 100 deg$^2$ to measure absorption line redshifts to $H < 24.5$, ensuring complete sampling for all galaxy types, and emission lines 4 times fainter than the Wide Survey to sample structure at higher redshifts. ATLAS Deep will survey 1 deg$^2$ to a continuum limit $H < 25.5$, achieving denser sampling of lower mass galaxies of all types at all redshifts, and unique 3D mapping of structure at $5 < z < 7$ with [OIII] line emitters. The ATLAS spectra, with $R=1000$, will be uniquely powerful in probing the physics of galaxy evolution.

    In the current cosmological paradigm, the initial conditions for galaxy formation come entirely from the quantum fluctuations that were imprinted on the dark matter distribution during inflation. The large-scale distribution of the seeds of galaxies results from the gravitational evolution and hierarchical clustering of these fluctuations. Over-densities collapse relative to the overall expansion, and the emergent properties of the dark matter structures depend on the nature of dark matter.  In today's era of precision cosmology, we believe that we understand the growth of structure in a universe of cold dark matter and dark energy, and sophisticated numerical simulations can map that evolution from early cosmic epochs to the present day.
  
    However, the galaxies that we see are not simply dark matter halos. Baryonic physics makes them far more complex, and we are still far from understanding how galaxies form and develop in the context of an evolving ``cosmic web" of dark matter, gas and stars. Gas flows along intergalactic filaments defined by the skeleton of the dark matter distribution. It cools and condenses into dark matter halos, forming stars that produce heavy elements that further alter the evolutionary history of the baryons, and which become the material out of which dust, planetary systems, and life itself ultimately emerge. Supermassive black holes form within massive galaxies, powering active galactic nuclei (AGN). The energetic output from star formation and AGN is believed to generate powerful feedback that can regulate, and even quench altogether, gas infall and subsequent star formation. Hydrodynamic models can make predictions for how 
these processes proceed, but no simulation can span the full dynamic range from cosmological
scales to the formation of individual stars and black hole accretion disks. A full understanding of 
galaxy evolution will not emerge until models are constrained by rich observational data from the era in which galaxies and cosmic structure were growing together.

\begin{figure}
\centering
\includegraphics[width=\columnwidth,clip]{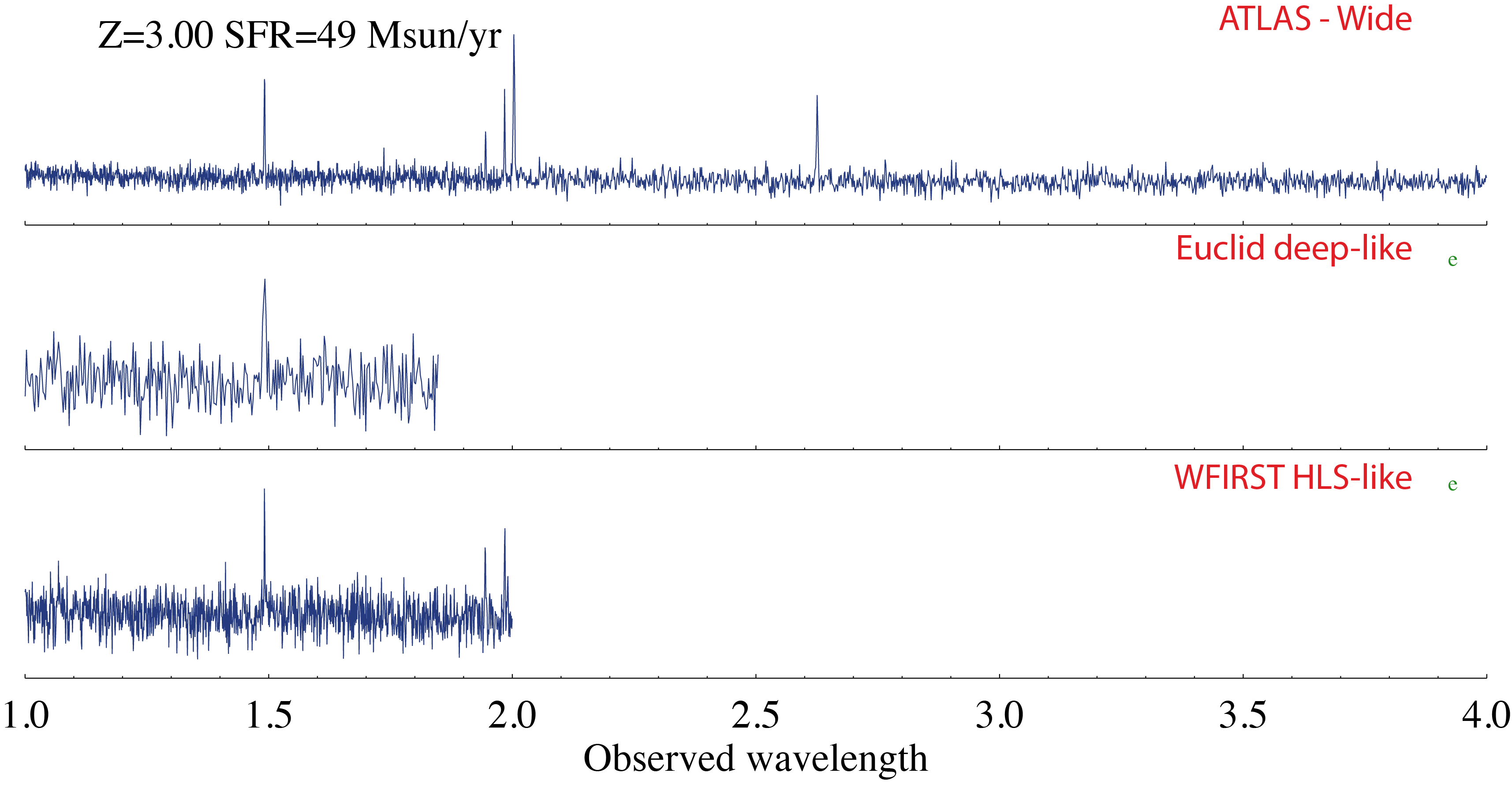}
\includegraphics[width=\columnwidth,clip]{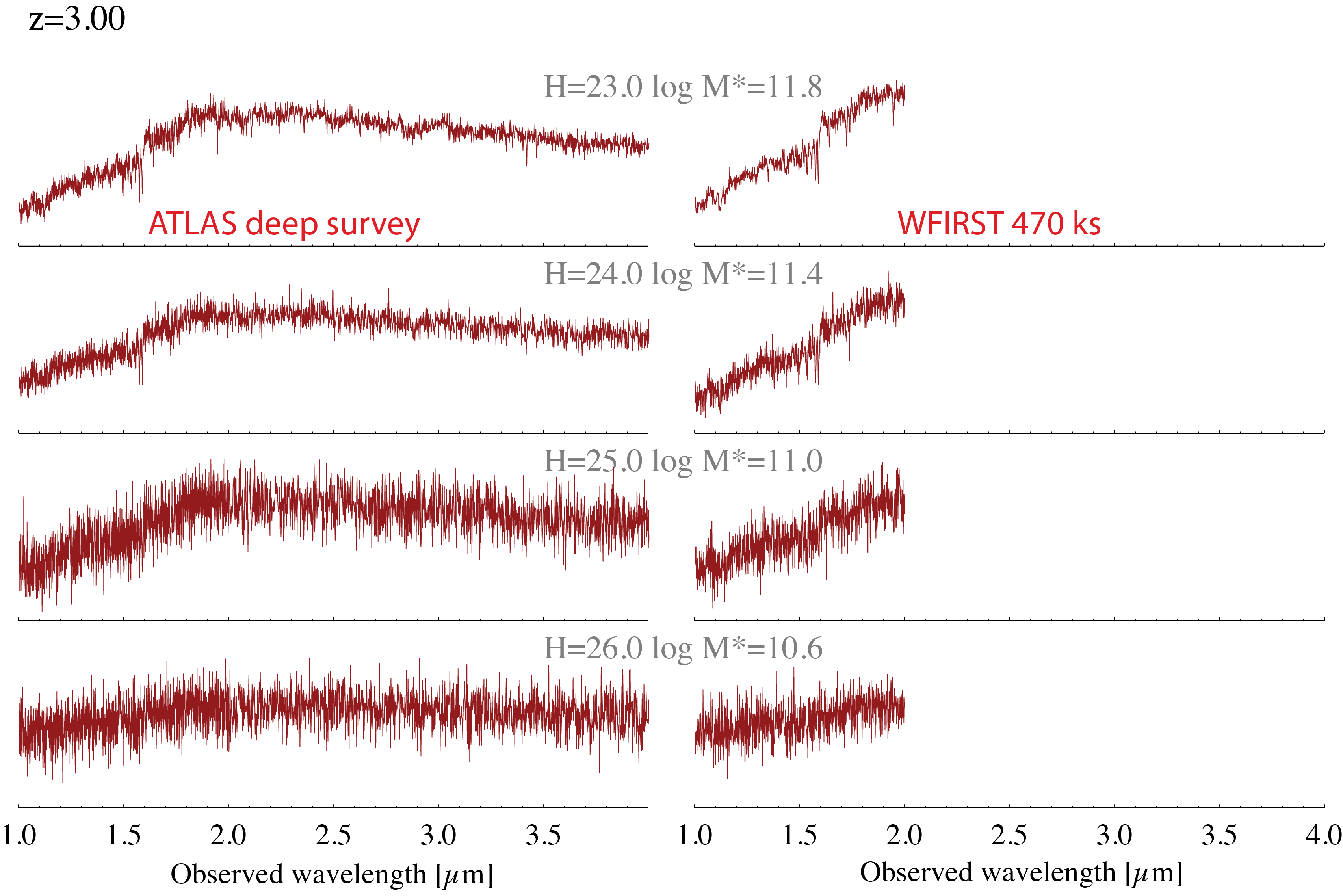}
\caption{Simulated ATLAS galaxy spectra at $z=3$ for a star-forming galaxy (f$_{\rm line}=8\times 10^{-17}$ erg/s/cm$^2$) (upper), and a passive galaxy (lower), compared to simulated galaxy spectra from WFIRST \citep{Spergel15} and Euclid \citep{Vavrek16}.}
\label{fig:spectra}
\end{figure}

    ATLAS is designed to watch galaxies emerge and grow within the cosmic web during the first half of cosmic history (Figs.\ref{fig:cosmic-web1}-\ref{fig:cosmic-web2}), with capabilities far exceeding any other missions and projects in the 2030 time frame. ATLAS wide-field, high-multiplex spectroscopy will map growth of structure in the galaxy distribution from $z=7$ to $\sim$1 with unprecedented detail, while uninterrupted spectral coverage from 1 to 4 $\mu$m will measure the evolution of essential properties of the gas, dust, stars, and active nuclei in galaxies, and their relation to local and large-scale environment. No other ground or space observatory now planned provides this combination of redshift range, survey volume, vast statistics, and spectral quality. A hierarchy of surveys (Table \ref{table:depths}) will probe cosmic structure on scales from Gpc to tens of kpc, observe $\sim$ 200M galaxies at $1 < z < 7$ (Fig. \ref{fig:overview}, right panel) spanning a broad range of luminosity and mass, and generate high quality spectra (Fig. \ref{fig:spectra}) enabling a tremendous variety of community science investigations. 
Extending the ATLAS Probe wavelength coverage to 0.5-1$\mu$m could further enhance ATLAS as a reionization probe (see Sec.\ref{sec:reionization}) and connect with ground-based studies; we will examine the mission design and cost implications of this possibility in the future.

\begin{figure}
\centering
\includegraphics[width=0.5\columnwidth,clip]{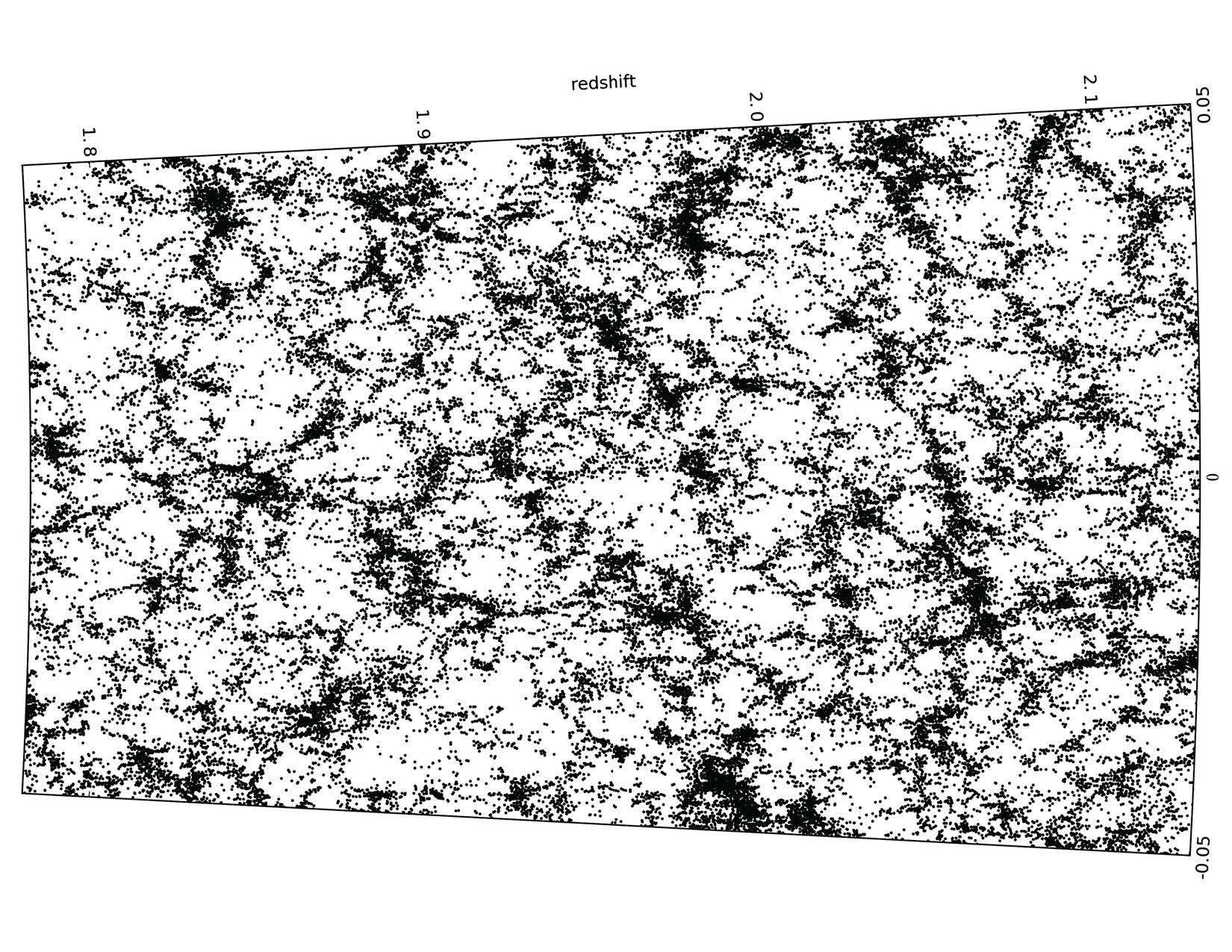}\\
\vspace{-1cm}
\includegraphics[width=0.5\columnwidth,clip]{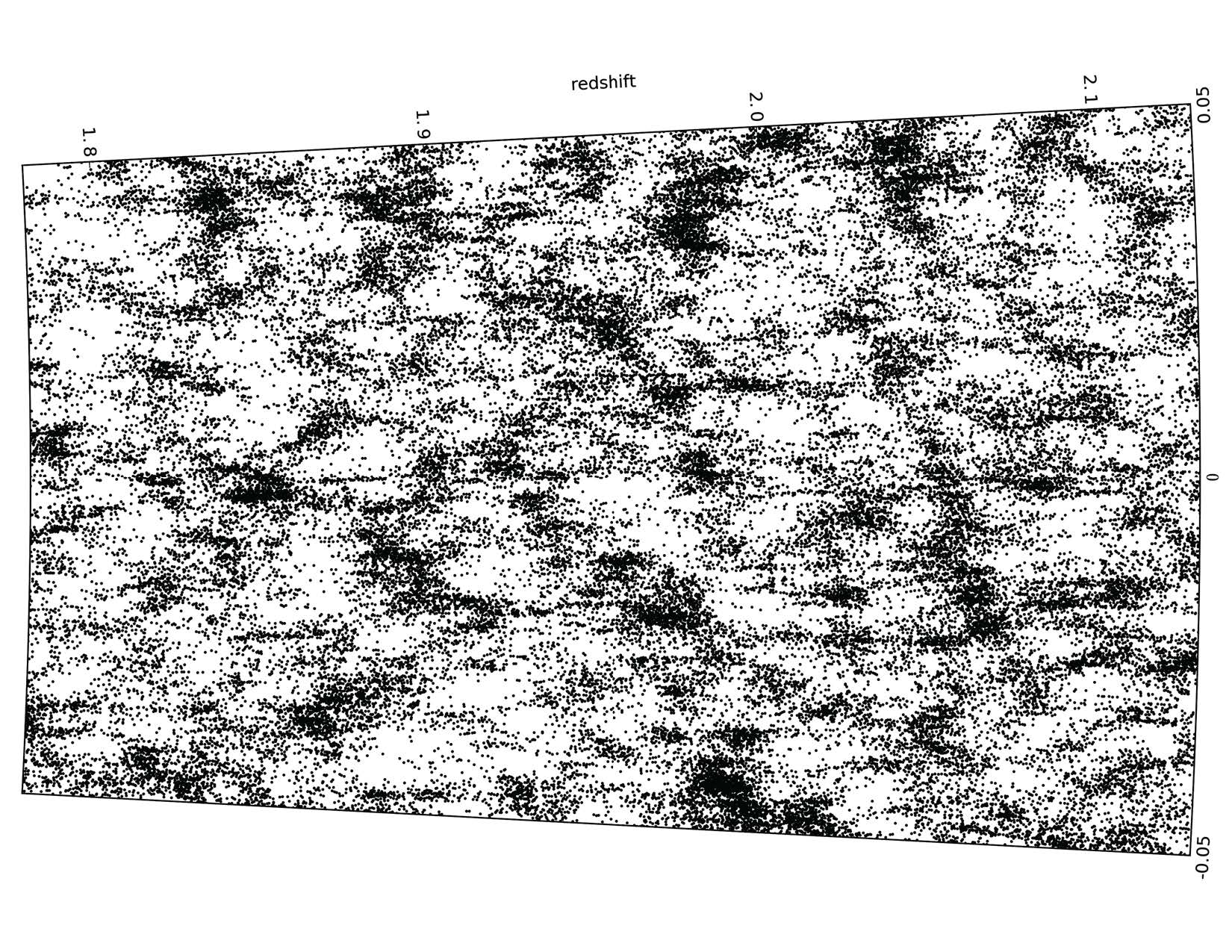}\\
\vspace{-1cm}
\includegraphics[width=0.5\columnwidth,clip]{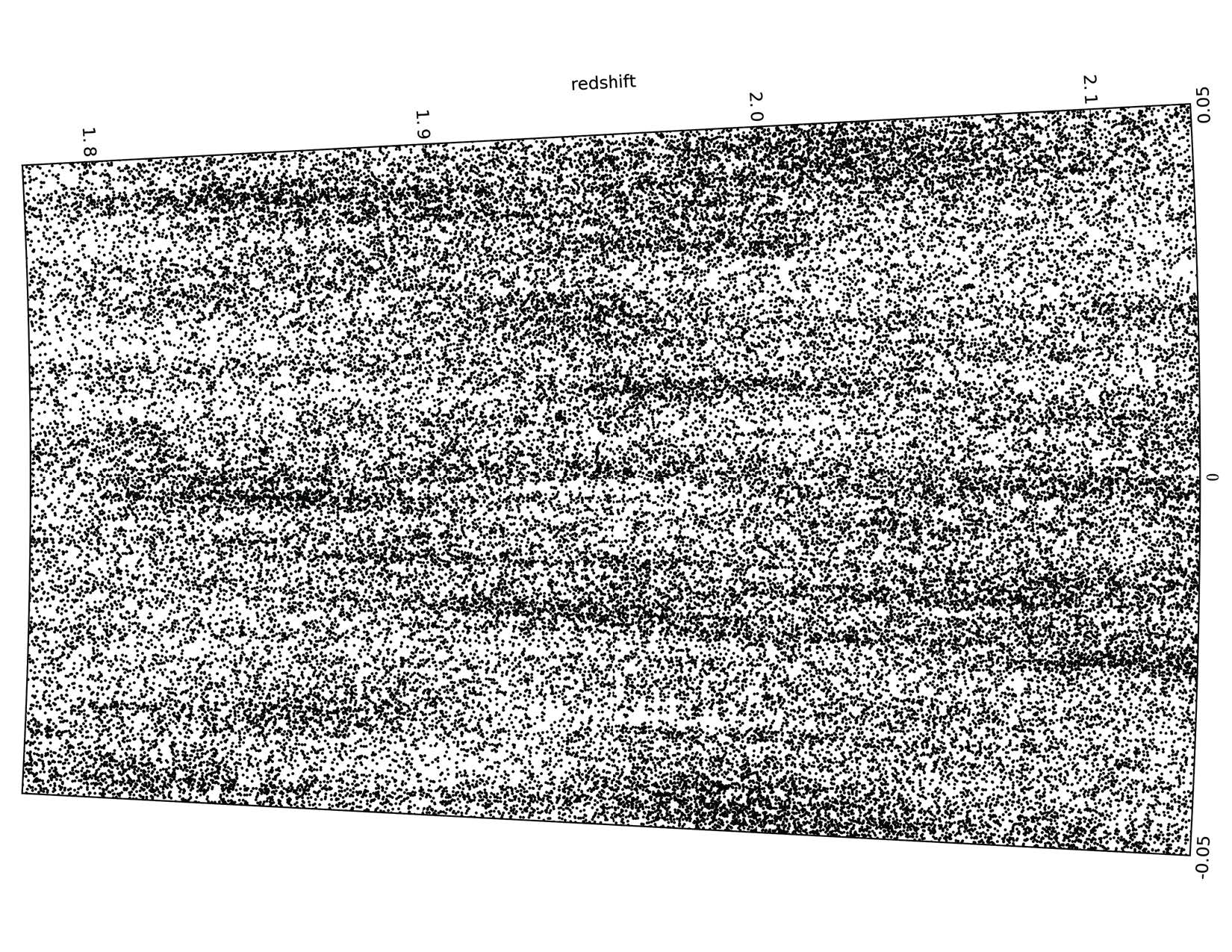}\\
\caption{The spatial distribution of H$\alpha$ emitting galaxies at $z=2$ from the semi-analytical galaxy formation model GALFORM. Each panel illustrates a different survey of the same galaxy distribution: a slit spectroscopic survey (with $\sigma_z/(1+z)=10^{-4}$, top panel), a slitless spectroscopic survey (with $\sigma_z/(1+z)=0.001$, middle panel), and a photometric survey (with the most optimistic assumption of $\sigma_z/(1+z)=0.01$, bottom panel). 
The spectroscopic surveys reveal the cosmic web, while the photometric survey does not. The slit survey traces the cosmic web in sharp detail, while the slitless survey does not. See Appendix B for details on the visualization.}
\label{fig:atlas_cosmic-web}
\end{figure}

    ATLAS Probe is designed to reveal the detailed structure of the cosmic web. This requires a slit spectroscopic survey in order to obtain redshift measurements with sufficient precision to trace cosmic large scale structure. Fig.\ref{fig:atlas_cosmic-web} shows the comparison of three galaxy redshift surveys: $\sigma_z/(1+z)=10^{-4}$ (slit spectroscopic survey --- ATLAS Probe,\footnote{The redshift precision of ATLAS Probe will depend on the signal-to-noise ratio of the galaxy spectra, as well as instrument parameters. Our preliminary estimates indicate that we should have $\sigma_z/(1+z)=10^{-4}$ for most of the galaxies, since they will have multiple emission lines detected with $S/N$ significantly higher than 5.} top panel) and $\sigma_z/(1+z)=0.001$ (slitless spectroscopic survey --- Euclid/WFIRST, middle panel), and $\sigma_z/(1+z)=0.01$ (the most optimistic assumption for a photometric survey, bottom panel). The spectroscopic surveys reveal the cosmic web, while the photometric survey does not.
The slit spectroscopic survey shows the detailed structure of the cosmic web, while the slitless survey shows a somewhat blurred picture. This contrast becomes even more pronounced at higher redshifts.
In particular, the high precision redshifts from a slit spectroscopic survey are required to explicitly map filaments in the cosmic web and enable their theoretical modeling. 

\subsection{Decoding the Physics of Galaxy Evolution by Mapping the Cosmic Web}

Galaxy properties should correlate with the underlying dark matter: the masses of the halos, their spins, and their positions within the cosmic web. Today, we have only approximate estimates of the stellar-mass/halo-mass ratio, with rather large uncertainties, from abundance matching, clustering, and weak lensing (\citealt{Behroozi13}, \citealt{Kravtsov13}, \citealt{Hudson15}). Direct clustering measurements allow us to estimate the stellar-mass/halo-mass ratio as a function of galaxy properties, which is impossible to do via abundance matching. It is also clear that the stellar-mass/halo-mass ratio does not capture the entire influence of dark matter on galaxy properties, as the poorly-understood "conformity" of galaxy properties is seen on scales much larger than the radius of even the most massive collapsed halos \citep{Hearin16}. 

\begin{figure}
\centering
\vspace{-1ex}
\includegraphics[width=0.65\columnwidth,clip]{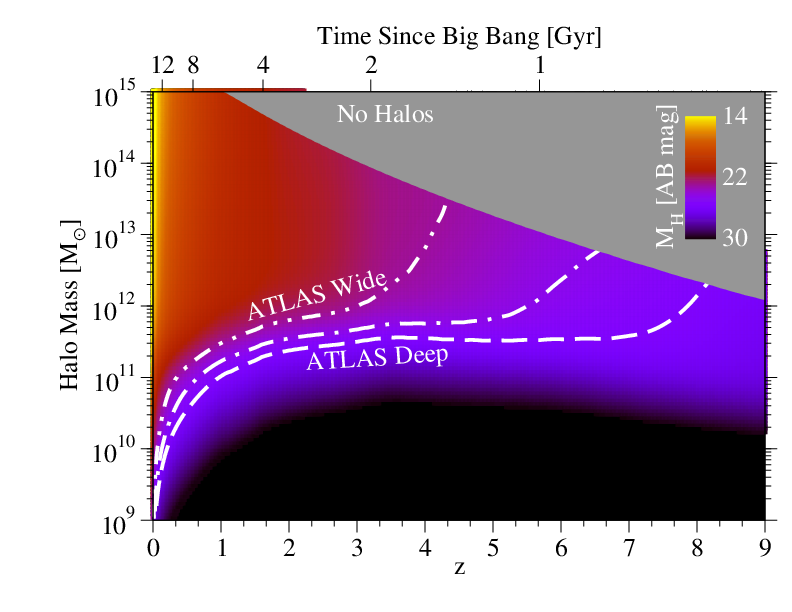}\\[-3ex]
\caption{ATLAS will resolve key galaxy formation physics over unprecedented halo mass and redshift ranges.  This figure shows limiting halo masses for each of the ATLAS Surveys as a function of redshift.  Typical H-band (F160W) apparent magnitudes calculated using FSPS \citep{Conroy09} from average star formation histories in \cite{Behroozi13}, assuming \cite{Charlot00} dust.}
\label{fig:galaxy-evolution-1}
\includegraphics[width=0.48\columnwidth,clip]{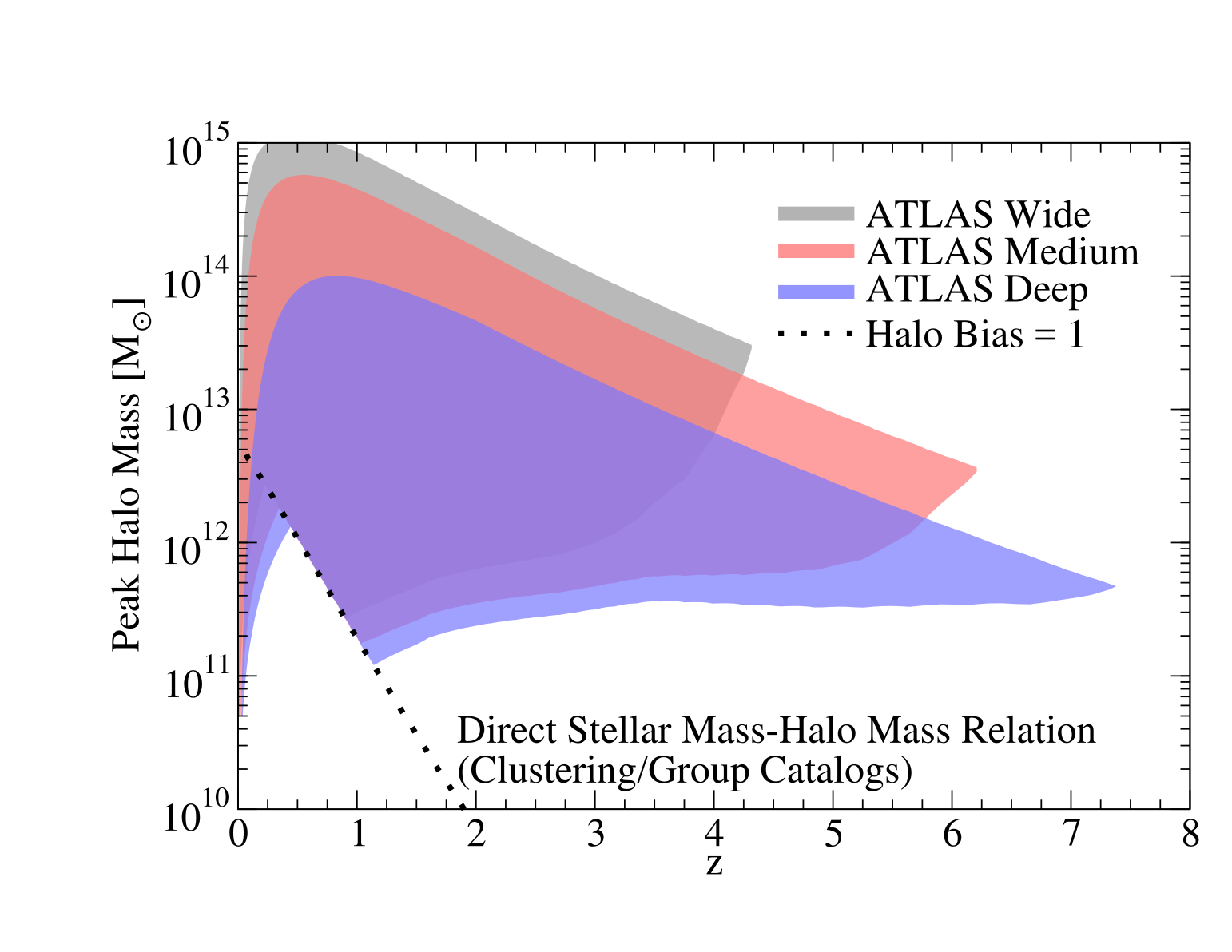}
\includegraphics[width=0.48\columnwidth,clip]{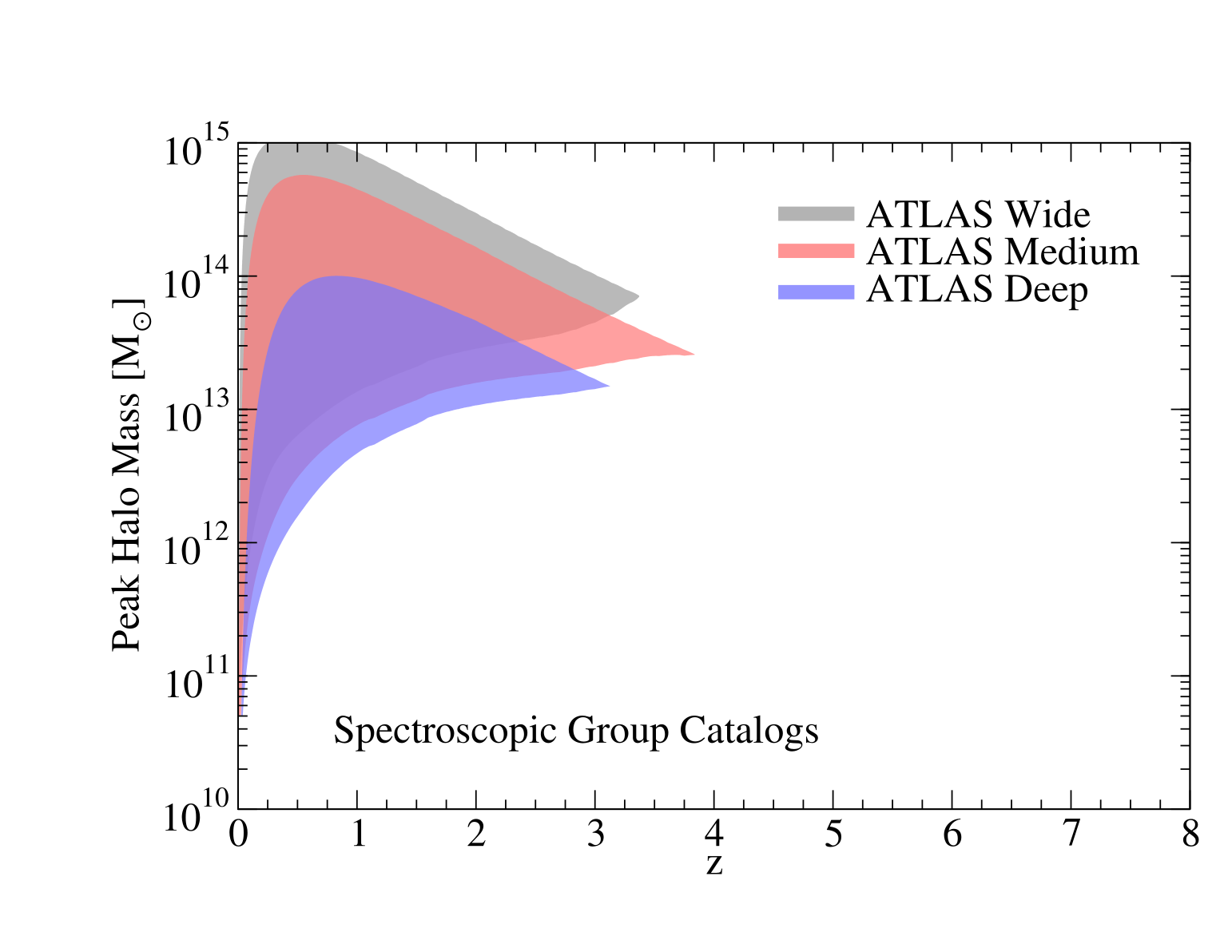}\\[-8ex]
\includegraphics[width=0.48\columnwidth,clip]{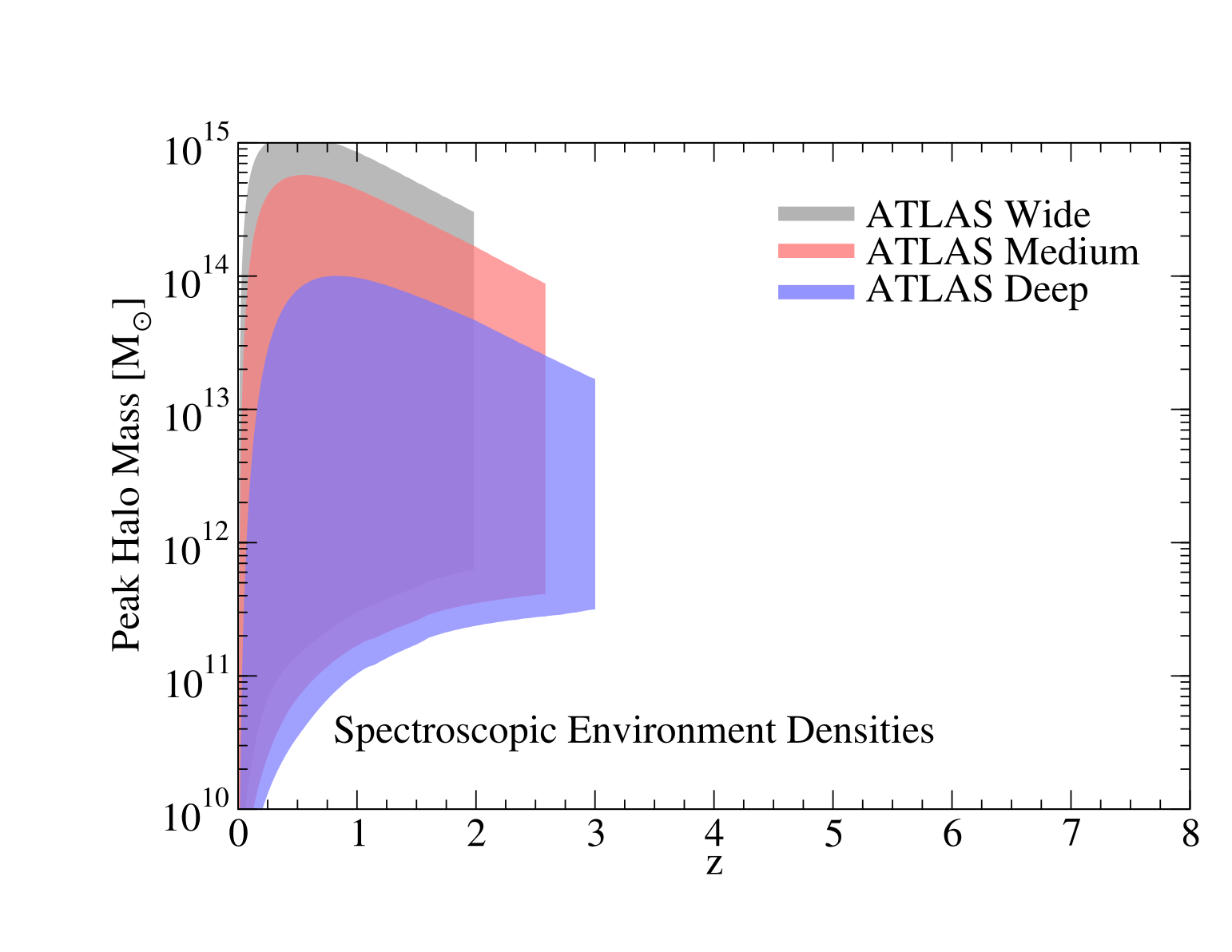}
\includegraphics[width=0.48\columnwidth,clip]{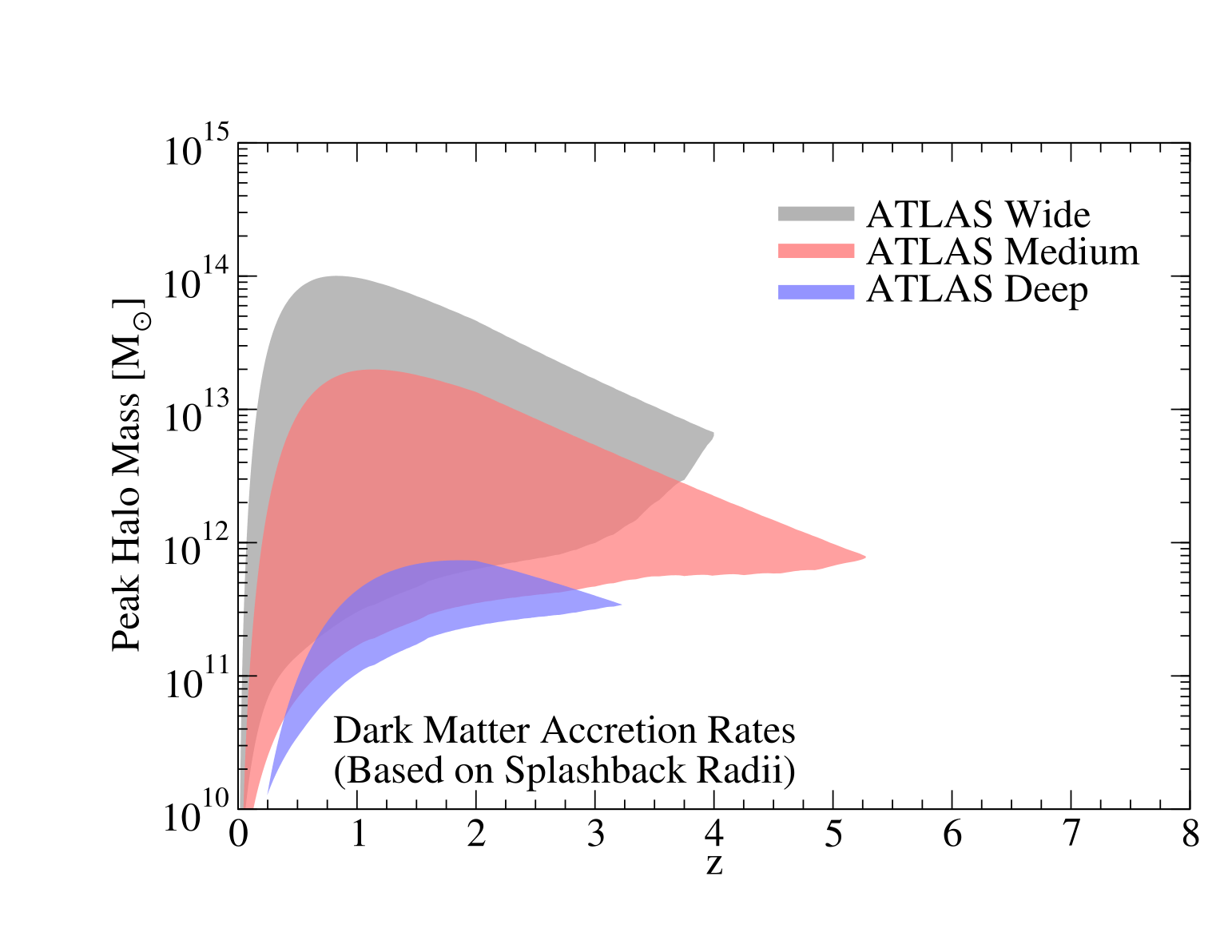}\\[-3ex]
\caption{ATLAS Surveys will allow direct measurements of the stellar mass---halo mass relationship (\textbf{top-left panel}), spectroscopic group catalogs (\textbf{top-right panel}), environmental densities (\textbf{bottom-left panel}), and dark matter accretion rates (\textbf{bottom-right panel}). See text for details on how halo mass and redshift ranges were estimated.}
\label{fig:galaxy-evolution-2}
\end{figure}

Spectroscopic redshifts over large areas are required to connect galaxy properties (e.g., stellar masses and star formation rates) to the underlying dark matter halo masses and environments that are key to understanding galaxy formation physics.  This has been demonstrated by the remarkable success of the Sloan Digital Sky Survey \citep[most recently,][]{SDSS14} at $z\sim 0$.  The ATLAS surveys will extend the redshift baseline of all SDSS-like galaxy science to $z \sim 3$ and in favorable cases to $z \sim 7$ and beyond (Figs.\ \ref{fig:galaxy-evolution-1} and \ref{fig:galaxy-evolution-2}).   Typical halo mass limits in Fig.\ \ref{fig:galaxy-evolution-1} are derived starting from empirical average star formation histories (SFHs) as a function of halo mass and redshift from \cite{Behroozi13}.  These SFHs are post-processed using FSPS \citep{Conroy09} to compute observer-frame F160W (H-band) apparent magnitudes, assuming \cite{Charlot00} dust and a \cite{Chabrier03} initial mass function.

We provide example halo mass and redshift ranges for four galaxy formation science cases in Fig.\ \ref{fig:galaxy-evolution-2}.  The \textit{stellar mass---halo mass relationship} \citep[e.g.,][]{Moster13,Behroozi13} measures the efficiency with which galaxies turn incoming gas into stars, and is a key probe of the strength of feedback from stars and supermassive black holes.  The spectroscopic clustering measurements provided by ATLAS will directly measure this relationship for massive halos (bias $b\gtrsim1$, Fig.\ \ref{fig:galaxy-evolution-2}).  Additional constraints will be provided by \textit{group catalogs}, wherein halo masses are measured for individual galaxies based on spectroscopically-identified satellite galaxy counts.  These catalogs will also provide information about whether a given galaxy is a satellite or not; this is important for interpreting whether galaxy properties arise mainly from internal processes or instead from interaction with a larger neighbor.  ATLAS will provide direct stellar mass-halo mass constraints out to $z\gtrsim 7$ and group catalogs to $z \gtrsim 3$, assuming a minimum of $3$ satellites per group (Fig.\ \ref{fig:galaxy-evolution-2}).

Besides halo mass, many open questions in galaxy formation concern how halo assembly affects galaxy properties and supermassive black hole activity.  Halo assembly is not directly measurable, but many assembly properties (e.g., concentration, spin, mass accretion rate) correlate strongly with \textit{environmental density} \citep[e.g.][]{Lee17}.  ATLAS will be able to measure environmental densities to $z \gtrsim 3$ (Fig.\ \ref{fig:galaxy-evolution-2}), with galaxies in median-density environments having at least $5$ neighbors within a 4 Mpc by 2000 km/s cylinder even at the highest redshifts.  Finally, ATLAS will be able to measure \textit{average dark matter accretion rates} for galaxies via the detection of the splashback radius (a.k.a., turnaround radius) of their satellites as in \cite{More16} out to $z\sim5$ (Fig.\ \ref{fig:galaxy-evolution-2}).  Here, we conservatively assume that a stack of $20,000$ galaxies is sufficient to detect the splashback feature; \cite{More16} achieved a $>6\sigma$ detection using only $8,000$ galaxies.

    At redshifts $z > 1$, most (but not all) galaxies were forming stars rapidly, and their spectra feature strong emission lines. With its 1 to 4 $\mu$m spectral range, ATLAS can observe the strongest optical rest-frame emission lines at high redshift (typically H$\alpha$ at $0.5 < z < 5$, and [OIII] at $1 < z < 7$) without the spectral gaps that result from OH airglow contamination and water vapor absorption and low S/N from thermal backgrounds that are unavoidable from the ground.
This continuous spectral coverage translates to continuous redshift sampling and very simple selection functions that cannot be achieved with ground-based observations. As the Universe ages, a growing number of galaxies cease star formation and evolve quiescently. This "quenching" is related to both galaxy mass and environment, with quiescent galaxies dominating the denser regions of galaxy clusters and groups today. ATLAS will measure redshifts for these galaxies too, detecting stellar absorption lines and spectral breaks thanks to the minimal sky background for slit spectroscopy in space.

    Today and in the next decade, ground-based spectroscopic surveys are measuring cosmic
structure with outstanding statistics at $0 < z < 1$, with more limited forays to higher redshifts, almost exclusively for strong emission line galaxies. The emission line flux limit of the ATLAS Wide Survey reaches 15 times fainter than the characteristic luminosity $L^*$(H$\alpha$) at $z=2.2$3 \citep{Sobral13,Pozzetti16} and L*([OIII]) at $z=3.24$ \citep{Khostovan15}, ensuring excellent sampling of large scale structure. The continuum limits of the Wide Survey reach 2 to 3.5 times fainter than optical rest-frame continuum $L^*$ at $z = 2$ to 3.5 (e.g., \cite{Marchesini07}), while the Medium and Deep Surveys reach 3.3 to 8.3 times fainter still, ensuring that ATLAS can measure absorption line and continuum break redshifts even for non-star-forming galaxies that might inhabit the densest cosmic structures. The continuum sensitivities of the Medium and Deep Surveys are tuned to match L* in the UV rest frame for Lyman break galaxies at $z \simeq 4$ and 6, respectively (e.g., \cite{Finkelstein16}). The Medium and Deep Surveys will uniquely map 3D clustering at $5 < z < 7$ via the [OIII] emission line, which is observed to become stronger at higher redshifts, where galaxy metallicities are lower, and with absorption line redshifts for $L^*$ galaxies. The "wedding cake" of ATLAS spectroscopic surveys will achieve outstanding spatial sampling of structure, from small to cosmological scales, and from $z = 7$ to 1. This is demonstrated in Appendix A, which gives the measurement errors on the galaxy two-point correlation function $\xi(r)$ as a function of the comoving separation $r$ for the ATLAS Medium and Wide galaxy surveys.

ATLAS Probe galaxy surveys will obtain galaxy evolution data at high $z$ that is equivalent to that from GAMA \citep{Driver2009} and SDSS\footnote{https://www.sdss.org/} at low $z$, revolutionizing our understanding of how galaxies have formed and evolved in the cosmic web of dark matter.

\subsection{Emergence of Galaxy Properties}

ATLAS will measure key diagnostics of galaxy properties through emission and absorption line spectroscopy. Its spectral range covers the most important optical rest frame nebular emission lines at $1 < z < 5$, including H$\alpha$, H$\beta$4861, [OIII], [NII]6549,6583 and [SII]6716,6731, with sufficient spectral resolution to resolve H$\alpha$+[NII] and the [SII] doublet. These lines trace star formation, gas excitation, metal abundance, electron density, and dust extinction, offering a wealth of information about the interstellar medium in distant galaxies. The sensitivity of the ATLAS Wide Survey ensures that both stronger and weaker lines can be measured in most typical, $L^*$ galaxies at $2 < z < 3.5$. The ATLAS Medium and Deep Surveys will measure line diagnostics for galaxies at higher redshifts and/or lower luminosities.  Moreover, in all surveys, spectra can be co-added for vast numbers of galaxies, grouped by other properties such as mass, star formation rate, excitation, and local environment density, in order to measure still weaker features, such as the auroral lines (e.g., [OIII] 4363, [OII] 7320, 7330) needed for direct electron temperature and metallicity determinations. The ATLAS Medium and Deep Surveys will measure stellar continuum and absorption lines for faint galaxies (individually and in stacked bins), providing diagnostics of stellar population ages, star formation histories, and chemical abundances. Observations of these features from terrestrial telescopes are limited by the Earth's atmosphere, leading to highly incomplete and non-continuous spectral coverage and redshift range. JWST will obtain outstanding near-infrared spectra for faint galaxies, but only ATLAS, with its wide field and high multiplex, and its dedicated use for large surveys, can obtain spectra for hundreds of millions of galaxies over large cosmic volumes, needed to tie the evolving properties of galaxies to the context of their environments.

\subsection{Black Holes, AGN, and Feedback}

The distribution function of galaxy stellar masses is strikingly different from that of dark matter halos. Its shape implies a characteristic mass at which galaxies most effectively convert gas into stars. At lower and higher masses, "feedback" is invoked to prevent gas from cooling onto galaxies, or to expel gas, thus suppressing star formation efficiency. The most massive galaxies today are predominantly quiescent, with little active star formation, and with dominant bulges that universally contain supermassive black holes. These, when fueled, can power active nuclei which may dump tremendous energy into their environments. AGN feedback, perhaps coupled with environmental effects, may be the dominant process regulating star formation and growth for massive galaxies. Like cosmic star formation, the luminous output of AGN and (by inference) the black hole growth rate, peaked at high redshift. ATLAS spectroscopy at 1-4 $\mu$m will be uniquely suited to identifying vast samples of AGN over a wide range of luminosity and redshift, using standard nebular excitation diagnostics (e.g., \citealt{Baldwin81}) and through detection of high-ionization emission lines (e.g., [NeV], HeII, CIV, NV). 
ATLAS will connect AGN activity to local and large-scale environment with exquisite statistical accuracy that is only possible today in the local universe, and [OIII] luminosities will provide a critical measure of the distribution of accretion luminosities and black hole growth rates. ATLAS spectroscopy will be highly complementary to surveys with Athena, the next major X-ray observatory, and uniquely suitable for redshift measurement of the most distant AGN candidates identified by Euclid, WFIRST or Athena. 

ATLAS will also enable spectroscopic redshifts to be measured for the majority of radio sources detected in wide-area surveys with the ASKAP such as EMU \citep{Norris11}. EMU will detect tens of millions of star-forming galaxies out to $z\sim3$ and several million quasars out to the earliest cosmic times. Although photometric redshifts of many of these galaxies will be possible through a combination of LSST and Euclid data, characterization of these objects, including metallicities and black hole masses and their location within the cosmic web, will require spectroscopic data which ATLAS will be able to measure.

\subsection{Reionization and Cosmic Structure}
\label{sec:reionization}

Current observations indicate that the intergalactic medium (IGM) completed its transition from neutral to ionized by $z=6.5$. This process is poorly understood: it is usually presumed that star-forming galaxies were responsible, but there is little evidence that sufficient ionizing radiation escapes from early galaxies to accomplish this.  Reionization may have been highly inhomogeneous as well, with expanding bubbles driven by strongly clustered young galaxies that are highly biased tracers of dark matter structure. In coming decades, new radio facilities (LOFAR, HERA, MeerKAT, SKA) will map (at least statistically) the distribution of neutral hydrogen in the epoch of reionization; ATLAS will provide essential complementary information about the spatial distribution of the (potentially) ionizing galaxies themselves over the same sky areas and redshift ranges. This requires accurate spectroscopy deep enough to detect $z=7$ galaxies over very wide sky areas --- exactly what ATLAS will deliver. At $5 < z < 7$, ATLAS will characterize the clustering of early galaxies with hard ionizing spectra that produce [OIII] emission, a signature of the low-metallicity population that may be the main driver of IGM reionization. While the redshift range falls mostly below the end of reionization, an accurate measurement of 3D clustering will strongly constrain theoretical models that can then be extrapolated to higher redshifts. At $z > 7.2$, ATLAS can observe Lyman $\alpha$. There is already evidence that Ly$\alpha$ may be inhomogeneously suppressed by the neutral IGM at those redshifts (e.g., \cite{Tilvi14}), and that its escape may correlate with galaxy overdensities that can more effectively ionize large IGM volumes \citep{Castellano16}. The ATLAS Medium and Deep Surveys can target galaxies at $z \simeq 7-8$ selected from deep WFIRST Guest Observer science programs over many square degrees, and measure (or set severe limits on) Ly$\alpha$ emission for statistical correlation with 21 cm surveys of the neutral IGM in the same volumes. JWST cannot survey areas wide enough to correlate large-scale structure with HI surveys, while the WFIRST slitless spectroscopy does not have suitable sensitivity for this. The ATLAS Wide Survey may also discover hundreds of examples of exceedingly rare, exceedingly luminous Ly$\alpha$ galaxies like "CR7" \citep{Sobral15}, which may also exhibit HeII 1640 emission, a signature expected from ionization by primordial Population III stars (but see \citealt{Bowler2017} and Sobral et al.\ (submitted)).

\subsection{The Circumgalactic Medium (CGM)}

The CGM is a key, but poorly understood component of the cosmic web. Containing the
bulk of baryons, it is the gas reservoir that cools to form stars and which is heated and enriched by feedback.
The nature of the CGM (density, temperature, metallicity) as a function of galaxy properties is crucial to
understanding galaxy formation, especially in the phase of rapid growth at $z > 2$. The high source density and
wide spectral band of ATLAS offers a powerful probe of the CGM at these redshifts. By stacking spectra of many
background galaxies, we can measure average CGM absorption around foreground galaxies with known redshifts as a
function of impact parameter, mass, star formation rate, metallicity, local environment, and other properties. In
the ATLAS Wide Survey, we can measure equivalent widths of 0.005\AA $\,$(3$\sigma$) for Ca H\&K absorptions (using
background galaxies at $z>2$) and 0.010\AA $\,$ for MgII (at $z>2.6$). In addition, we can reach a 5$\sigma$ sensitivity of
8$\times 10^{-22}$ erg$\,$ s$^{-1}$cm$^{-2}$ arcsec$^{-2}$ for detecting H$\alpha$ emission from gaseous halos at
$z\sim$2, probing the CGM to radii $>100$ kpc.

\subsection{Protoclusters}

ATLAS Medium Survey will have the survey sensitivity and volume to discover forming clusters and protoclusters, all the way from $z=2$ to 6 and thus to unveil the crucial phases of cluster formation and galaxy formation in cluster environments. 
Based on cosmological N-body simulations, cluster at $z>2$ are extended enough to utilize the wide-field high spectroscopic multiplex capability of ATLAS (see e.g., \citealt{Orsi16}). 
In addition, the ATLAS spectral resolution of R=1000 is sufficient for obtaining useful measurements of cluster galaxy velocities to probe cluster dynamics.
ATLAS will provide a unique probe of clusters and their cosmic web environments.

The first collapsed structures hosted within a single dark matter halo with masses of few times of 10$^{13}$ $M_\odot$ are expected to be in place in reasonable numbers at $z \sim$ 2-3, with space densities ranging from 10$^{-5}$ to $10^{-6}\,$Mpc$^{-3}$, depending on the exact mass cut, corresponding to several per square degree over $2<z<3$ (so on average one in each ATLAS field of view). ATLAS will be able to obtain in depth spectroscopic identifications of a dozen members down to low levels in the mass function in less than 1 hour observations (line fluxes down to few 10$^{-18}$ erg/s/cm$^2$). Many hundreds will be found, for example, over a systematic survey of 100 square degrees by ATLAS Medium. Pointed observations of similar objects found by Euclid/Athena and/or other facilities will be very efficient. Observations of the first structures ($z=1.99$: \citealt{Gobat13}; $z=2.50$: \citealt{WangT16}) show that these objects host spectacular action: from strong star formation activity to quenching, from AGN feedback to the morphological transformation of galaxies and the formation of ellipticals. At those redshifts, the AGN activity will become much more prominent, following in parallel the rise of SFR and gas content in the Universe. In these conditions, interactions between the ICM and the cluster's environment through the inflow of pristine cold gas, the outflows originated from stellar and AGN's winds, and the deposition of warm plasma will shape the baryon distribution, the metal content and the energy budget of the structures (e.g., \citealt{Valentino15,Valentino16}).

At higher redshifts,  the first lower mass group-like halos (masses of 10$^{13}\,M_\odot$ or below) as well as protoclusters (larger scale overdensities not yet collapsed) will be within reach of spectroscopic identification by ATLAS. Groups of $10^{13}\, M_\odot$ have an expected density of 1 per tens of ATLAS fields at $5<z<7$, but several dozens could be found over the 100 square degree ATLAS Medium Survey. A basic characterization with identification of most massive galaxies (few $10^9 \,M_\odot$) will require a few hours of integration. Detailed characterization down the stellar mass function will require many dozens hours of observations with ATLAS (fluxes of a few $10^{-19}$ erg/s/cm$^2$). Observing such objects to $z\sim 6$ will be key to explore the early phases of environmental effects on galaxy formation and evolution as well as the beginning of structure formation (see e.g., \citealt{Orsi16}, and \citealt{Izquierdo17}). 

\subsection{Galaxy Kinematics}

With the spectral resolution of $R$=1000, ATLAS Probe will allow us to obtain accurate kinematic measurements that will be essential in several galaxy evolution cases. For example:  
\begin{itemize}
\item{Absorption line velocity dispersions, which enable the measurement of scaling relations and, in combination with galaxy sizes estimated from imaging,
dynamical masses of spheroidal/quiescent galaxies. \citet{Belli17} and \citet{VanDokkum09} have used such measurements to demonstrate the evolution towards higher stellar velocity dispersion at fixed mass at the highest redshifts where such measurements have been performed to date (i.e., $z\sim 2$).}

\item{Emission-line virial kinematics, which can be interpreted in concert with WFIRST imaging and structural properties to infer dynamical masses \citep{Price16,Alcorn18}. Such dynamical masses can be compared with inferred baryonic
masses to probe the amount of dark matter within galaxy effective radii. It will be possible to infer baryonic masses from the sum of gas masses estimated from dust-corrected star-formation rates, assuming the Schmidt-Kennicutt relation, and stellar masses estimated from WFIRST galaxy photometry. At the same time, such comparisons place constraints on the allowed range of stellar initial mass functions \citep{Price16}.}

\item{Velocity shifts of interstellar absorption lines (e.g., Na I 5890,5896) relative to galaxy systemic velocity, to track outflows/inflows. Rest-optical nebular emission lines such as H$\alpha$ and [OIII] 5007 can be used to establish the galaxy systemic velocity, as they are dominated by the surface-brightness weighted ensemble-averaged emission from H~II regions in galaxies. Over a wide range of redshifts, kinematic evidence for gas flows around galaxies has been assembled based on the small differences in redshift measured for interstellar absorption features relative to the systemic redshift. Most kinematic evidence of this type suggests outflows \citep[e.g.,][]{Shapley03,Steidel10,Weiner09,Chen10}, but there are rare instances of kinematic evidence of infalling gas \citep[][]{Rubin12,Martin12}.}

\item{Probe the dark matter content of disk galaxies as a function of redshift for $0.5<z\la 7$, by obtaining galaxy rotation curves from stacking thousands of spectra of galaxies with similar colors in redshift bins, scaled using stellar disk sizes measured by WFIRST. 
This approach is independent of rotation curve models, and shown to be unbiased \citep{Tiley19}.
We will use WFIRST imaging to select nearly-edge-on disk galaxies at $0.5<z\la 7$, and obtain several spectra along the major axis out to a few arsec for each galaxy. This will allow us to construct a rotation curve for each galaxy. The spectra from many galaxies, after being scaled using the observed disk size, can be stacked to obtain a well sampled rotation curve \citep{Tiley19}. ATLAS Probe can obtain disk galaxy rotation curves out to z=7 this way, since it can detect both [OIII] and [OII] lines (both strong emission lines) out to z=7. At the highest redshifts (z$\gg$ 3), the very compact sizes of galaxies  will likely make the measurement of galaxy rotation curves challenging, but ATLAS Probe measurements will at least set interesting limits. 
ATLAS complements the ground-based projects by extending the redshift reach of rotation curve measurements to high redshifts not possible from the ground.}
\end{itemize}

In addition, ATLAS Probe galaxy kinematic measurements can be used to study galaxy pairs and the evolution of merger fraction and merger rate,  measure virial masses of clusters and groups based on galaxy velocity dispersion, and better deblend spectral lines, leading to better accuracy in the measurement of equivalent widths, metal abundances (e.g. ISM absorptions), stellar lines, etc.

\subsection{Physics Enabled by Emission Lines}

\begin{figure}
\centering
\includegraphics[width=1\columnwidth,clip]{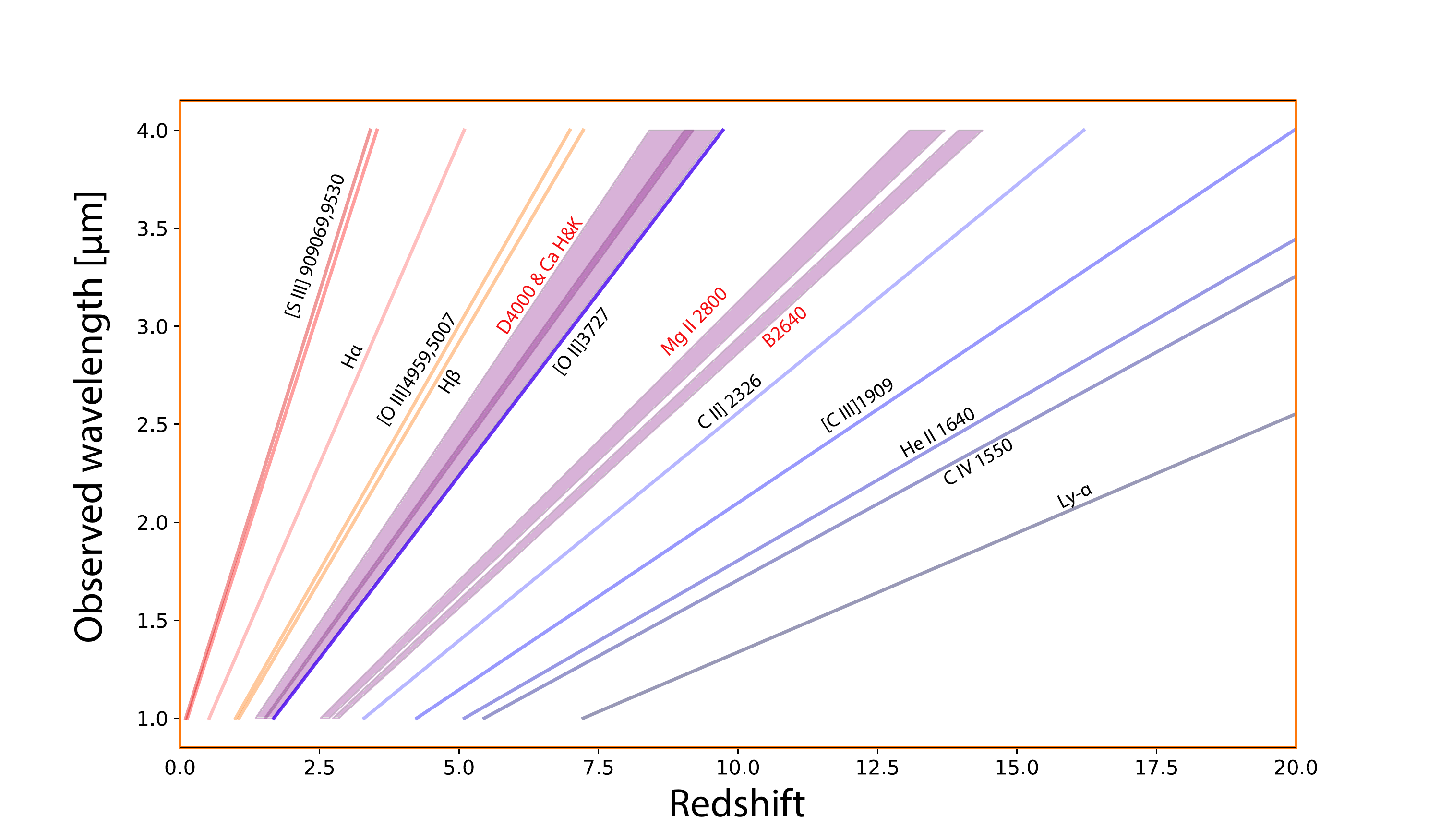}
\caption{The wavelength at which different spectroscopic features fall as a function of redshift, for the ATLAS wavelength range of 1-4$\mu$m. The thinner lines denotes emission lines as indicated in the figure. Also indicated with broader bands are the D4000 and B2640 absorption features as well as the region including the Mg \textsc{ii} absorption lines. 
}
\label{fig:lines}
\end{figure}

A wide range of physical information is encoded in a galaxy's spectrum. The main strong optical lines are particularly well-studied and start to shift into the observer frame of ATLAS at $z > 0.5$. Figure \ref{fig:lines} shows when various key diagnostic features become accessible to ATLAS with emission lines shown as lines and absorption features as shaded regions. 

The optical emission lines ([N II]6548,6584, H$\alpha$, [S II] 6717,6731, H$\beta$, [O III]4959,5007 and [O II]3727 in particular) have a long history of being used to estimate star formation rates, gas-phase metallicities and ionization conditions.  At $1.7 < z < 5.0$, all of these strong, optical rest-frame lines are accessible to ATLAS.  Metallicity and ionization conditions can be studied in large samples of galaxies over a wide range of cosmic epoch, drastically reducing the current systematic uncertainties that are caused by patching together different indicators at different redshifts.  The hardness of the ionizing spectrum can be traced by weaker optical lines, such as [O I]6300, while in the UV the [C III]1909, [OIII] 1661,1666, and C IV 1548,1550 doublets and the He II 1640 line have been shown to be present frequently in low-mass star-forming galaxies with low metallicity/high redshift (e.g. \citealt{Senchyna17}; \citealt{Stark15}) but they are also prominent tracers of AGN activity \citep{Feltre16}. 

Finally, the Ly$\alpha$ line offers a potential cornucopia of information on the properties of galaxies. It is related to the star formation rate in galaxies, modulated by resonant scattering and dust absorption through the neutral interstellar and circumgalactic medium. The shape of the Ly$\alpha$ profile can be used to constrain the distribution and kinematics of outflowing and inflowing gas around galaxies \citep{Verhamme06,Steidel10}. Recent research on low-redshift galaxies with Lyman continuum escape has shown that the separation of the blue and red peaks in the emission line also correlate with the escape fraction of ionising photons, $f_{\mathrm{esc}}$ (\citealt{Verhamme17}; \citealt{Izotov17}). At $f_{\mathrm{esc}}>0.05$ the typical separation is $<400$km/s which is marginally resolvable by ATLAS with $R=1000$. When combined with a study of the systematic changes in the overall line profiles this offers ATLAS the possibility to trace changes in the escape fraction with redshift at the very highest redshifts. 

\newpage

\section{Cosmology}
\label{sec:cosmology}
ATLAS Science Objective 2, ``Measure the mass of dark-matter-dominated filaments in the cosmic web on the scales of $\sim$ 5-50$\,$Mpc$\,h^{-1}$ over 2,000 deg$^2$ at $0.5<z<3$," flows down to the requirement for the Wide Survey, which obtains spectroscopic redshifts for 70\% of the galaxies in the 2,000 deg$^2$ WFIRST High Latitude Survey (HLS) Weak Lensing (WL) sample at $0.5<z<4$. The WFIRST WL sample has a continuum depth of $H=24.7$ (the shape measurement limit), a redshift range of $0<z<4$, and a galaxy number density of 44.8/(arcmin)$^2$ (combining the filters). 80\% of these galaxies are at $z>0.5$, of which 87\% have emission line fluxes $f_{\rm line}>5\times 10^{-18}$erg$\,$ s$^{-1}$cm$^{-2}$ (see Fig.\ref{fig:emission-line} in Sec.\ref{sec:source}), setting the 5$\sigma$ depth of the ATLAS Wide Survey, which will require 1.6 years of observing time, and obtain 183M galaxy spectra (an order of magnitude larger than that from Euclid or WFIRST), with a completeness of 70\% (percentage of the WFIRST WL sample at $z>0.5$ with $S/N>5$ spectra). Note that the continuum limit of ATLAS Wide is only AB=23.0 (3$\sigma$), since we are targeting emission line galaxies of the WFIRST WL sample. A significant fraction of the galaxies with spectra from ATLAS Wide will be fainter than AB=23.0. The left panel of Fig.\ref{fig:atlas_wide} shows the expected redshift distribution of galaxies from the ATLAS Wide Survey. 
These galaxies trace the cosmic web densely enough on scales $\sim$5-50$\,$Mpc$\,h^{-1}$ at $0.5<z<3$ (see Fig.\ref{fig:atlas_wide}, right panel),  so that their ellipticities can be stacked for filament detections \citep{Epps17}.
The number of galaxies at $z>3$ may be too few for the detection of the dark-matter-dominated filaments. 

ATLAS Probe and WFIRST together will produce a 3D map of the Universe with $\sim$Mpc resolution in redshift space over 2,000 deg$^2$, opening a new window into the dark Universe by weighing the dark matter dominated filaments in the cosmic web, and precisely measuring the cosmic expansion history and growth rate of large scale structure for redshifts of 0.5 to 3.5.  Although the requirements for the ATLAS Wide Survey is driven by the detection of dark-matter-dominated filaments connecting massive galaxies, they lead to a spectacular data set for other cosmological studies as well. 

\subsection{Weighing Dark-Matter-Dominated Filaments in the Cosmic Web}

\begin{figure}
\centering
\includegraphics[width=0.49\columnwidth,clip]{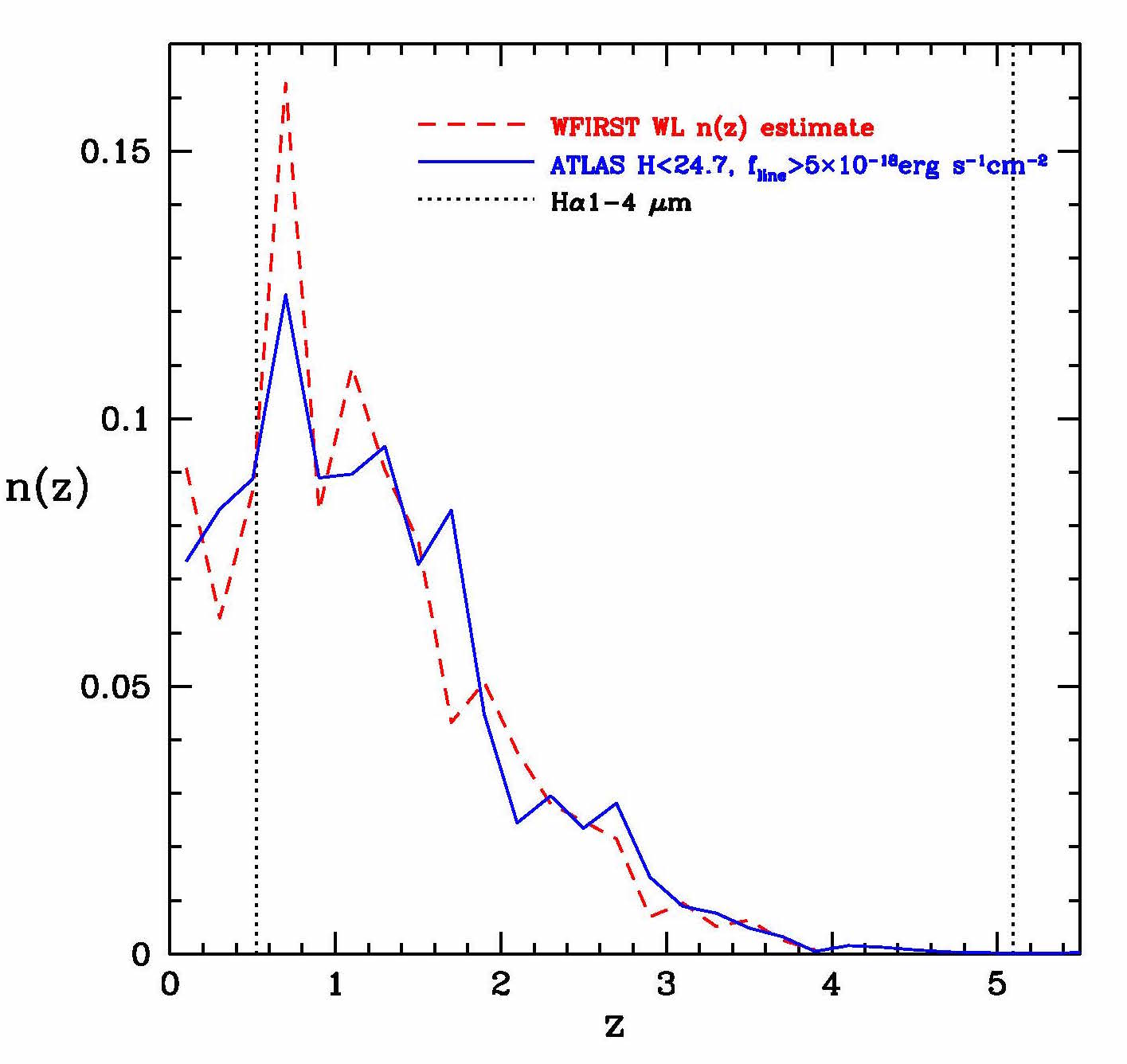}
\includegraphics[width=0.49\columnwidth,clip]{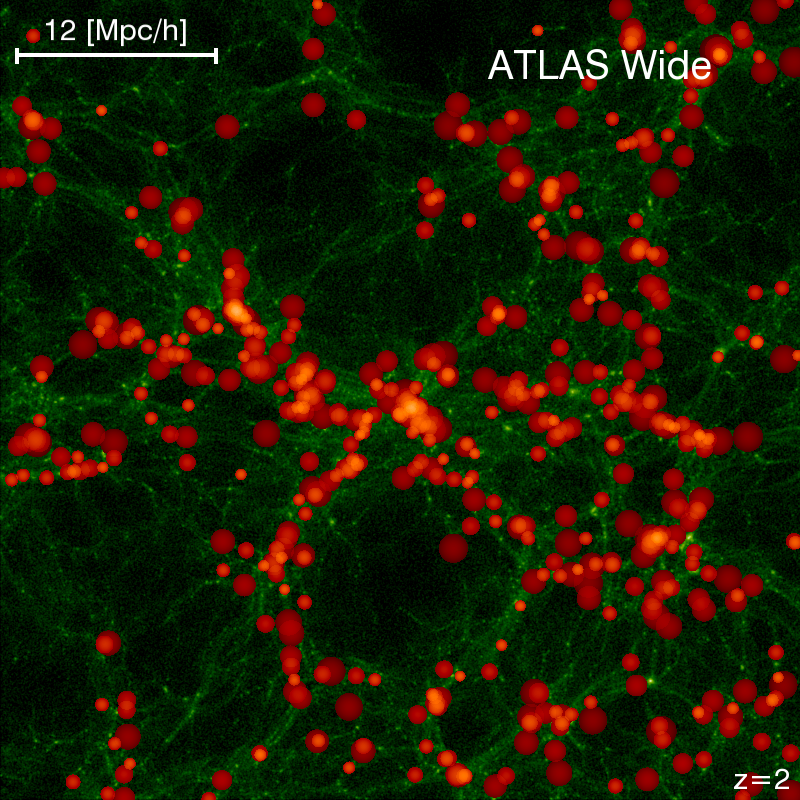}
\caption{Left panel: Expected redshift distribution of galaxies from the ATLAS Wide Survey.
Right panel: cosmic web of dark matter (green) at z=2 traced by galaxies from ATLAS Wide Survey (red). See Appendix B for details on the visualization.}
\label{fig:atlas_wide}
\end{figure}

A key prediction of the cold dark matter model is the existence of the "cosmic web", a network of low-density filaments connecting dark matter halos.
{\it Spectroscopic redshifts} are required to locate the galaxy groups and clusters that are connected by filaments in redshift space.
The uncertainty associated with photometric redshifts scatters the true physical pairs and smears the filamentary structure (see
Fig.\ref{fig:atlas_cosmic-web}).
\cite{Epps17} recently demonstrated that the stacked dark-matter-dominated filament connecting pairs of massive galaxies can be detected when spectroscopic redshifts along with weak lensing data are available for most of the source galaxies lensed by the massive galaxy pairs. 
ATLAS Probe spectroscopy and WFIRST imaging together provide the ideal data set for such detections.
The detection of these filaments by ATLAS Probe over a significant cosmic volume will test the cold dark matter model for structure formation, 
and provide key insight into large-scale structure in the Universe.

\cite{Epps17} used the overlap in the CFHTLenS imaging and BOSS spectroscopic data over 105 deg$^2$ \citep{Miyatake15} to select a sample of $\sim$ 20,400 luminous red galaxies (LRGs), from which they constructed a catalog of LRG pairs by selecting pairs that were separated by $\Delta z_{\rm spec} < 0.002$ ($\sim \, 5\, h^{-1}$Mpc in comoving separation) and by 6 to 10 $h^{-1}$Mpc in the transverse direction. This yielded a sample of $\sim$ 23,000 pairs of LRGs, with a mean separation of $\sim$ 8.23 $h^{-1}$Mpc at a mean redshift of 0.42, and a mean stellar mass of 10$^{11.3}M_\odot$ (with the expected halo mass of $10^{13.04}M_\odot$, corresponding to galaxy groups). 
\cite{Epps17} constructed shear and convergence maps by stacking all the LRG pairs in their sample, and obtained a $5\sigma$ detection of the stacked filament connecting the LRG pairs.

\begin{figure}
\centering
\includegraphics[width=0.6\columnwidth,clip]{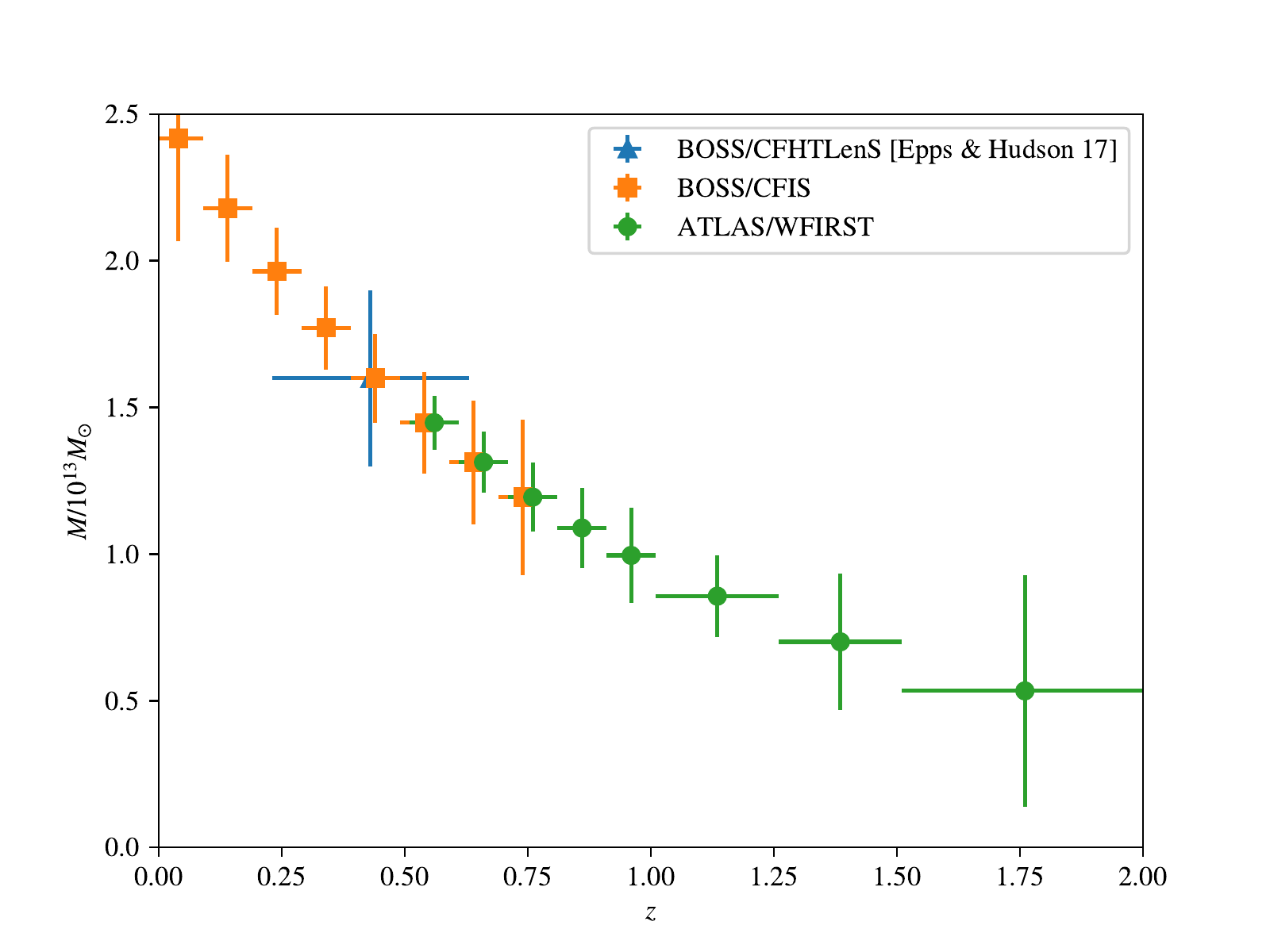}
\caption{Conservative estimate of the measurement errors of filament mass from ATLAS Wide, for a 6-10 Mpc/$h$ cosmic web filament between $10^{13}\, M_{\odot}$ halos. See text for details.}
\label{fig:filament}
\end{figure}

The Canada-France Imaging Survey\footnote{http://www.cfht.hawaii.edu/Science/CFIS/}  \citep[CFIS;][]{Ibata17}  is a CFHT Large Program that has been allocated 271 nights from 2017 to 2021. It will provide imaging to complement the spectroscopy from BOSS and DESI, to enable dark matter filament measurements over a wide sky area at $z \lesssim 0.8$. ATLAS Probe complements this by enabling dark matter filament measurements over 2,000 deg$^2$ at $0.5<z<3.0$, with much higher galaxy number densities (see Fig.\ref{fig:atlas_wide}).

Fig.\ref{fig:filament} shows the estimated uncertainties of filament mass measurements for slices in redshift $z$ of width indicated by the horizontal error bars (so averaging over larger redshift range at higher $z$). The large blue cross is the CFHTLenS result from \cite{Epps17}. Orange symbols at low $z$ are forecasts for CFIS (imaging) with BOSS (spectroscopy),  and green represents conservative forecasts for ATLAS Probe (spectroscopy) with WFIRST (imaging). Note that the signal drops at high $z$ due to structure growth and the noise increases at high $z$ due to lack of source galaxies behind the higher lens redshifts.

We have assumed that filaments with mass $\sim$ 1.5$\times 10^{13}\, M_\odot$ connect halos that are $\sim$ $10^{13} \,M_\odot$ at $z \sim 0.4$, and have a number density of $\sim 3\times 10^{-4} (h/\mbox{Mpc})^3$ (BOSS-LRG-like in density but the halos do not have to be passive, just massive).
To derive galaxy bias $b(z)$, we assumed constant clustering, i.e., $b(z)G(z)$ is constant, where $G(z)$ is the linear growth factor. Since the galaxies have $b=2$ at $z \sim 0.43$, we find $b(z)=b(0.43)G(0.43)/G(z)$. The filament mass is assumed to scale with redshift like the three point correlation function.

The mass estimate uncertainties in Fig.\ref{fig:filament} are conservative because it should also be possible to detect filaments between lower mass halos. In future work, we will carry out calculations based on realistic simulations to increase the fidelity of our estimates, and study the measurement of filament mass at higher redshifts.

The massive galaxies that mark the galaxy groups can be identified through color selection in the WFIRST HLS imaging data. We expect that $\sim$ 70\% of the source galaxy redshifts will be measured by ATLAS Wide. The lens galaxies are expected to be massive galaxies, some of which may be passive ellipticals with very low emission line flux. However, these luminous red galaxies (LRGs) are very bright. \cite{Zhai17} showed that the LRGs from eBOSS (0.6<z<0.9) all have $i \la 22$, with the majority at $i <21$, with very mild redshift evolution. Since ATLAS will reach a continuum depth of AB = 23 at 3$\sigma$ in 5000s (the exposure time per target for ATLAS Wide), we expect to obtain redshifts for nearly all of the LRGs used for the dark matter filament mass measurements.

The dark matter filament mass measurements by ATLAS Probe are synergistic with galaxy evolution studies using the three-tiered ATLAS Probe galaxy surveys. Using the GAMA spectroscopic survey in the redshift range $0.03\leq z\leq 0.25$, \cite{Kraljic17} have shown that the cosmic web plays an important role in shaping galaxy properties. Key galaxy properties (such as stellar mass, dust corrected color, and specific star-formation rate) are found to be correlated with galaxy distances to the 3D cosmic web features such as nodes, filaments, and walls. ATLAS Probe is unique in providing samples of different galaxy types, i.e., with different bias factors. These enable the ATLAS Probe galaxies to trace the cosmic web in unprecedented details and precision. ATLAS Probe measurements of dark matter filaments can be used to constrain models of the cosmic web, and enable the calibration of the reconstruction of the underlying dark matter distribution. This will enable transformative progress in our understanding of galaxy evolution and the nature of dark matter.

\subsection{Differentiating Dark Energy and Modification of General Relativity} 

The observed cosmic acceleration is one of the biggest unsolved mysteries in cosmology today \citep{Riess98,Perlmutter99}. We don't even know if it is due to an unknown energy component in the Universe (i.e., dark energy), or a modification of General Relativity (i.e., modified gravity). The measurement of both the cosmic expansion history and the growth history of cosmic large-scale structure is required to solve this mystery (see, e.g., \citet{Guzzo08,Wang08}). WL and galaxy clustering both probe cosmic expansion history, and probe different ways in which gravity can be modified. The ATLAS Wide Survey will provide the ultimate data set to remove systematic effects in our quest to discover the cause for cosmic acceleration.

\begin{figure}
\centering
\includegraphics[width=0.49\columnwidth,clip]{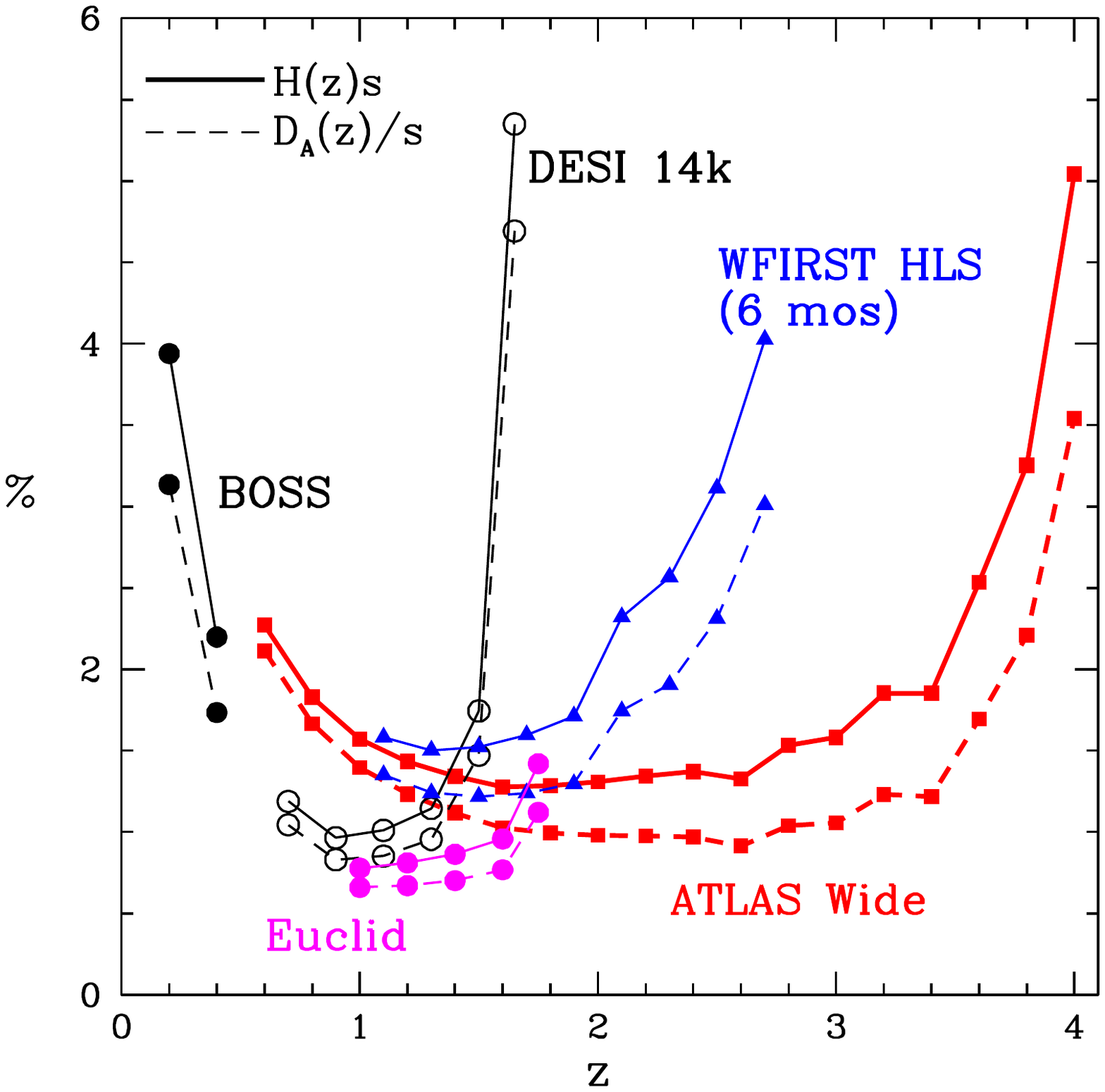}
\includegraphics[width=0.49\columnwidth,clip]{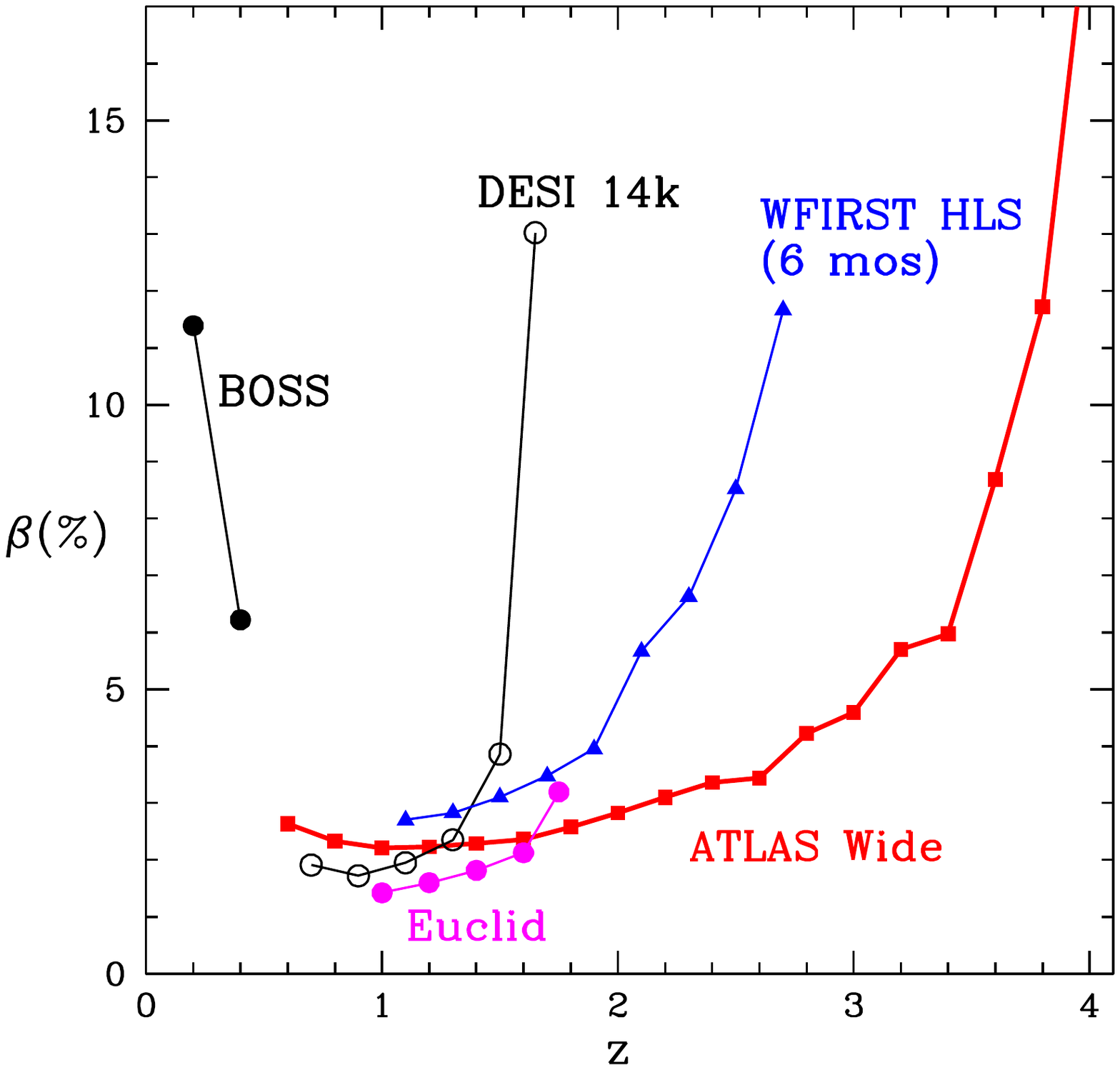}
\vspace{-1.2in}
\caption{ATLAS vs. WFIRST and Euclid comparison of the measurement errors from BAO/RSD on the cosmic expansion history $H(z)$, angular diameter distance $D_A(z)$, and the growth rate of cosmic large-scale structure $f_g(z)=\beta(z)b(z)$ (where $\beta(z)$ is the linear redshift-space distortion parameter, and b(z) is the bias factor). 
The BOSS survey and the future 14,000 deg$^2$ ELG survey from the ground-based project DESI \citep{DESI16} are also shown for comparison.
$H(z)$ and $D_A(z)$ are scaled with the BAO scale $s$ (the comoving sound horizon at the drag epoch), which is measured precisely by Planck.}
\label{fig:BAO/RSD}
\end{figure}

{\bf (i) Multi-Tracer and Ultra High Density BAO/RSD}. Galaxy clustering data from 3D distributions of galaxies is the most robust probe of cosmic acceleration. The baryon acoustic oscillation (BAO) measurements provide a direct measurement of cosmic expansion history $H(z)$ and angular diameter distance $D_A(z)$ \citep{Blake03,Seo03}, and the redshift-space distortions (RSD) enable measurement of the growth rate of cosmic large-scale structure $f_g(z)$ \citep{Guzzo08,Wang08}. The ATLAS Wide Survey will obtain spectroscopic redshifts of 183M galaxies over 2,000 deg$^2$, $\sim$ 20 times higher in galaxy number density compared to WFIRST. These provide multiple galaxy tracers of BAO/RSD (red galaxies, different emission-line selected galaxies, and WL shear selected galaxies) over $0.5<z<4$, with each at high number densities. These enable robust modeling of BAO/RSD (e.g., the removal of the nonlinear effects via the reconstruction of the linear density field), and significantly tightens constraints on dark energy and modified gravity by evading the cosmic variance when used as multi-tracers \citep{McDonald09}. ATLAS measures $H(z)$, $D_A(z)$, and $f_g(z)$ over the wide redshift range of $0.5< z < 4$ (see Fig.\ref{fig:BAO/RSD}), with high precision over $0.5<z<3.5$, improving the constraints from WFIRST by a factor of three or more at $2<z<3$ (beyond the reach of Euclid), and extending constraints to $3<z<4$, beyond the reach of both WFIRST and Euclid. The 14,000 deg$^2$ ELG survey from the ground-based project DESI \citep{DESI16} is complementary to the space-based surveys, as shown in Fig.\ref{fig:BAO/RSD}. 

We have followed the methodology of \cite{Wang13} in deriving the constraints shown in 
Fig.\ref{fig:BAO/RSD}, with 100\% nonlinear effects, a cutoff of $k_{max}=0.25h/$Mpc,
and peculiar dispersion of 290$\,$km/s. We assumed a bias function $b(z)=0.84G(0)/G(z)$, following the fitting of the \cite{Mostek13} results from \cite{DESI16}, where $G(z)$ is the growth factor. This corresponds to constant clustering. It is appropriate for the majority of the WFIRST WL source galaxies, most of which are emission line objects at some level (though the lines are fainter than the ones used for slitless spectroscopy on Euclid and WFIRST).
For the slitless surveys, we assumed an efficiency of 75\% for WFIRST, and 40\% for Euclid, since the signal-to-noise thresholds for WFIRST and Euclid are 7$\sigma$ and 3.5$\sigma$ respectively. The galaxy number density for ATLAS is shown in the left panel of Fig.\ref{fig:atlas_wide} (normalized for a total of 183M galaxies). For WFIRST and Euclid, we use the forecasts shown in \cite{Merson17}. The galaxy number density for DESI is from \cite{DESI16}.

{\bf (ii) 3D WL with Spectroscopic Redshifts}. The ATLAS Wide Survey replaces photometric redshifts with spectroscopic redshifts for $\sim$ 70\% of the lensed galaxies in the WFIRST HLS WL sample. This eliminates the photo-z calibration ladder as a major source of systematic uncertainty in WL. Furthermore, one can identify every pair of galaxies that are near each other in 3D space for 70\% of the sample, dramatically suppressing the systematic error from intrinsic galaxy alignments. This improves the measurement of the growth rate index (used to parametrize the gravity model) by $\sim$ 50\%. In future work, we will investigate the cosmological constraints from the ATLAS spectroscopic subsample of the WFIRST WL sample.

{\bf (iii) Joint Analysis of Weak Lensing and Galaxy Clustering}. Less than 10\% of the galaxies in the WFIRST WL sample will have WFIRST spectroscopy; $\sim$ 70\% will have ATLAS spectroscopy from the Wide Survey at $z>0.5$. This super data set of both WL shear and 3D galaxy clustering data for the same 183M galaxies over $0.5<z<4$ enables a straightforward joint analysis of the data. This facilitates the precise modeling of the bias between galaxies and matter $b(z)$, and provides the ultimate measurement of $H(z)$ and $f_g(z)$. We expect this to result in definitive measurements of dark energy and modified gravity for $0.5<z<4$, with the removal of photo-z errors as the primary systematic for weak lensing, and detailed modeling of bias systematic for BAO/RSD.

{\bf (iv) Type Ia Supernovae (SNe Ia)}. 
ATLAS Probe will obtain the host galaxy redshifts of nearly all 30,000 SNe Ia that will have
lightcurves measured by the LSST DDF and WFIRST SN surveys, over the redshift range of $0.2 < z <  2.0$. This will provide a powerful measurement of $H(z)$ that is independent of cosmic large scale structure.  The WFIRST and LSST SN surveys will cover $\sim40$ sq. deg. each.  There are currently no plans in place to obtain host galaxy redshifts for LSST SNe with $z>0.5$.  For WFIRST, there may be an Integrated Field Channel on board to obtain spectra of SNe and hosts; however, for $z>1.0$, the number of spectra acquired will be $\sim 10\%$ of the full discovery sample \citep{Hounsell17}.  Other plans, like a joint effort with Prime Focus Spectrograph will obtain host galaxy redshifts to $z\sim1.0$, but a small and largely biased sample for $z>1.0$.  Multiple studies (e.g. \cite{Sullivan10}) have shown that there is a relation between SN luminosity and host galaxy properties, so a large effort should be made to create an unbiased sample of host galaxies to as high a redshift as possible.  

With host galaxy redshifts up to z=2.0 from ATLAS Probe, the statistical distance precision of WFIRST and LSST SNe should be sub-1\% out to z=2.0 \citep{Hounsell17}.  This is comparable to the constraints shown in Fig.\ref{fig:BAO/RSD} from the ATLAS measurement errors from BAO/RSD.  Without the host galaxy redshifts, SN analyses will be forced to rely on photometric redshifts, which for the full SN sample degrades the cosmological precision by introducing a series of systematic uncertainties and reducing the statistical precision as well.

{\bf (v) Clusters}. 
The abundance of mass-calibrated galaxy clusters provides a complementary measurement of cosmic expansion history and growth rate of large-scale structure. In the ATLAS Wide Survey, we will carry out spectroscopy for the 40,000 clusters with $M>10^{14}M_\odot$ expected to be found by WFIRST HLS imaging \citep{Spergel15}. The ATLAS/WFIRST cluster sample will have mass accurately measured by the deep 3D WL data, and cluster membership precisely determined by spectroscopic redshifts. This will provide a powerful crosscheck to constraints from BAO/RSD and 3D WL, and a unique data set of spectroscopic clusters for studying cluster astrophysics.

{\bf (vi) Higher-order Statistics.}
The very high number density galaxy samples from ATLAS Probe provides the ideal data set for studying higher-order statistics of galaxy clustering, which will boost the power to probe dark energy and gravity by a significant factor. 
While the use of galaxy clustering 2-point statistics is now standard in cosmology, the use of the 3-point statistics is still limited due to a number of technical challenges. Since the galaxy 3-point statistics provides additional information to that from the galaxy 2-point statistics, the combination of these is needed to optimally extract the cosmological information from galaxy clustering data (see, e.g., \cite{Gagrani17}).

{\bf (vii) Calibration of Photometric Redshifts.}
In addition to providing redshifts for $\sim$70\% of the WFIRST HLS lensing sample at $z>0.5$, ATLAS should greatly improve photometric redshifts for those WFIRST galaxies which it does not obtain a robust redshift determination for.  This will be true for two reasons.  First, the redshifts ATLAS obtains will all be helpful for improving our knowledge of the relationship between galaxy colors/magnitudes and redshift (or, equivalently, for constraining models of galaxy SED evolution); the Medium survey should be particularly well-suited for this, especially if the survey area is distributed amongst multiple fields to mitigate the effects of sample/cosmic variance (in contrast, variance in the 1 sq. deg. deep field will be sufficient to limit its use for this purpose).  If the survey is appropriately designed, ATLAS should greatly exceed the requirements for a Stage IV photometric redshift training sample established in \cite{Newman15}.  

Second, the large sample of galaxy redshifts over a large volume provided by the ATLAS Wide survey would provide a high-precision {\it calibration} of photometric redshifts, even in the presence of systematic incompleteness, via cross-correlation between ATLAS galaxies with secure redshifts and photometric-only samples \citep{Newman08}.  Gains would be particularly great at $z\ga 1.5$, where the DESI and 4MOST samples will be limited to QSOs (and hence relatively sparse).  The resulting calibration would ensure that accurate dark energy inference can be done even if spectroscopic redshift samples are systematically incomplete.
We note that not only WFIRST but other Stage IV (or future) photometric dark energy experiments (e.g., LSST and the weak lensing imaging survey on Euclid) would benefit from the improved photometric redshift training and calibration provided by ATLAS.

\subsection{Particle Cosmology: Inflationary Physics, Neutrino Mass, and Particle Dark Matter}

{\bf Probe Inflationary Physics}. Current observational data are consistent with inflation having occurred when the Universe was ~10$^{-34}$s old. However, we have very little insight on the physics of inflation. While the detection of primordial gravity waves by future CMB observations would provide the definitive proof that inflation occurred, the combination of CMB data from Planck and the ultra high precision matter power spectrum $P(k,z)$ (with $0.5<z<4$) measured from the ATLAS Wide Survey will characterize the primordial matter power spectrum $P(k)_{\rm in}$ with definitive precision and accuracy. The measured $P(k)_{\rm in}$ allows us to probe inflationary physics, e.g., whether inflation occurred in one stage, or in multiple stages, and whether inflation occurred smoothly, or in spikes \citep{Wang99}. These help determine the particle physics required for inflation, e.g., properties of the inflaton, including its coupling to other particles, which will help in its identification. We expect ATLAS to provide an order of magnitude improvement in measuring $P(k)_{\rm in}$ over other planned galaxy redshift surveys due to its extremely high galaxy number density over a broad redshift coverage, opening up in particular the high redshift window where the nonlinearities are less important and the modeling more robust.

{\bf Measure the Total Mass of Neutrinos}. Solar and atmospheric neutrino data indicate that neutrinos have a small but nonzero mass (see, e.g., \cite{Lesgourgues06}). This has profound implications for particle physics and cosmology. Measuring the tiny mass of neutrinos is a major enterprise in experimental particle physics, but cosmological observations will likely provide the most stringent constraints on the total mass of neutrino species, and possibly be sensitive enough to the mass hierarchy \citep{Carbone11}. By covering $\lambda>2\mu$m and obtaining spectra for faint galaxies, the ATLAS galaxy surveys probe structures at early times and is not limited to the most massive/biased objects. ATLAS thus has the potential to be the least affected by systematic errors in modeling nonlinear structure evolution, allowing us to measure $P(k,z)$ ($0.5<z<4$) down to very small scales. This will allow us to measure the total mass of neutrinos to high precision, as the effect of massive neutrinos increases at small scales. The ATLAS Medium Survey will provide a unique new angle to measure neutrino mass by providing precise measurement of galaxy clustering on the scales of 1-50 $h^{-1}$Mpc for $0.5<z<7$, with sub-percent measurement of the galaxy correlation function on 1-20 $h^{-1}$Mpc. This will help tighten the ATLAS neutrino mass measurement by probing the effect of massive neutrinos to even earlier times. 
If the neutrino mass $m_\nu \geq 0.06\,$eV, ATLAS Probe will be able to detect it with a significance exceeding 3$\sigma$ (Ballardini et al., in preparation). Because ATLAS Probe will have the tightest possible control of systematic uncertainties in both galaxy clustering and weak lensing, we expect it to lead to the most accurate measurement of $m_\nu$ within the next few decades. In particular, ATLAS Probe data will enable the modeling of galaxy bias on small scales via the joint analysis of 2-point galaxy statistics with 3-point galaxy statistics and weak lensing data, removing the main systematic uncertainty in the neutrino mass measurement. 

{\bf Constrain Axions as Dark Matter}. Axion is the leading alternative to WIMPs (weakly interacting massive particles) as a particle candidate for dark matter. Axions are motivated by known particle physics, as it solves the strong CP problem. If dark matter were made of axions, or a mixture of axions and WIMPS, they would give rise to specific features in the cosmic large scale structure \citep{Hlozek15,Cedeno17}, which can be measured to exquisite statistical precisions and accuracy using ATLAS Wide data. We will explore these constraints in future work.

\section{Milky Way Science}
\label{sec:Milky-Way}

ATLAS Science Objective 3, "Measure the dust-enshrouded 3D structure and stellar content of the Milky Way to a distance of 25kpc," flows down to a Galactic Plane Survey that obtains $SNR>30$ spectra for 95M sources with $AB<18.2\,$mag, covering 700 deg$^2$ in 0.4 years of observing time.
This is complemented by ATLAS Wide Survey, with its high Galactic latitude, to probe the low-mass stellar content of the Milky Way.
Within our Galaxy, ATLAS will advance our understanding of the structure, star-forming history, and content of the Milky Way.

In addition, the ATLAS Solar System Survey will cover a total of 1,200 deg$^2$ in 3000 fields of 0.4 deg$^2$ each, to obtain the spectra of KBOs. Since the ATLAS Probe can obtain 5,000-10,000 spectra simultaneously, this will enable a spectroscopic survey of stars to AB=22.53 at 3$\sigma$ (corresponding to AB=21.88 at 5$\sigma$, and AB=18.96 at 30$\sigma$), over 1,200 deg$^2$. This will provide a powerful data set for stellar astrophysics, to supplement the ATLAS Galactic Plane survey.

\subsection{The 3-D Structure of the Hidden Milky Way} 

\begin{figure}
\centering
\includegraphics[width=0.99\columnwidth,clip]{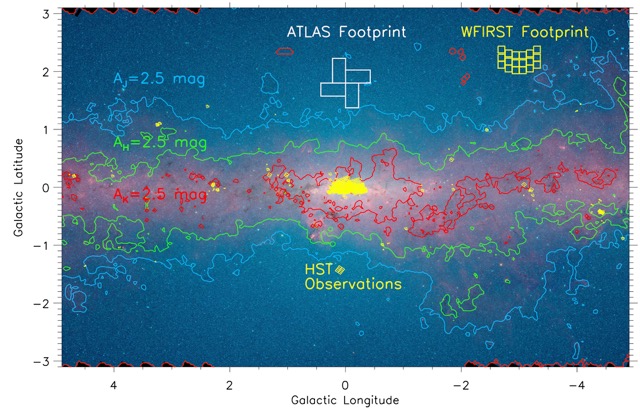}
\caption{A Spitzer/IRAC color image of the inner Galactic plane/bulge from 3.6 (blue) to 8.0 (red) $\mu$m shows the dramatic reduction of extinction from the $J$ band (1.2 $\mu$m; blue contours) to the $K$ band (2.2 $\mu$m; red contours) to the Spitzer 3.6 and 4.5 $\mu$m bands (background image) using the extinction map of \cite{gonzalez2012}. Extinction in the infrared bands compared to the visual are $A_{band}/A_{V}$=0.27, 0.17, 0.10, 0.06, and 0.03 for the $J$, $H$, $K$, $L$ (3.4 $\mu$m) and $M$ (4.8 $\mu$m) bands, respectively (\citealt{draine2011}). Stars with 30 magnitudes of visual extinction in the midplane will only suffer one magnitude of extinction for the longest wavelengths covered by ATLAS. The respective footprints of WFIRST and ATLAS are shown as are all of the Galactic plane fields that have been imaged by HST.
}
\label{fig:extinction}
\end{figure}

Currently, we know more about the structure and the spatially resolved star formation histories of galaxies in the Local Group (and even beyond) than we do about our own Galaxy.  Results from Gaia will revolutionize our understanding of the 3D Milky Way, but Gaia is limited, especially in the inner Galaxy, due to it being an optical survey.  Figure~\ref{fig:extinction} illustrates the extinction challenge.  
For the inner Galactic plane ($|l|<65^{\circ}$ and $|b|<1^{\circ}$), 98.5\% of of the Galactic plane has a minimum extinction of 2.5 mag in the $J$ band. This drops to only 10.4\% of the Galactic plane in the $K$ band. The infrared range of ATLAS opens up large regions of the Galaxy for spectral investigation.

\begin{figure}
\centering
\includegraphics[width=0.45\columnwidth,clip]{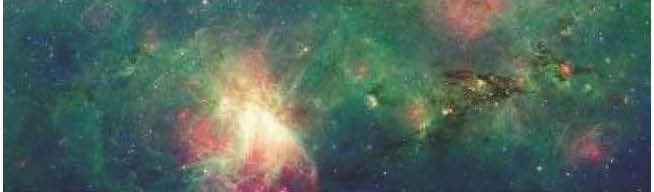}
\includegraphics[width=0.5\columnwidth,clip]{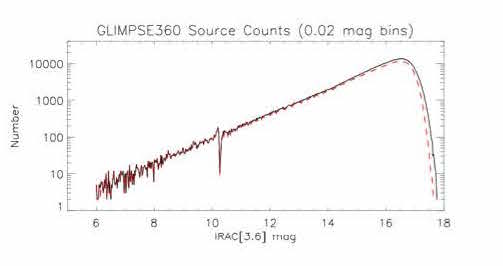}
\caption{Left panel: GLIMPSE image. Right panel: Magnitude distribution at 3.6$\mu$m of the several hundred million GLIMPSE catalog sources. Reprinted from \cite{Meade14}, ``GLIMPSE360 Data Description GLIMPSE360: Completing the Spitzer Galactic Plane Survey", https://irsa.ipac.caltech.edu/data/SPITZER/GLIMPSE/doc/glimpse360$_{-}$dataprod$_{-}$v1.5.pdf}
\label{fig:glimpse}
\end{figure}

ATLAS will acquire the spectra in a strip of $\pm 1^\circ$ centered on the Galactic Equator, corresponding to the approximate area of the family of Spitzer/IRAC 3.6 $\mu$m and 4.5 $\mu$m surveys: GLIMPSE-I ($|l|=10^{\circ}-65^{\circ}$), GLIMPSE-II ($|l|<10^{\circ}$); GLIMPSE-360 ($|l|> 65^{\circ}$) 
\citep{Churchwell09,Hora07,Carey08,Majewski07,Whitney08}\footnote{Glimpse survey data:  http://irsa.ipac.caltech.edu/data/SPITZER/GLIMPSE/}. These cover both the inner and outer Galaxy, and thus very different regions in terms of spiral structure, star formation, evolved stars and interstellar dust. As an example, Figure \ref{fig:glimpse} illustrates some of the 720 deg$^2$ footprint of the Spitzer/GLIMPSE surveys.  A future Galactic Plane imaging survey with WFIRST will improve our knowledge of the stellar populations, especially in the inner Galaxy where GLIMPSE is several magnitudes shallower than in the outer Galaxy, and would provide the ideal source catalog for ATLAS.

The ATLAS Galactic Plane Survey will represent a formidable follow-up to the GLIMPSE surveys, and complement shallower existing Galactic plane surveys: the near-infrared UKIDSS ($JHK$) and optical IPHAS ($r$, $i$ and H$\alpha$), as well as the even larger PanSTARRS optical, WISE 3-23 $\mu$m mid-infrared, and SCUBA-2 sub-mm survey, the FCRAO surveys of CO in the outer Galaxy, the Herschel Hi-GAL survey and Planck, plus all future major optical/infrared surveys like Gaia, ZTF, LSST, and WFIRST. ATLAS spectra will be crucial to interpreting WFIRST photometry in particular, in which photospheric temperatures and extinction effects will be almost completely degenerate due to the filter choices.  ATLAS spectra will also complement the planned SDSS-V spectroscopic survey programs covering brighter Milky Way objects ($H<11.5$ mag).

The ATLAS Galactic Plane Survey will probe galactic structure, study star forming regions, and measure interstellar extinction.

{\bf (a) Galactic Structure.} By taking advantage of the low extinction in the 1-4~$\mu$m region and the relatively high angular resolution and sensitivity of ATLAS compared to Spitzer, a spectroscopic survey of the inner Galaxy can produce unique information on the bar(s) of the Galaxy, the nuclear region, the stellar disk, and spiral arms. Existing and planned photometric surveys can provide useful results using star counts techniques and statistical analysis of color-color and color magnitude diagrams. However, only spectroscopic information allows firm derivation of the extinction and effective temperature of the sources, and therefore their position in the HR diagram. Gaia will determine 1 (10)\% distances for $10^6$ ($10^8$) stars out to 2.5 (25) kpc, as faint as $V\sim 19$ mag or $K\sim 11-19$ mag, depending on intrinsic color. 
Not only will ATLAS be able to provide spectral information on all of the (nearby) stars observed by Gaia in the optical, it will be able to see many additional stars that are not detectable in the optical. The calibration of "spectroscopic parallax" using Gaia for nearby stars will allow us to use improved spectroscopic parallax for the much more distant stars that will be detected by ATLAS. \cite{hogg2018} shows an example of this, using Gaia observations of nearby red giants to improve distance estimates of APOGEE selected red giants across the Galactic disk.
Reconstructing the 3D structure of the Galaxy will allow progress on a number of fundamental questions, e.g. regarding the scale-length of the Galactic disk, whether the stellar warp and the gas warp coincide, and existence of stellar streams across the Galactic plane. ATLAS will address these issues by reaching disk regions well beyond the extinction-limited range sampled by SDSS, PanSTARRS, Gaia, and LSST.

\begin{figure}
\centering
\includegraphics[width=0.32\columnwidth,clip]{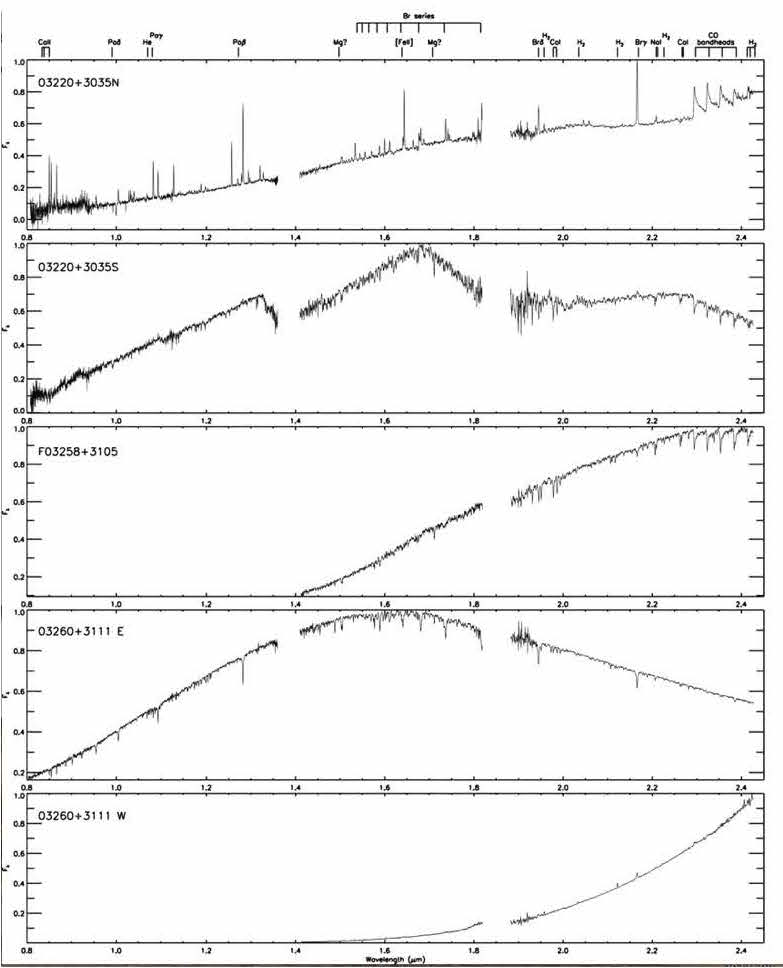}\includegraphics[width=0.31\columnwidth,clip]{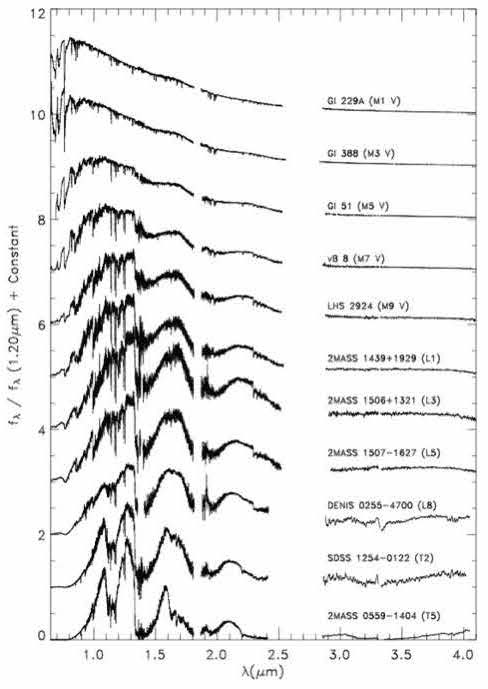}
\includegraphics[width=0.35\columnwidth,clip]{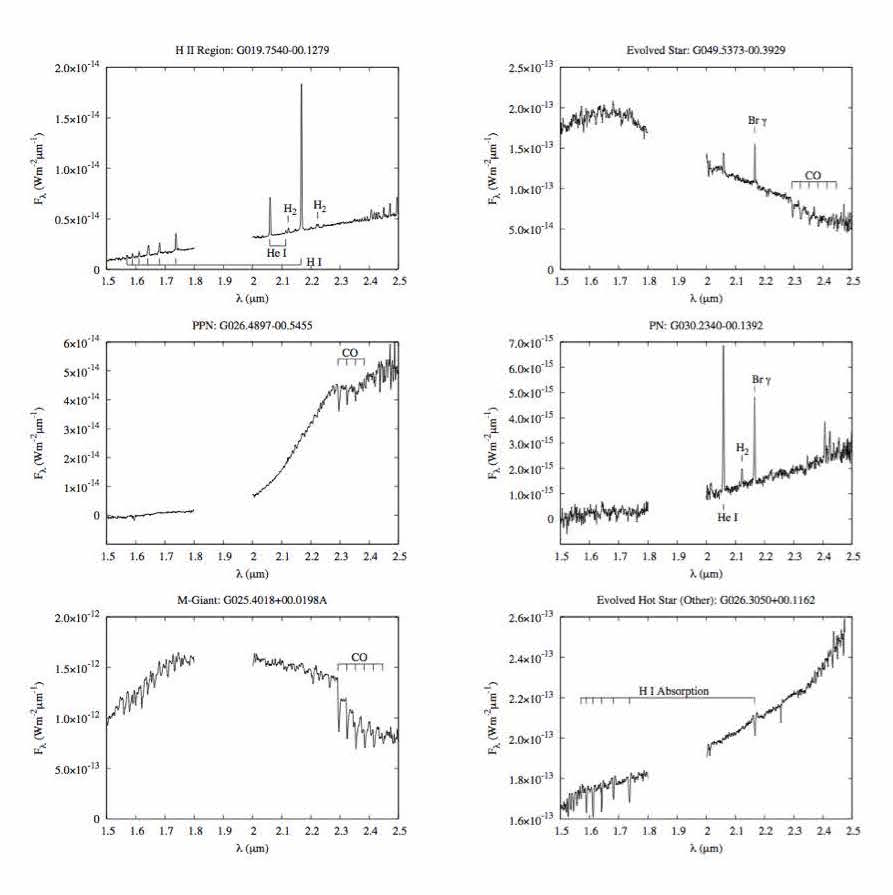}
\caption{Example published spectra of: young stars (left panels out to 2.5$\mu$m only; \cite{Connelley10}),  
low-mass stars (middle panels, covering 1-4$\mu$m; \cite{Cushing05}), and evolved objects (right panels out to 2.5$\mu$m only; \cite{Cooper13}). Figures reprinted with permission from the authors.}
\label{fig:stellar-spectra}
\end{figure}

{\bf (b) Star Formation.} Spitzer/GLIMPSE surveys produced an unprecedented picture of star formation in our Galaxy by unveiling hundreds of new star forming regions. ATLAS will quantify the Star Formation Rate (SFR) of the Galaxy, its variation with Galactocentric radius, and its association with various dynamical features in the Galaxy, testing theories of star formation both on a global scale and at the molecular cloud level. Moreover, as the metallicity declines in the Outer Galaxy (reaching values intermediate between that of the LMC and in the inner Galactic disk), ATLAS will allow measurement of variations in the SFR representative of typical conditions at $z=2-3$, the peak of the cosmic star formation efficiency. ATLAS will characterize Young Stellar Objects (YSO) and their surrounding environments, analyzing the energy budget and spatial correlation between ionization (through recombination lines), photodissociation (through PAH emission), and high-velocity outflows (from shocked $H_2$). The specific early type (<B2) stars that excite Galactic Plane H II regions out to 10 kpc will be identified via their distinguishing 1-4$\mu$m spectroscopic features, revealing important details regarding the spatially resolved star formation history of the Galaxy over the past <50-100 Myr. Among lower mass Young Stellar Objects (left panels of Fig. \ref{fig:stellar-spectra}), accretion and outflow rates can be measured across individual star forming regions, probing accretion histories and quantifying burst vs steady accretion scenarios.

{\bf (c) Interstellar Extinction.}
By measuring the extinction law over tens of millions of lines of sight, ATLAS will provide a unique dataset to study in detail the variation of the extinction curve out to the edge of the Galactic disk. Variations with Galactic longitude have been identified and attributed to small variations in ISM density, mean grain size, or disk metallicity gradient. Variations of chemical composition of dust grains reflect the abundance/depletion of metals in the ISM, and hence the cooling mechanisms that control the efficiency of star formation.
Characterizing the variations in infrared extinction is important for enabling extinction corrections of Galactic plane sources from photometric surveys and understanding the nature of dust grains.  A particular puzzle from Spitzer photometric studies was the fact that extinction in the 3-8 $\mu$m range was unchanging ("flat")  in contradiction to expectation from models by \cite{draine2001}. It has been hypothesized that this indicates a population of micron-sized grains which accounts for 15\% of the total dust mass (\citealt{wang-li-jiang2015}). The spectra from ATLAS of stars with known intrinsic spectra, e.g. red clump giants, will allow one to characterize how the infrared extinction varies with line of sight and search for deviations from the flat extinction curve indicated by current photometric data.

\subsection{Low-mass stars, Brown Dwarfs, and Exoplanets}

ATLAS Wide Survey (the spectroscopic follow-up to the WFIRST HLS imaging) , with its high Galactic latitude, complements the ATLAS Galactic Plane Survey. It will probe the low-mass stellar content of the Milky Way.

Brown dwarfs cool through the L-T-Y spectral sequence as they age (\citealt{burrows1997, kirkpatrick2005, cushing2011}). These objects provide important information on the shape and low-mass cutoff of the field mass function, and they also stand as proxies for exoplanets, with their atmospheres having similar effective temperatures but without the complicating effect of an irradiating host star. Based on the theoretical predictions of \cite{baraffe2003}, at the 3$\sigma$ limit of J$\approx$23.5 mag AB ATLAS will be able to detect 5-Gyr-old field brown dwarfs as low in mass as $10 M_{Jup}$ ($T_{\rm eff} \sim 300$K, an early-Y dwarf) at 10 pc and $35 M_{Jup}$ ($T_{\rm eff} \sim 700$K, late-T) at 100 pc. For a robust, well-characterized sample in which ATLAS spectra of SNR=100 are required, corresponding to $J \approx 18.5$ mag AB, these limits are $35 M_{Jup}$ ($T_{\rm eff} \sim 700$K, late-T) at 10 pc and $75 M_{Jup}$ ($T_{\rm eff} \sim 1800$K, mid-L) at 100 pc. In terms of spectral type, ATLAS can detect and characterize mid-L dwarfs to 250 pc and mid-T dwarfs to 100 pc at $J = 21.0$ mag AB ($SNR \sim 10$).

For low-mass stars, ATLAS can detect and characterize sources at the distance of the Galactic Bulge. These are important for a variety of Milky Way research topics (see section above) as well as providing valuable data for the interpretation of microlensing light curves when these objects are lensed by an intervening system. Spectra with $SNR > 100$ are obtainable at a distance of 8.0 kpc for types earlier than $\sim$F8 V and with $SNR > 10$ for types earlier than $\sim$K7 V. Detections at the $SNR = 3$ limit are possible down to a type of $\sim$M3 V.

\subsubsection{The Mass Function at Low Metallicities}

Low-mass stars and brown dwarfs are brightest in the near-infrared (see Fig.\ref{fig:stellar-spectra}, middle panels), but the nearest examples to the Sun almost all have metallicities near solar. Is the mass function at lowest masses dependent upon the composition of the gas from which the objects formed or is it invariant to metallicity, as early theoretical investigations suggest \citep{bate2014}? \cite{gizis1999} determined that the local M star population is comprised of 99.8\% dwarfs with [M/H] $\sim 0.0$ and only 0.2\% subdwarfs with [M/H] $< -0.5$. Therefore, to find large collections of these metal-poor objects with which to study population statistics, deeper surveys in the near-infrared are needed.

The WFIRST HLIS provides a seed list with which to answer this question. Subdwarf candidates will be selected photometrically using the WFIRST data themselves, with ATLAS providing spectroscopic verification or refutation. The ATLAS data will be further used to establish both the spectral type and metallicity class of those that are verified. These, combined with WFIRST HLIS photometric measurements, will establish space densities of these objects with which we can determine mass functions of the stellar subdwarf populations as a function of metallicity. 

\begin{figure}
\includegraphics[scale=0.3,angle=0]{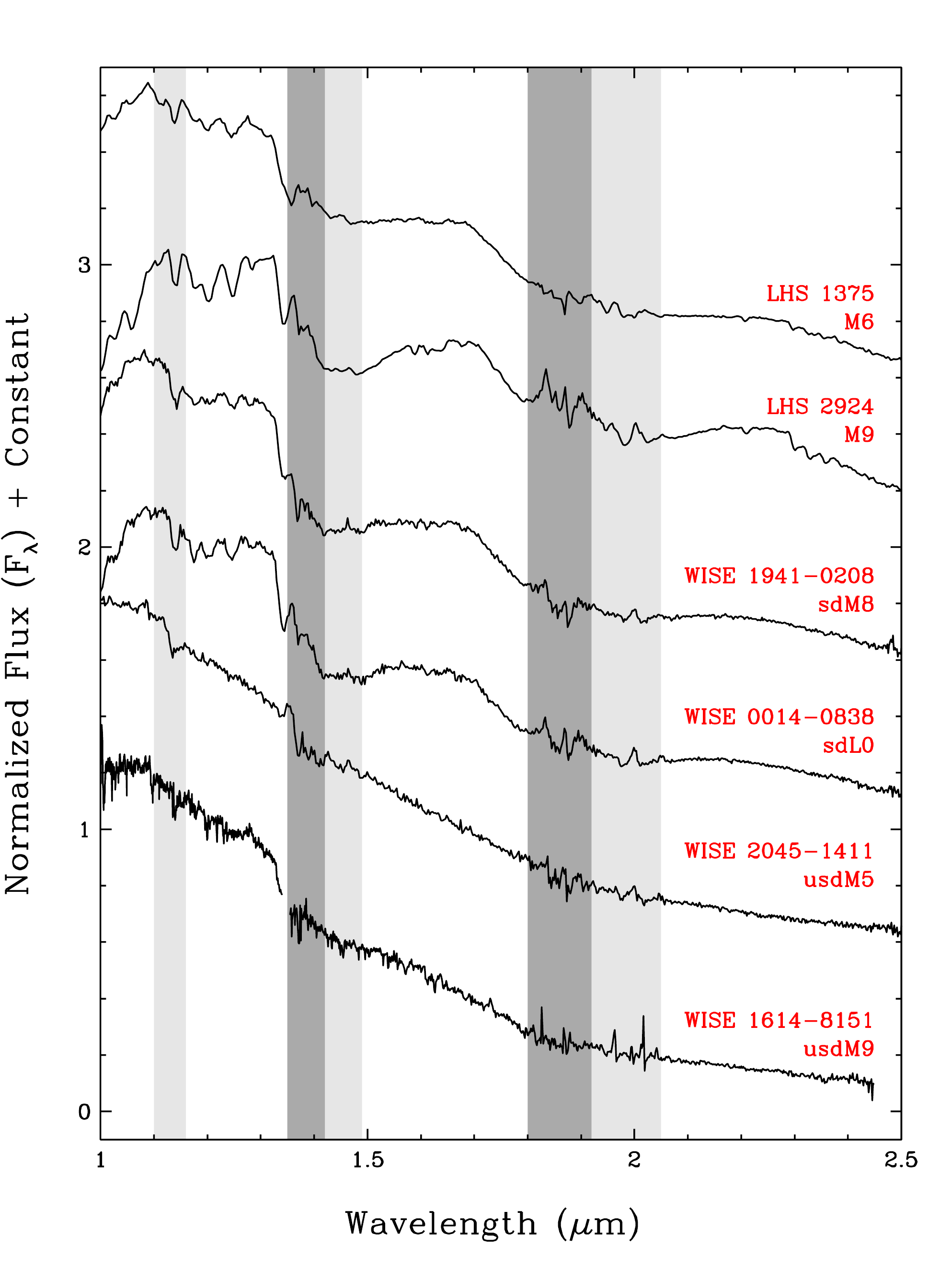}
\includegraphics[scale=0.65,angle=0]{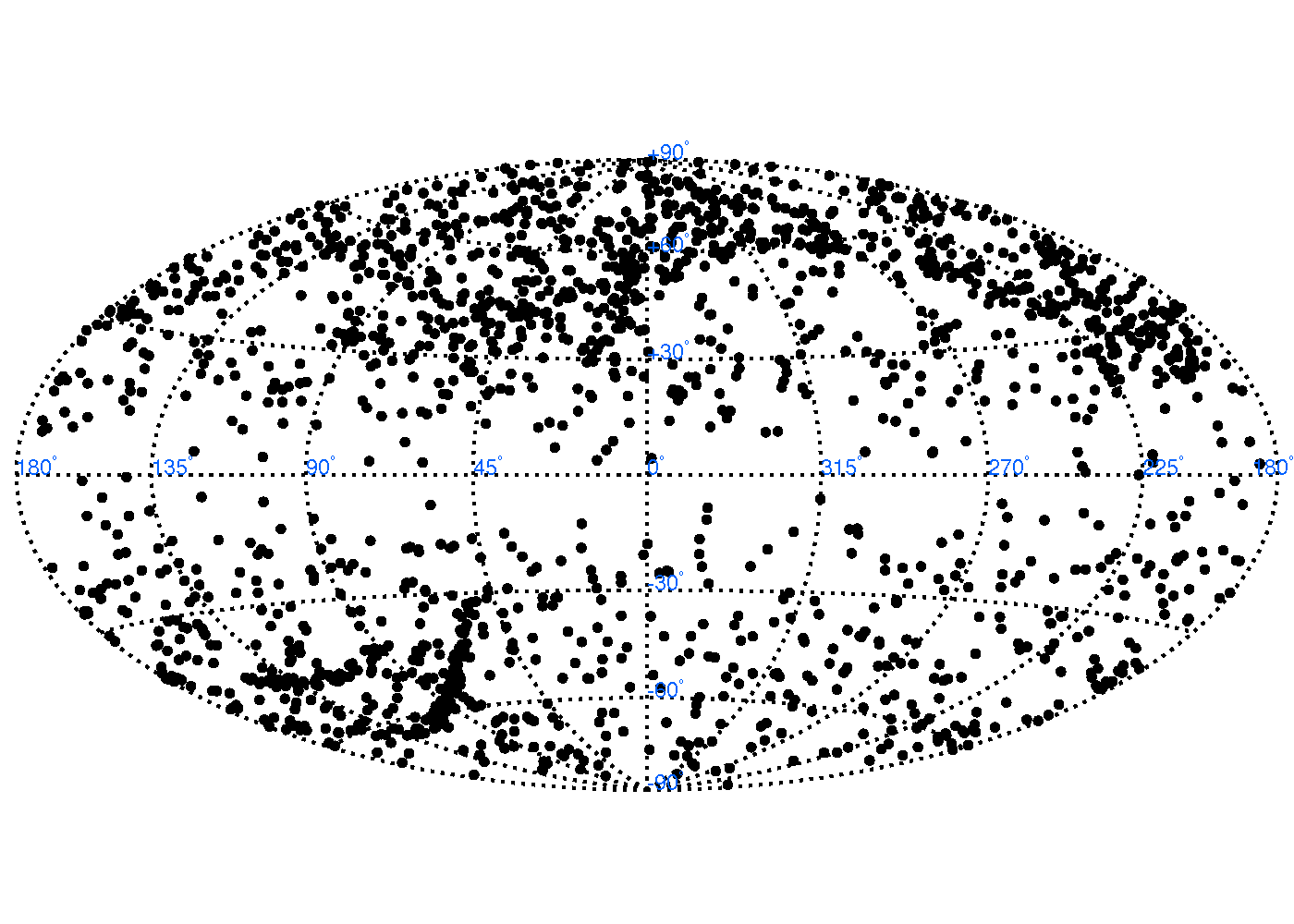}
\caption{{\bf Left panel:} The 1-2.5 $\mu$m spectra of two normal late-M dwarfs (top), two metal-poor late-M/early-L subdwarfs (middle), and two even more metal-poor late-M ultrasubdwarfs (bottom) from \cite{kirkpatrick2010, kirkpatrick2016}. For metal-poor objects, the broad trough caused by collision-induced absorption by H$_2$ flattens out the entire near-infrared spectrum, with the strongest effect at $K$-band. This flattening along with the concomitant bluer $J-K$ colors are the hallmarks of low metallicity. In this plot, regions of telluric absorption are shown by the dark gray bands if the atmospheric transmission is less than 20\% and by light gray bands if the atmospheric transmission is between 20\% and 80\% (\citealt{rayner2009}). Spectra are normalized at 1.28 $\mu$m and offsets added to separate the spectra vertically. 
{\bf Right panel:} A Hammer projection in Galactic coordinates showing the sky distribution of known L, T, and Y dwarfs (black points), taken from Dwarf Archives\footnote{See {\url http:/dwarfarchives.org}} and the List of All UltraCool Dwarfs\footnote{See {\url https://jgagneastro.wordpress.com/list-of-ultracool-dwarfs/}}. Note the lack of brown dwarfs in a wide band centered along the Galactic plane.
\label{fig:subdwarfs}}
\end{figure}

Only a few dozen late-M and early-L subdwarfs have been uncovered to date in the immediate Solar Neighborhood, thanks primarily to 2MASS, which probes to $J \sim 17$ mag AB. For sources with $J < 21.0$ mag AB (spectra with $SNR > 10$), ATLAS follow-up of candidates selected from the 2000-sq-deg HLIS can reach a volume $\sim$15 times larger than the all-sky survey of 2MASS. 
The spectroscopic hallmark of these objects falls within the ATLAS observing window; specifically, strong collision-induced absorption by H$_2$ is the dominant shaper of the continuum, as Figure~\ref{fig:subdwarfs} (left) shows.

Rarer still are late-L and T subdwarfs that populate the brown dwarf side of the low-metallicity mass function. Less than a dozen of these objects are known near the Sun (\citealt{kirkpatrick2014, kirkpatrick2016, pinfield2014}) so identifying more of these objects in the ATLAS Wide Survey would enable a mass function analysis below the substellar break. As more of these objects are uncovered, the width of the temperature gap between the stellar and substellar sequences at a particular metallicity can be measured, offering an independent measurement of the age of these populations.

\subsubsection{Toward Measuring the Masses of Isolated Low-mass Stars and Brown Dwarfs}

As Figure~\ref{fig:subdwarfs} (right) shows, our knowledge of nearby low-mass stars and brown dwarfs is sorely lacking along the Galactic plane. Although there have been some spectacular finds along the Zone of Avoidance in recent years -- the T6 dwarf WISE J192841.35+235604.9 at 13.0 pc ($b = +3.1$ deg; \citealt{kirkpatrick2012}), the T8 dwarf WISE J200050.19+362950.1 at 7.5 pc ($b = +3\fdg3$; \citealt{cushing2014}), the M6.5 dwarf WISEA J154045.67$â$510139.3 at 5.9 pc ($b = +3\fdg4$; \citealt{kirkpatrick2014}), and the T9 dwarf UGPS J072227.51$-$054031.2 at 4.1 pc ($b = +4\fdg3$; \citealt{lucas2010}) -- none has been as eye-opening as the L7.5/T0.5 binary WISE J104915.57$â$531906.1 at 2.0 pc ($b = +5\fdg3$; \citealt{luhman2013, burgasser2013}), which, at 2.0 pc, is now recognized as the third closest ``stellar'' system to the Sun.

Finding such systems against a rich stellar background is challenging but also replete with possible future rewards. Any nearby low-mass star or brown dwarf will likely exhibit high proper motion, meaning that there is a much higher probability of microlensing a background source if the object lies against the Galactic plane than if it lies at a higher Galactic latitude. As \cite{cushing2014} has shown, a hypothetical brown dwarf located at ($l$,$b$) = ($45\degr$,$0\degr$) and with a distance of 10 pc and proper motion of $0\farcs$5 yr$^{-1}$ has an observable astrometric microlensing event every two decades. Thus, finding even a few dozen low-mass stars and brown dwarfs near $b = 0\degr$ suggests that microlensing events could be studied every few months and could be predicted in advance with proper motion and parallax measurements accurate to the milliarcsecond level. 

Monitoring such events enables a direct measurement of the mass of the lensing system. In fact, for single low-mass stars and brown dwarfs, microlensing is the only means available to perform a direct measurement of the mass, which is why other researchers have been  trying, so far in vain, to predict events in advance (e.g., \citealt{lepine2012}). Spectroscopic follow-up by ATLAS of Spitzer-selected high-motion candidates from multi-epoch GLIMPSE data and of to-be-microlensed sources in the background provides a new tool to improve the chances of predicting such occurrences. 
Identifying L and T dwarfs in the Galactic Plane can be challenging because of source crowding, but the unique spectral signatures of water and methane absorption (middle panel of Fig.~\ref{fig:stellar-spectra}) in these objects can aid in their discovery even in crowded fields, assuming that the contaminating stars have comparable or dimmer magnitudes. Low-resolution, near-infrared spectroscopy has used these features to disentangle the components in unresolved, L+T dwarf and T+T dwarf binaries in the past (\citealt{burgasser2007, burgasser2016}), and similar techniques can be used to identify an L or T dwarf superimposed on a background object with a distinctly different spectral morphology.

\subsubsection{Measuring the Star Formation History of the Galaxy}

Low-mass stars and brown dwarfs represent fossilized records of star formation across all Galactic epochs. Identification of these objects in the ATLAS surveys will enable measurements of their scale heights as a function of spectral type, which in turn can be used to track the rate of star formation over time. 

ATLAS spectra can be used to identify and classify M, L, and T dwarfs using the depths of their intrinsic water bands and overall colors. Specifically, the H$_2$O band depth can be measured from the spectra by using a narrow synthetic filter centered on the water band near 1.38 $\mu$m and comparing that to the pseudo-continuum in two synthetic bands centered near 1.28 and 1.58 $\mu$m. The measured fluxes can be used to classify M, L, and T dwarfs found by ATLAS and provide distances to each. Based on the Galactic models of \cite{ryan2016}, the ATLAS Wide survey is expected to uncover $\sim$18,000 M8-L0 dwarfs, $\sim$21,000 L0-L5 dwarfs, $\sim$2,100 L5-T0 dwarfs, and $\sim$600 T0-T5 dwarfs to distances extending to the thin disk scale height and beyond. The Medium and Deep surveys are able to probe to even further distances, although with diminished statistics: $\sim$4,400 M8-T5 dwarfs for the Medium survey and $\sim$70 for the Deep survey.

The vast majority of field late-M dwarfs are hydrogen-burning stars with a mixture of ages. The brown dwarfs that overlap the stars in the late-M to mid-L regime and continue into the substellar-only regime at late-L and T also have a mixture of ages. However, the ages are not uniformly distributed as a function of spectral type because brown dwarfs cool with time. The upper-temperature end of the brown dwarf sequence is dominated by young, high mass brown dwarfs; the lower-temperature end is comprised of old, high-mass brown dwarfs which have had time to cool to those temperatures as well as younger, lower mass brown dwarfs with cold birth temperatures. Measuring the scale height across many spectral subtype bins therefore provides information about the ages as a function of temperature. When combined with evolutionary models and the functional form of the underlying mass function, the star formation history over the age of the Milky Way can be deduced.

\subsection{Evolved Stars and Other Objects}

{\bf Evolved Stars.} Because of their absolute luminosity, evolved stars can be traced across a large fraction of the Galactic disk. At 3 $\mu$m, ATLAS Galactic Plane Survey will detect Red Giant Clump stars standard candles with $L_{\rm mag}=-1.75$ (0.71 in AB mag) at 10kpc with $A_v=30\,$mag, reaching the heavily obscured regions of the Galactic Center or the outer edges of the Milky Way over at least the two outer quadrants. Clusters of Red Supergiants, $\sim\,$100 times brighter, can be studied via spectroscopy across the entire Galactic disk. Late-type stars tend to produce significant amount of dust and may have colors similar to YSOs and galaxies; IR spectroscopy is needed to disentangle their nature (Fig.\ref{fig:stellar-spectra}, right panels). The more massive AGB stars (initial mass 4-8$\,M_\odot$) are sufficiently short lived that they can be used to trace the spiral arms of the Galaxy. Their number and distribution provides a fossil record of the recent history of star formation in the Milky Way. 

{\bf Other Objects.} ATLAS will provide an exceptional database for studying a wide variety of sources: bubbles and young supernova remnants, Infrared Dark Clouds, heavily embedded high-mass clusters, Planetary Nebulae, Old Supernova Remnants, Variable Stars including outbursting YSOs, extremely metal-poor halo stars, and even galaxies and clusters hiding beyond regions of high extinction, identified as such by their spectra.

\section{Solar system science}
\label{sec:solar-system}

ATLAS Science Objective 4, "Probe the formation history of the outer Solar System through the composition of $~$3,000 comets and asteroids," flows down
to a Solar System Survey with 3000 individual pointings, covering a total area of 1,200 sq deg. 
ATLAS can provide valuable insight into the properties and composition of Solar System objects from comets and asteroids to Trans-Neptunian objects.
ATLAS Solar System Survey will focus on the little explored Kuiper Belt Objects.

Astronomers have a good handle on the icy surface compositions of large volatile dominated Kuiper Belt Objects, such as Eris, or Pluto. This knowledge has largely originated from ground-based spectroscopy \citep{Barucci2011,Brown2012} and the New Horizons flyby of Pluto. Despite more than 2 decades of spectroscopic observations however, exceptionally little is known about the overall bulk compositions of small, $D\lesssim500$~km KBOs which dominate the mass of the population (Fraser et al. 2014). To date, the only features identified in the IR spectra of small KBOs are those absorption features associated with water-ice \citep[$1.5~\micron$ and $2.0~\micron$, eg.,][]{Barkume2008, Brown2012}, methane \citep[eg.][]{Alvarez-Candal2011}, and methanol \citep{Cruikshank1998}.
Otherwise, no materials have been robustly detected. In particular, we have been entirely unable to identify the silicate components in KBOs -- even from New Horizons -- which from density considerations, must make up a large portion of these bodies \citep{Stansberry2008}. Our ignorance is predominantly the result of the lack of identifying absorption features in the $\lambda\lesssim 2.5$~\micron, the current wavelength limit beyond which telescopes lack sufficient sensitivity to gather quality spectra for the majority of KBOs. Longer wavelength spectra have gone virtually undetected. 
The 1-4~$\mu$m range is key, as many icy and non-icy components exhibit broad absorption features in this range \citep{Parker2016}. Over its lifetime, the JWST will provide IR spectra of exceptional quality for a few tens of KBOs at most. While comparatively of a lower typical SNR and of a smaller wavelength range, ATLAS will gather 1-4~$\mu$m spectra of many more targets. ATLAS Wide Survey will gather spectra of $\sim300$ KBOs. The focused survey, ATLAS Solar System Survey, will target individual KBOs discovered by imaging surveys, and gather spectra of  $\gtrsim 3,000$ KBOs brighter than the practical depth limit of r'<23.2. 

No tracking capability is required for the spacecraft by the ATLAS Solar System Survey.
KBOs are distributed at distances 30-50 AU from the Sun; they move $\sim 1^{\prime\prime}$-4$^{\prime\prime}$ per hour. Most KBOs will move across multiple micro-mirrors on the DMD during each 1000s exposure, since each micro-mirror forms a 0.75$^{\prime\prime}\times$0.75$^{\prime\prime}$ slit. 
ATLAS can take slit spectra of moving targets with regular guiding, as the micro-mirrors can either form long slits or switch on in sequence as the target drifts across the field. 
For faint targets such as KBOs, the main source of uncertainty is the stability of the system response; a change of a few mK or 1/f electronic noise would bias the counts. Since ATLAS can take spectra of hundreds of other sources in parallel to the KBO being observed, these spectra can be used to monitor and correct detector drifts. 
Our estimates of survey depth in this section do not include any losses due to trailing of the target across the FoV, which will be studied in future work.

\subsection{KBO model}

\begin{figure}
\centering
\includegraphics[width=0.8\columnwidth,clip]{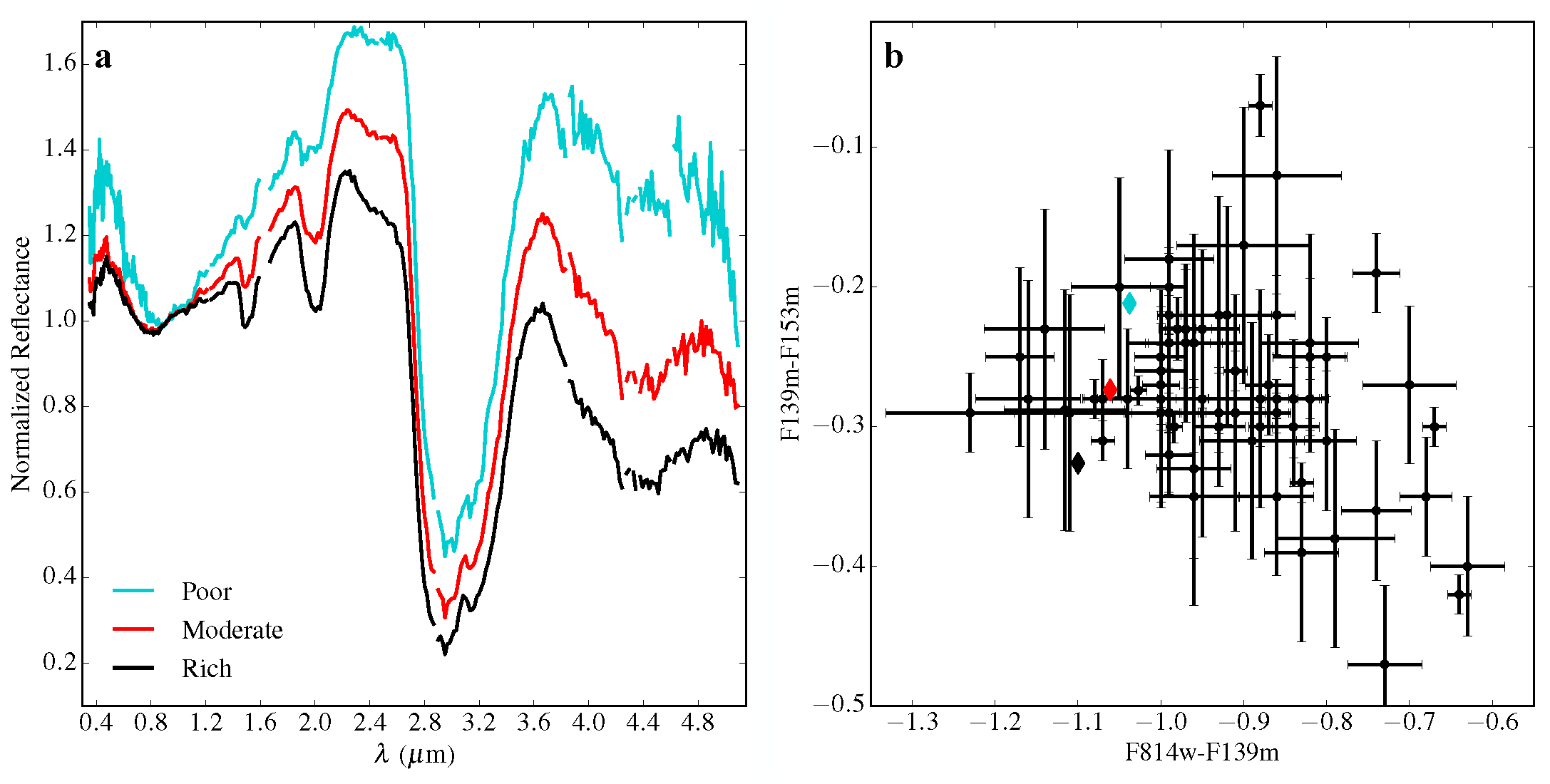}
\caption{\textbf{a:} Spectra of the different spectra units found on Phoebe's surface.  These spectra will appear in \citet{Fraser2018}. "Poor", "Moderate", and "Rich" represent different levels in the water absorption depth. In our observable KBO density estimates, we adopt the moderate spectrum which exhibits water-ice absorption typical of most KBOs. 
\textbf{b:} NIR colors of KBOs gathered with the Wide Field Camera 3 on the Hubble Space Telescope (black points; \cite{Fraser2015}. Simulated colors of Phoebe of the spectra presented in Figure~\ref{fig:phoebe-spectra} are shown by the colored points. The red point is that drawn from the moderate spectrum shown in panel a.}
\label{fig:phoebe-spectra}
\end{figure}

To estimate the 1-4~$\mu$m brightness of a non-volatile rich KBO, we avoid the specific compositional modeling approach presented in Parker et al. (2015), but rather, turn to other well studied planetesimals as examples of what we might find in the Kuiper Belt. Our best analogy for the IR behaviour of a KBO is the spectrum of Phoebe (see Figure~\ref{fig:phoebe-spectra}), which we believe has originated from the same primordial population that also populated the Kuiper Belt \citep{Nesvorny2007, Levison2008}. Phoebe has likely avoided temperatures that would cause water-ice to become volatile, and therefore likely presents a spectral behavior similar to that of KBOs in the water dominated $3-4~\micron$ region.
While the optical color is significantly less red than many KBOs, possibly due to increased volatile loss of the organic compounds that possibly contribute to the development of a red refractory or organic coating on KBOs \citep{Sheppard2010}, Phoebe still exhibits water-ice absorption depths and 1-4 $\mu$m colors compatible with small KBOs \citep{Fraser2018}. As far as available observations suggest, Phoebe appears as an excellent KBO analog. For our modeling, we adopt the spectrum of moderate water absorption depths, but we point out that from the range of spectra exhibited on Phoebe, the global variations in 1-4 $\mu$m reflectance exhibited by KBOs are possibly as large as exhibited across Pheobe. 

KBOs are known to exhibit two compositional classes \citep[eg.][]{Fraser2012,Peixinho2012}, with distinctly different optical colors. We model the 1-4 $\mu$m detectability of each class using two separate optical spectral 
slopes\footnote{percent increase in reddening per 100 nm normalized to 550 nm}: 10\% for the \emph{neutral} class, and 30\% for the \emph{red} class \citep{Peixinho2015,Pike2017, Fraser2017}. We model the 1-4 $\mu$m reflectance as approximately flat, as KBO spectra are known to be roughly neutral for $\lambda\gtrsim1$~\micron. In estimating the KBO 1-4$\mu$m flux then, we simply scale the relative reflectance of the Phoebe spectrum at $1~\micron$ according to the optical spectral slopes of spectral class. We also adopt an intrinsic \emph{neutral:red} population ratio of 7:1 (Schwamb et al. in prep.).

\subsection{ATLAS Solar System Survey}

ATLAS Solar System survey is a pointed survey that targets individual known and tracked objects, in order to 
maximize the overall number of KBOs observed. We consider the AB=22.5 (3$\sigma$) depth achieved in 2500~s exposure time, since a spectrum of sufficient quality is gathered that not only enables the detection of water-ice absorption, but also allows the differentiation between water-ice absorption and hydroxyl absorption. Phoebe's $3~\micron$ hydroxyl absorption feature is a $\sim50\%$ absorption feature, while the ice feature is much shallower, with depth of only $\sim10\%$. Thus, the latter sets our SNR requirement at roughly SNR$>10$. From resolution considerations, the instrumental spectra can be binned to a lower resolution by a factor of $\sim15$ while still preserving sufficient spectral resolution to detect each of these broad-band features. With the above color considerations, translating Phoebe's 2.5 micron reflectance to r-band, the single exposure limiting magnitudes are r'=22.7 and r'=23.2 for the \emph{neutral} and \emph{red} classes, respectively. 

At the AB=22.5 (3$\sigma$) continuum depth limit, the number of detectable KBOs over the entire sky is $\gtrsim3000$. The vast majority of these will be observed and tracked by the LSST, allowing direct pointed observations by ATLAS for these targets. To observe each target with 2500~s exposure would require roughly 2083~hours of on-target observations. Assuming 95\% observing efficiency, a dedicated KBO ATLAS survey that is \textbf{brightness complete} to AB=22.5 would require roughly 0.25 observing years.

\subsection{KBOs from ATLAS Wide Survey}
Now we estimate the number of KBOs that will be observed in the ATLAS wide survey. For this, we consider the (\emph{stacked}) 3$\sigma$ depth of 23.4 AB mags of ATLAS Wide. 
Each field in ATLAS Wide will be observed for 10,000s (10$\times$1000s exposures). Since the KBO targets are known and small in number, we will keep these on the target lists for ATLAS Wide, and stack the spectra gathered over 10 exposures of each KBO target in the wide survey. The stacked depths for the \emph{neutral} and \emph{red} classes become r'=23.6 and r'=23.1, respectively. These depths represent the practical limit achievable by the ATLAS probe. Conveniently, all KBOs with r'$<23.6$ will have already been detected and tracked by the LSST, providing fully sufficient ephemerides for direct acquisition of spectra with ATLAS.

To determine the number of KBOs that will be observed in the ATLAS Wide Survey, we execute a survey simulation. We consider the nominal WFIRST wide survey region as presented in Holler\footnote{\url{https://arxiv.org/pdf/1709.02763.pdf}}. As a Kuiper Belt model, we make use of the Canada France Ecliptic Plane Survey (CFEPS) model v7 \citep{Petit2011}, which is the most up to date outer Solar System model. We note that this model does not include certain highly populated Kuiper Belt dynamical classes, such as the scattered disk and objects in the 5:2 mean-motion resonance with Neptune \citep{Gladman2008}. As such, detectability estimates drawn from this model likely represent lower bounds on the true detectable number of targets. 

From the CFEPS model, at the \emph{stacked} depths of ATLAS Wide, the number of KBOs detected will be $\simeq250$ and $\simeq20$ for the \emph{neutral} and \emph{red} classes, respectively.

\section{Source predictions}
\label{sec:source}

To maximize spectroscopic efficiency, ATLAS Probe will measure the majority of galaxy redshifts based on their emission lines.
We have implemented a methodology to predict line fluxes from photometric data, simply assuming that H$\alpha$ is primarily emitted by H II regions in normal galaxies that follow the Main Sequence correlation between star formation rate and stellar mass \citep{Noeske07,Elbaz07,Daddi07}. Below we summarize the salient points of this technique, while we refer the reader to \cite{Valentino17} for a full description of the method, performances and calibration on actual observations at high redshifts, based in part on the same sample that has been used for ATLAS Probe predictions here.

We first selected a sample of $\sim30,000$ star-forming galaxies in the GOODS-S field based on their $U-V$, $V-J$ rest-frame colors \citep{Williams09}, covering the redshift space up to $z_{\rm phot}\sim7$, based on the CANDELS catalog by \cite{Guo13}.
GOODS-S benefits from an ultra-deep \textit{HST} coverage and an exquisite multi-wavelength dataset, allowing for good constraint on the SEDs and thus of the star formation rates (SFRs) down to low levels. This is important for estimating the tail of \ha\ emitters down to the deep flux limits probed by ATLAS Probe. We modeled the multi-wavelength spectral energy distributions of our galaxy sample with FAST \citep{Kriek09}, assuming \cite{Bruzual03} stellar population synthesis models, a constant star formation history, a \cite{Calzetti00} reddening law, solar  metallicity, and a \cite{Salpeter55} initial mass function. We adopted a constant star formation history since at $z>1$ it reconciles the SFR estimates derived independently from different indicators (e.g., \cite{Rodighiero14}). We further checked that our modeling consistently retrieves the Main Sequence of SFGs at different redshifts, comparing with the analytical parametrizations by \cite{Sargent14} and \cite{Schreiber15}, independently derived from the far-infrared emission. We then converted the SFRs into \ha\ fluxes following \cite{Kennicutt98},
and applying a reddening correction based on the SED-derived $E_{\mathrm{star}}(B-V)$ for the stellar component. Note that these predictions represent the \textit{total observed} H$\alpha$ fluxes, i.e, without taking into account any aperture corrections, but reproducing the effect of dust attenuation. The latter fully takes into account the possibility that $E_{\mathrm{neb}}(B-V)$ for the nebular emission is different from $E_{\mathrm{star}}(B-V)$ for the stellar continuum, assuming that $E_{\mathrm{neb}}(B-V)=E_{\mathrm{star}}(B-V) / f$ (Calzetti et al. 2000).
For the purpose of this section, $f$ assumes the role of a fudge factor to reconcile SFRs estimated from different tracers and empirically
predict \ha\ fluxes as close as possible to observations. The real physical meaning of $f$ is uncertain, being strongly affected by sample selections, SED modeling, UV-to-H$\alpha$ timescale variations, or shape and
normalization of the reddening curves over the wavelength range from
the Lyman continuum to H$\alpha$ \citep{Puglisi16}. Eventually, we adopted $f=0.57$ as derived in \cite{Valentino17}, where they extensively compared the results of a SED modeling identical to ours and spectroscopic data from the FMOS-COSMOS survey at $z\sim1.55$ \citep{Silverman15}. This value of $f$ is higher than the one found in local galaxies ($f=0.44$) with identical reddening laws 
(\cite{Cardelli89,Calzetti00}, and see the discussion in \cite{Steidel14}), but characterized by a substantial scatter ($\sigma_{f}\sim0.23$). This result is consistent with recent findings of slightly higher values of $f$ at high redshift 
\citep{Kashino13,Pannella15}. Here we assume the validity of $f=0.57$ over the entire stellar mass and reddening ranges covered by the GOODS-S photometric sample. 
These calculations are appropriate for the bulk of the population, but would lead to over-estimating emission line fluxes for the most extreme population of merging driven starbursts with large column densities of dust, even for  near-IR rest frame Paschen lines (Calabr\`o et al. 2018). Such objects are only a few percent among massive star forming galaxies (Rodighiero et al. 2011).

\cite{Valentino17} estimate a typical uncertainty on the final H$\alpha$ flux predictions of $0.2$~dex, derived from the direct comparison between fluxes forecast and spectroscopic observations. Note that this uncertainty estimate was largely dominated by the errors on the aperture correction of the FMOS spectroscopic data ($\sim0.17$~dex), while the rest is due to SED modeling and/or intrinsic scatter due to variations within the population of Main Sequence galaxies, and the different timescales probed by H$\alpha$ and rest-frame UV light as star formation tracers. We will still conservatively assume this uncertainty is representative for our predictions, too. With assumptions discussed in \cite{Valentino17} we also extended our computations to other lines, including [OII]$\lambda$3727, [OIII]$\lambda$5007, Paschen lines, etc.  

\begin{figure}
\centering
\includegraphics[width=0.4\columnwidth,clip]{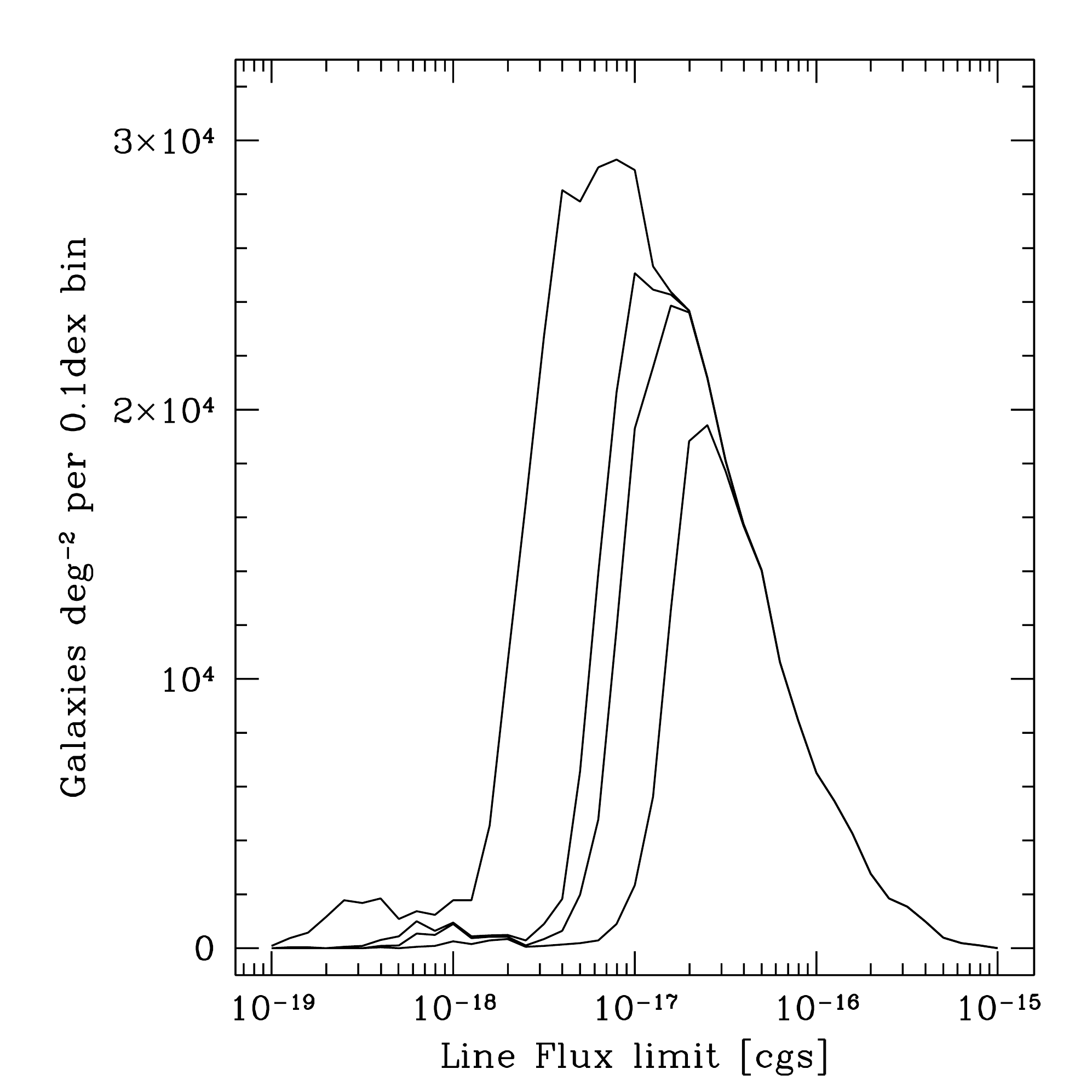}
\includegraphics[width=0.4\columnwidth,clip]{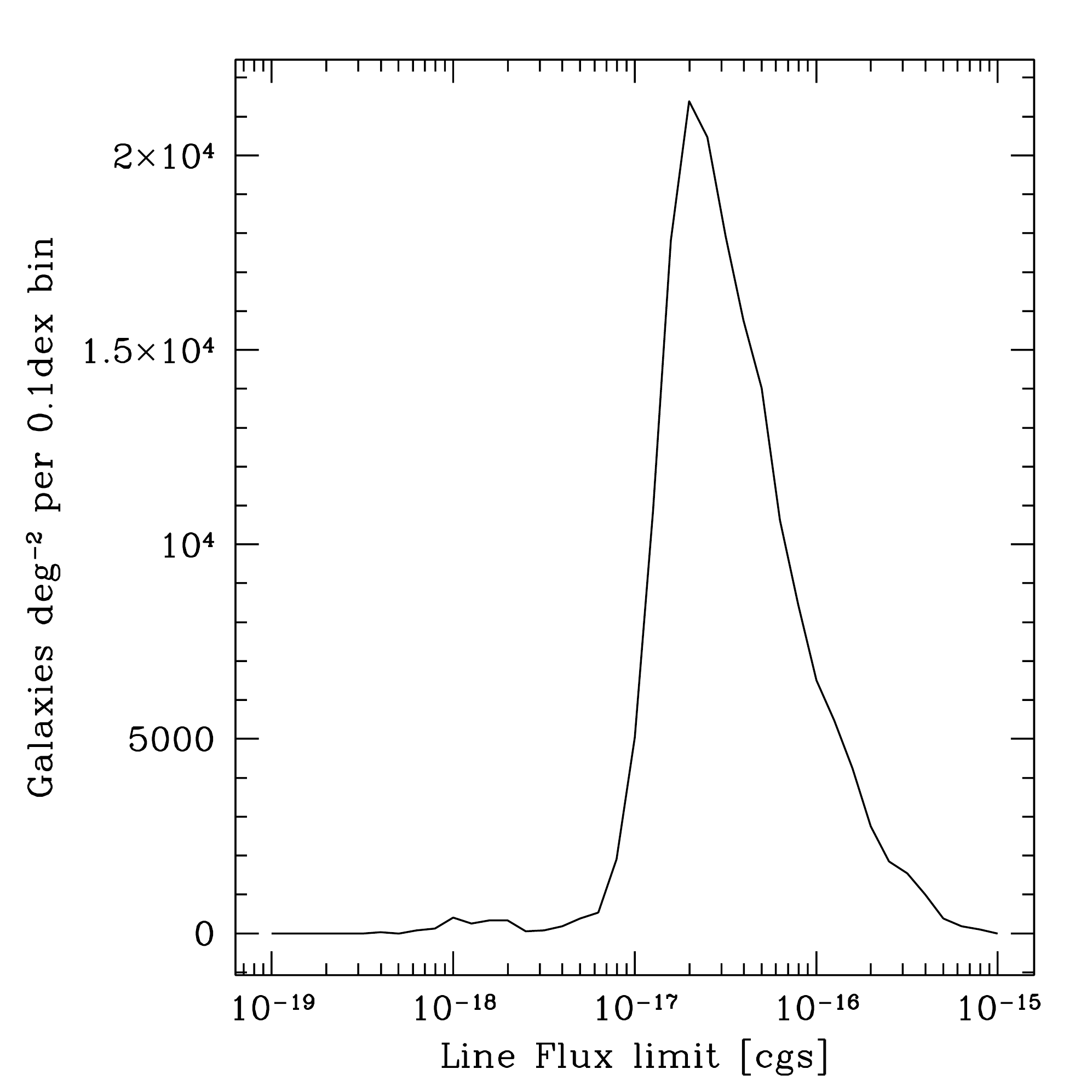}
\caption{The expected distribution of H$\alpha$ line flux of galaxies.
Left: AB=24.5, 25.2, 25.5 and 26.5 (the nominal and cumulative continuum depths for ATLAS Medium and Deep. Right: $H=24.7$ (depth of the WFIRST WL sample, which provides the target list for ATLAS Wide).}
\label{fig:emission-line}
\end{figure}

We have obtained source forecasts for ATLAS Probe by applying emission line models to the galaxy colors and magnitudes from CANDELS data (as discussed above).
Fig.\ref{fig:emission-line} shows the expected distribution of H$\alpha$ line flux of galaxies to $H=24.7$, the depth of the WFIRST WL sample, to which ATLAS Wide Survey provides the spectroscopic follow-up,
Based on this flux distribution, we expect 87\% of the WFIRST WL sample (targets for ATLAS Wide) to have emission line fluxes $f_{\rm line}>5\times 10^{-18}$erg$\,$ s$^{-1}$cm$^{-2}$ at $z>0.5$.

The right panel of Fig.\ref{fig:overview} shows the galaxy number counts per 0.5 redshift interval in the three ATLAS Probe galaxy redshift surveys: ATLAS Wide, ATLAS Medium, and ATLAS Deep. Table \ref{table:depths} in Sec.\ref{sec:overview} gives the details of these three surveys. 
ATLAS Wide is the spectroscopic follow-up of the WFIRST WL sample. It will cover 2000 square degrees, and is defined by the line flux limit of 5$\times 10^{-18}$erg/s/cm$^2$ (5$\sigma$), which will include $\sim$ 70\% of the galaxies in the WFIRST WL sample at $z>0.5$. ATLAS Wide has a continuum limit of AB magnitude 23.0 (3$\sigma$), and will yield the redshifts of $\sim$ 183M galaxies with $0.5<z<4$. 
ATLAS Medium will cover 100 square degrees, and is defined by the continuum limit of AB magnitude 24.5 (3$\sigma$), driven by key galaxy evolution science goals. It has a line flux limit of 1.2$\times 10^{-18}$ erg/s/cm$^2$ (5$\sigma$), and will yield the redshifts of $\sim$ 17M galaxies with $0<z<7$.
ATLAS Deep will cover 1 square degree, and is defined by the continuum limit of AB magnitude 25.5 (3$\sigma$), driven by additional galaxy evolution science goals. It has a line flux limit of 4.6$\times 10^{-19}$ erg/s/cm$^2$ (5$\sigma$), and will yield the redshifts of $\sim$ 0.31M galaxies with $0<z<7$.
ATLAS Probe may discover a significant number of very bright emission line galaxies at $z>4$ in the Wide Survey and $z>6$ in the Medium Survey, but we do not have enough statistics from current data to make those predictions.
The actual number of such galaxies discovered by ATLAS Probe will help shed light on the physics behind galaxy evolution.

\section{Spectroscopic multiplex}
\label{sec:multiplex}

We have carried out simulations of DMD multi-object spectroscopy (MOS) target selection for ATLAS in order to estimate the number of galaxies that can be targeted per DMD field of view (FOV) for various assumptions about the telescope and instrument parameters, and about the sky surface density of potential targets.  

\subsection{Assumptions and requirements}

We assume that the $2048\times 1080$ DMD projects onto a $4096\times 4096$ detector with $2\times$ magnification, and that spectra are dispersed along detector columns (see Sec.~\ref{sec:DMD} and Fig.~\ref{fig:DMD}).  To avoid overlap of adjacent spectra, targets must be separated by at least one column of ``closed'' DMDs. Thus, a maximum of 1024 DMD columns (per quadrant) may be used to observe targets.  At low spectral resolution, the projected length of spectra on the detectors is short enough that more than one target per DMD column can be observed without spectral overlap.  

Spectra are assumed to span one octave in wavelength, e.g., from 1 to 2 microns, or 2 to 4 microns.  We consider a range of instrument spectroscopic resolutions, $R \equiv \lambda/\Delta\lambda$, from $R = 400$ to 1600, defined at the mid-point of the spectral range, with the instrument producing spectra with constant dispersion in nm/pixel.  Note that the baseline ATLAS design assumes $R = 1000$.  The dispersion scale (nm/pixel) and total spectral length on the detector (pixels) is set assuming that the resolution element is sampled by 2.5 pixels in the dispersion direction. The total spectral length in detector pixels is $n_{\rm spec} = 5R/3$, e.g.,  $n_{\rm spec} = 1667$~pixels for $R=1000$.  The maximum spectral length that ensures that the full spectral range falls onto the detector for any DMD mirror is 1936 pixels.  Therefore, for spectral resolutions $R > 1162$, spectra for targets located in the top rows of DMD mirrors will not fall entirely on the detector.  In the present simulations, target positions are limited to the portion of the DMD where full spectral coverage is possible. 

For spectral resolution $R < 1162$, targets may be placed anywhere within a DMD column, whereas at higher resolution the top rows must be avoided in order for the entire dispersed spectrum to fall on the detector.  The maximum column multiplex, i.e., the maximum number of galaxies that can be targeted per DMD mirror column without spectral overlap in the dispersion direction, is:
\begin{equation}
   M^C_{\rm max} = 1 + \mathrm{int}[j_{\rm max} \times 6 / (5R)]
\end{equation}
where $j_{\rm max}$ is the maximum DMD row allowed for target placement, i.e., $j_{\rm max} = 1080$ for $R < 1162$, and $j_{\rm max} = 2048 - (5R/6)$ for $R > 1162$.  Table~\ref{tab:multiplex} reports the maximum column multiplex ($M^C_{\rm max}$) and total DMD multiplex ($M^T_{\rm max} = 1024 M^C_{\rm max}$) for various ranges spectral resolution $R$.  In practice, the achieved multiplex is generally lower and depends on the underlying density of potential targets.  When that density is high, there is a greater chance that potential targets will be found in the right places to fit several spectra into a given DMD column.  

\begin{table*}[h]
\begin{center}
\begin{tabular}{ccc}
\hline
Resolution  &  $M^C_{\rm max}$   & $M^T_{\rm max}$ \\
\hline
$ 324 < R <  432$ &  4  &   4096 \\
$ 432 < R <  648$ &  3  &   3072 \\
$ 648 < R < 1228$ &  2  &   2048 \\
$1228 < R < 2458$ &  1  &   1024 \\
\hline
\end{tabular}
\caption{Maximum target multiplex as a function of spectral resolution}
\label{tab:multiplex}
\end{center}
\end{table*}

We further assume that targets should be fairly well centered on a DMD mirror in order to be observed spectroscopically.  If not, a significant fraction of their light may be lost outside the mirror and not directed to the spectrograph.  Placing a target near the mirror edge may also have deleterious consequences for the effective PSF of the spectrum recorded on the detector.  In the present simulations, we require that a target falls within the central 25\% surface area of a DMD mirror;  i.e., if the fractional x,y position of a source within the mirror ranges from 0 to 1 in each axis, a target can only be selected if it falls in the range $0.25 < (x,y) < 0.75$.

The mirror centering criterion and the use of alternate DMD columns to avoid spectral overlap lead naturally to an observing strategy that cycles through a dithered pattern of 4 telescope pointings. Defining the angular coordinates of an arbitrary pointing as (0,0), the 4 pointings would have coordinates of (0,0), (0, 0.5$\theta_{DMD}$), (0.5$\theta_{DMD}$, 0) and (0.5$\theta_{DMD}$, 0.5$\theta_{DMD}$) where $\theta_{DMD}$ is the angular size of a DMD mirror projected on the sky (0.754$^{\prime\prime}$ for f/2.5). Additionally, 2 exposures will be taken at each pointing with a different set of micro-mirrors turned on to allow observation of galaxies in DMD columns that were previously excluded in order to avoid overlap of adjacent spectra. Half-step dithering in x and y combined with the 2 exposures per pointing ensure that all galaxies will fall within the central 25\% surface area of one of the DMD mirrors. A minimum of 8 exposures will then be taken, more if there are still targets to observe, most probably in that case in multiples of 8 exposures.

\subsection{Target densities}

\begin{figure}
\centering
\includegraphics[width=0.55\columnwidth,clip]{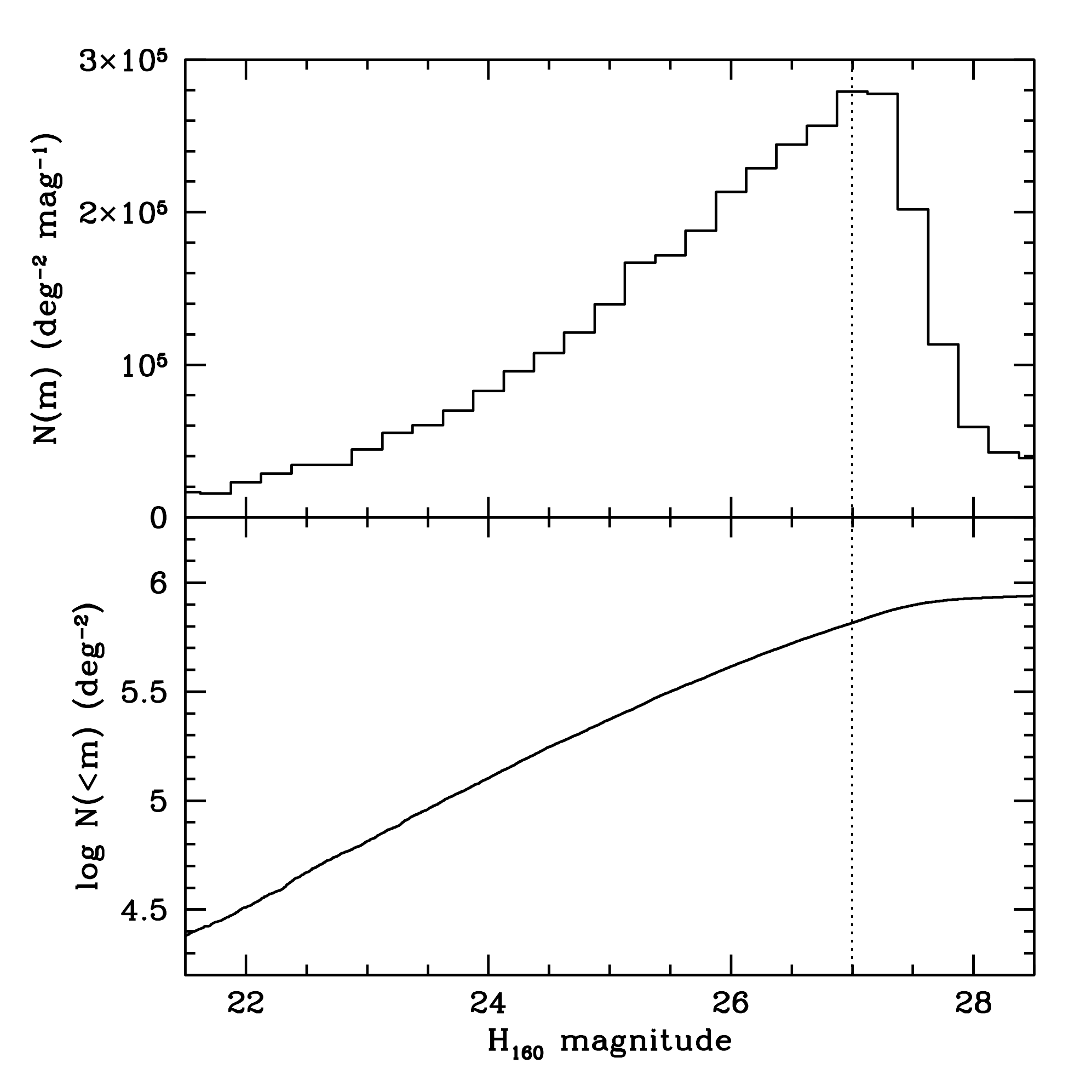}
\caption{
1.6 micron (F160W) source counts from the CANDELS ultra deep region of GOODS-S.  Upper panel: differential source counts, $n(m)$, with no correction for incompleteness.  The raw counts become significantly incomplete at $m_{\rm F160W} > 27$~mag.  Lower panel:  cumulative source counts, $n(<m)$.
}
\label{fig:numbercounts}
\end{figure}

We assume that a specified number ($N_{\rm targ}$) of potential targets are randomly distributed (i.e., with no clustering) over the DMD FOV.  The present simulations consider target densities in the range log $N_{\rm targ}$ = 3.4 to 4.8 per DMD FOV.  For a 1.5m telescope with f/2.5, the DMD FOV is 0.0969 deg$^2$, and the simulations correspond to galaxy surface density from $10^{4.4}$ to $10^{5.8}$~deg$^{-2}$.  From near-infrared (1.6 micron, F160W) source counts in the GOODS-S CANDELS-Deep region (Figure~\ref{fig:numbercounts}), these cumulative densities would approximately correspond to magnitude limits from $m_{\rm F160W} = 21.5$ to 27.  Actual ATLAS targeting strategies could preselect ranges of photometric redshifts, etc., so for the present purposes we represent the simulation results as a function of the number of potential targets per DMD FOV, rather than by limiting magnitude.   

\subsection{Simulating target selection}

In the present simulations, targets are selected for spectroscopy following a very simple procedure that is not optimized in any formal sense.  Starting at DMD column $i$=0 (zero indexed), all galaxies that fall within that column and which meet the mirror centering criterion are identified.  A viable target falling in the lowest row (index $j$) is selected for observation.  Then, the $j$ index is incremented by the spectral length in DMD mirror units ($n_{\rm spec}/2$), in order to avoid selecting any other galaxies whose spectra would overlap that of the previous object.  If there are viable targets that fall within the remaining rows of that DMD column, the object in the lowest-$j$ row is also selected as a target, and the process is repeated until the maximum DMD row ($j$=1079) is reached, or, for higher spectral resolutions, the maximum row for which the full spectral range would be projected onto the detector (see above).   Next, if one or more targets are selected for a given DMD column, then the column index is incremented by two, i.e., an empty column of mirrors is left as a buffer, as described above. However, if no targets are selected in a given DMD column, then the column index is incremented by one.  This process is repeated over all DMD columns, until targets have been selected over the full DMD FOV. 
In practice, the target selection process will be optimized to remove selection biases and to measure the sky through a small number of slits distributed across the field.

\subsection{Results}

\begin{figure}
\centering
\includegraphics[width=0.55\columnwidth,clip]{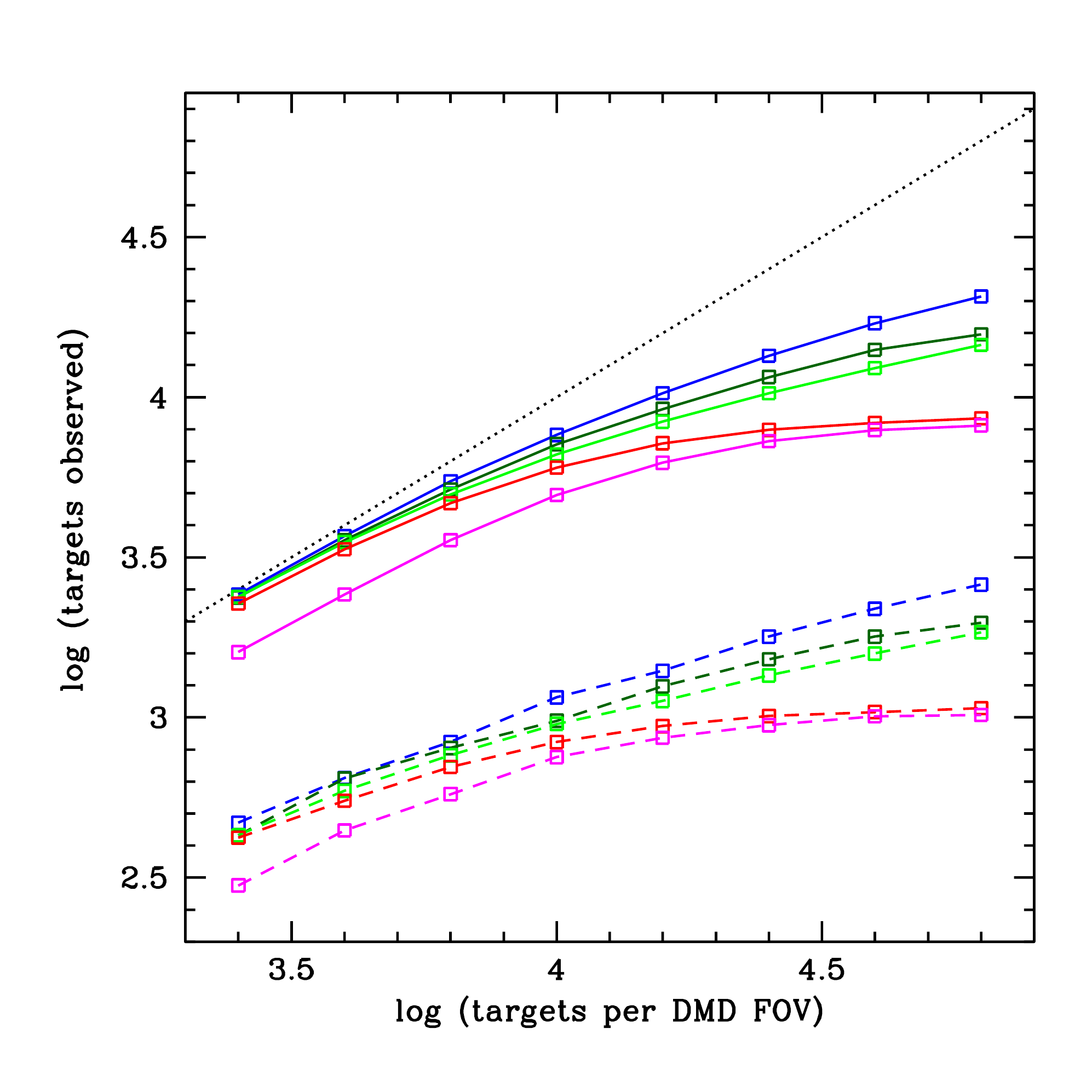}
\caption{
Number of observed spectra as a function of the input surface density of potential targets. Colors indicate spectral resolutions, in order of decreasing value (top to bottom): $R$=400 (blue); $R$=600 (dark green); $R$=800 (green); $R$=1200 (red); and $R$=1600 (magenta).  The dashed lines show results for a single telescope pointing, while the solid lines show results for a dithered sequence of 8 exposures. The dotted line represents 100\% target observation efficiency.  The ATLAS instrument design uses 4 DMDs and 4 spectrographs, and hence the target numbers should be multiplied by 4 for the full instrument FOV.
}
\label{fig:DMDresults}
\end{figure}

Figure~\ref{fig:DMDresults}  presents simulation results showing the total number of targets per DMD FOV, and the number of targets selected for spectroscopic observation, for a range of target densities and spectral resolutions.  Results are shown for a single telescope pointing, and for a dithered sequence of 8 exposures, as described above.  The number of galaxies selected for spectroscopic observation increases with the number density of potential targets within the instrument FOV, but the {\it fraction} of selected targets is lower at higher galaxy densities.  For a full sequence of eight exposures, the fraction of targeted galaxies is greater than 90\% at relatively low target densities, but at high densities (and at higher spectral resolutions) the multiplex approaches a saturation limit set by the need to avoid overlap between adjacent spectra.

Spectroscopic multiplex decreases at higher spectral resolution, as the dispersed spectra get longer and fewer targets can be ``stacked'' in a single DMD column without spectral overlap. For $R$=1000, although the theoretical multiplex per DMD column is 2 (see Table~\ref{tab:multiplex}, our simulations find that the practical upper limit on actual, achieved multiplex in the high-density limit is about 1.5 objects per column $\times 1024$ columns $\approx 1540$. Note that we have assumed 2.5 pixels per resolution element sampling, which is likely overly conservative.
Our preliminary optical design (see Sec.\ref{sec:instrument}) assumes 2 pixels per resolution element sampling, for which the multiplex factor is significantly larger.
We will study the optimization of this in our future work. 

The present simulations treat all targets as equal, i.e., there is no prioritization.  Moreover, there is no consideration of scaling exposure times with source brightness.  An optimized ATLAS observing strategy might include repeated visits to a given region of the sky.  Fainter objects could be observed repeatedly to build up total integration time, while brighter targets could be swapped out after some specified integration time or SNR has been achieved.

\section{Exposure time calculations}
\label{sec:ETC}

\begin{figure}
\centering
\includegraphics[width=0.45\columnwidth,clip]{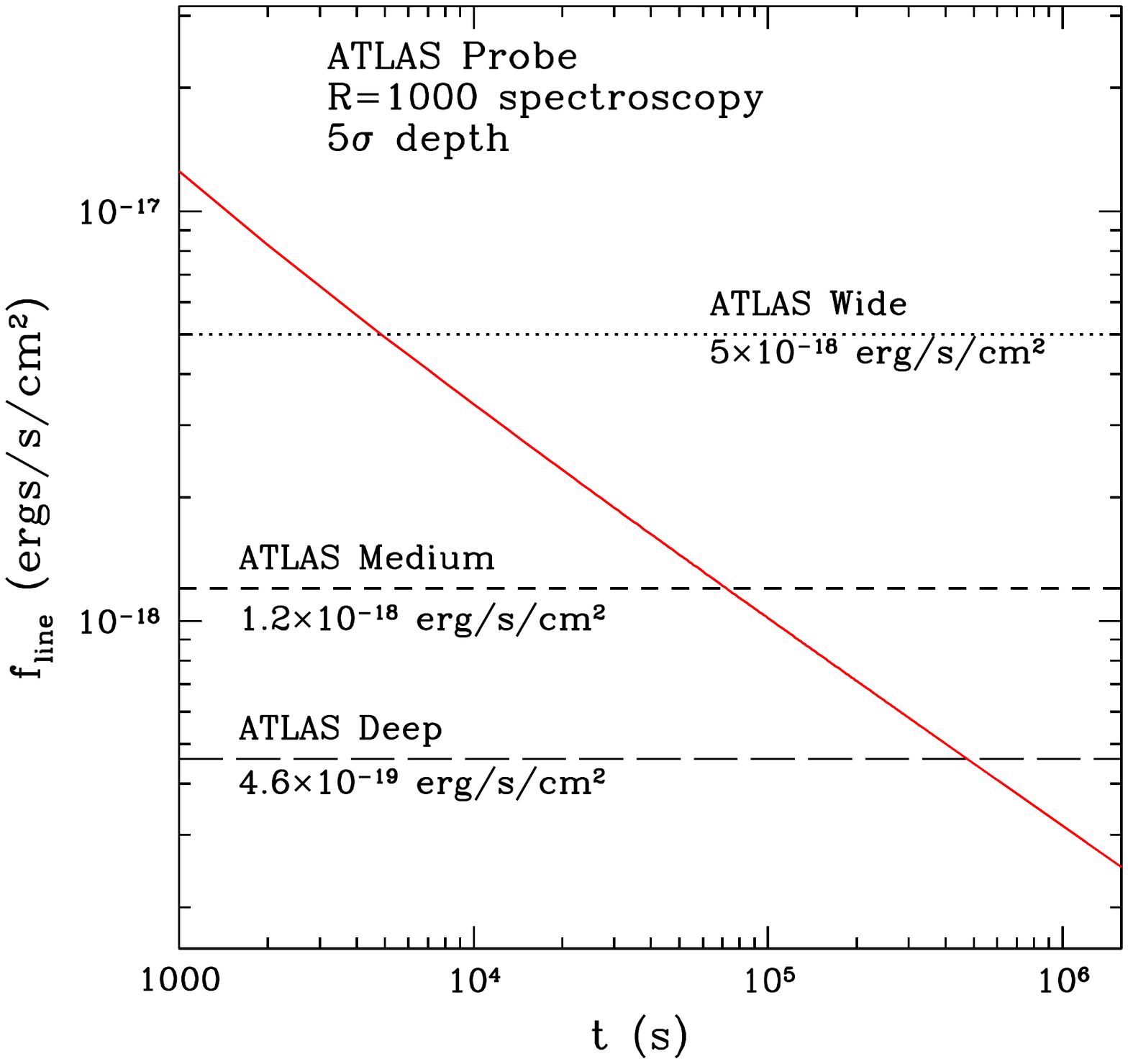}
\includegraphics[width=0.45\columnwidth,clip]{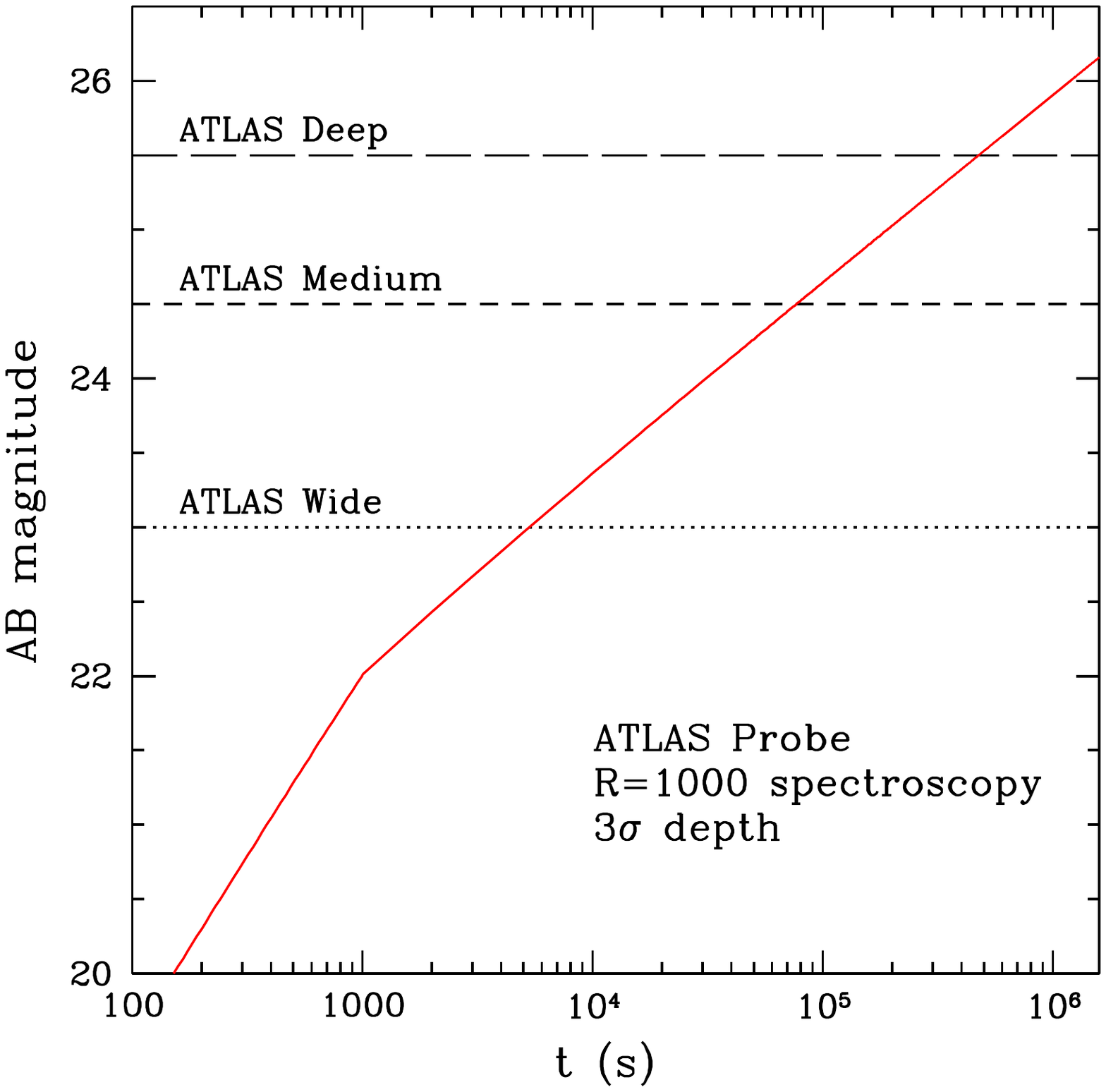}
\vspace{-1in}
\caption{The 5$\sigma$ emission line flux (left) and 3$\sigma$ continuum flux (right) limits versus exposure time for ATLAS Probe.
}
\label{fig:ETC}
\end{figure}

We have built an exposure time calculator from first principles, and modeled after that of JWST, with the assumed parameter values given in Table \ref{tab:etc}. We assume a telescope aperture of 1.5m, with 20\% obscuration, and a focal ratio
(the focal length divided by the aperture) of 2.5, which corresponds to a FoV of 0.4 sq deg, and a slit size of 0.75$^{\prime\prime}$. For simplicity, we assume a fixed spectral resolution of $R=1000$, telescope throughput of 0.3, a mean sky background of 0.14 MJy/Sr (zodiacal), and a read noise 5e$^-$ per pixel. The total read noise is 10e$^-$, for 4 pixels of extraction area per spectral resolution element. The exposure times for galaxies and stars have been estimated at $\lambda=2.5\,\mu$m (the mean of the ATLAS spectral range of 1-4$\,\mu$m). 
Long exposure times are segmented into 1000 second exposures for cosmic ray rejection.
We have not included slit loss in this estimate. Fig.\ref{fig:ETC} shows the 5$\sigma$ emission line flux (left) and 3$\sigma$ continuum flux (right) limits versus exposure time for ATLAS Probe.
\begin{table}[ht]
\centering
\begin{tabular}{|l|l|}
\hline
aperture & 1.5m \\
\hline
obscuration factor & 0.8\\
\hline
spectral resolution & 1000 \\
\hline
throughput (telescope+QE+instrumentation) & 0.3\\
\hline
fiducial wavelength & 2.5$\mu$m \\
\hline
read noise per pixel & 5 e$^{-}$\\
\hline
number of pixels in the measuring aperture $N_{\rm pix}$ & 4\\
\hline
sky background $F_{\rm sky}$ & 0.14 MJy/sr \\
\hline
slitsize & 0.75$^{\prime\prime} \times 0.75 ^{\prime\prime}$\\
\hline
\end{tabular}
\caption{The assumed parameter values for the ATLAS Probe exposure time calculator.}
\label{tab:etc}
\end{table}

\section{ATLAS Probe Instrument}
\label{sec:instrument}

ATLAS has only one instrument, compact and modular; it fits below the primary mirror structure into a cylindrical envelope only slightly larger than 1.5~m in diameter (the size of the primary) and $\approx$65~cm in height. The size can be reduced in a future improvement to the design.
To achieve our large FoV, the telescope beam is diverted into 4 arms, each composed by reimaging fore-optics, a DMD, and a dual-channel (blue/red) $R=1000$ spectrograph. ATLAS science objectives require a spectroscopy multiplex of $\sim\,$3000 per module, which can only be met by using DMDs with $\sim$2K$\times$1K micro-mirrors. We have selected the Cinema 2K DMD produced by Texas Instrument, with 2048$\times$1080 micro-mirrors, each of 13.7$\times$13.7 micron size. For a 1.5m telescope at f/2.5, this corresponds to a plate scale on the DMD of about 0.75$^{\prime\prime}$/mirror, and a field of view of about 0.10 deg$^2$, or 0.4~deg$^2$ for the four arms. The field with the dispersed spectra is finally projected on a 4k$\times$4k IR H4RG Sensor Chip Assembly.

The camera optics image a square 0.75$^{\prime\prime}$ field onto a little less than 2 x 2 pixels on the detector, i.e. they deliver a scale of about 0.385$^{\prime\prime}$/pixel, i.e. the footprint on the DMD on each NIR detector is about 4,000 x 2,100 pixels. The scale can easily be changed in the spectral direction to at least 2.1 pixel/micromirror because of the unused space on the detector.

The main source of scattering will likely be the surface of the DMDs. We are planning to perform laboratory test to measure the loss of contrast that one may expect from a DMD at mid-IR wavelengths. Assuming that the instrument is maintained at low enough temperatures, thermal noise should not represent a significant problem (i.e. the noise should be dominated by zodiacal light).

\subsection{Preliminary Optical Design}

Our preliminary optical design is based on the design originally developed by \cite{Content08a,Content08b} for the SPACE mission proposal \citep{Cimatti09} and the following pre-Phase-A and Phase-A studies of the ESA Euclid mission, which resulted from the merger of the SPACE and DUNE proposals. Euclid did not adopt the concept of slit mask Multi-Object Spectroscopy preferring the scientifically less ambitious slitless spectroscopy approach that can be more simply added to its NIR camera. 

\begin{figure}
\centering
\includegraphics[width=0.6\columnwidth,clip]{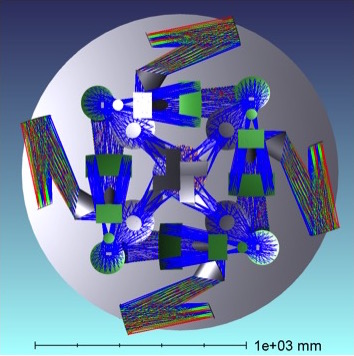}
\caption{A full view of the preliminary optical design for the ATLAS Probe instrument. The large gray circle is the back of the primary.
}
\label{fig:optical-full}
\end{figure}

\begin{figure}
\centering
\includegraphics[width=0.8\columnwidth,clip]{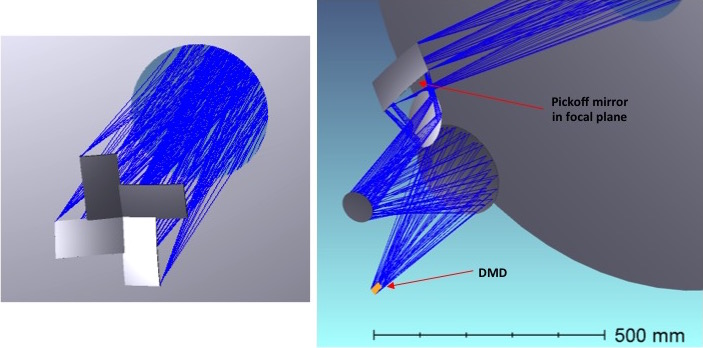}
\caption{Preliminary optical design for the ATLAS Probe instrument. Left panel: Pickoff mirrors in focal plane; positions and shapes are approximate.
Right panel: Shade layout of the fore-optics.
}
\label{fig:optical-1}
\end{figure}

The ATLAS telescope is a 1.5-m f/11.2 modified Ritchey-Chr\'{e}tien design similar to the SPACE telescope design. 
The telescope focal plane falls on a four-faceted pyramid mirror, a solution adopted e.g. by the WFPC2 camera on the Hubble Space Telescope.  
A full view of the ATLAS Probe instrument is shown in Figure \ref{fig:optical-full}.
The 4 faces of the pyramid mirror act as pickoff mirrors (Figure \ref{fig:optical-1}, left panel) sending the light to a fore-optics system . The 4-petals shape of the field allows for almost perfect tessellation of the sky, except for 2 small squares totaling only 0.07\% of the area, due to the aspect ratio of the DMD which is different from the ideal 2:1 ratio (full field coverage will be achieved through dithering). Our approach also permits the split of the instrument into 4 identical and separable blocks, which helps simplify the assembly, integration and testing. 

The 3 mirrors fore-optics (Figure \ref{fig:optical-1}, right panel) receive the f/11.2 telescope beam and release a $\simeq$ f/2.5 $\times$ f/2.3 beam on the DMD, accommodating for the 12$^\circ$ micro-mirror tilt along the diagonal. 
The DMD acts as an addressable multi-slit mask, reflecting to the spectrograph the light of the science targets ($+12^\circ$, "on" beam) and sending the rest of the field ($-12^\circ$, "off" beam) to a light dump. 

\begin{figure}
\centering
\includegraphics[width=0.6\columnwidth,clip]{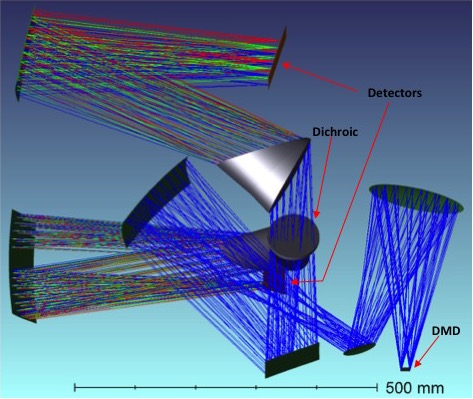}
\caption{Preliminary optical design for the ATLAS Probe instrument. Shade layout of the spectrograph; the short wavelength camera is on the opposite side.
}
\label{fig:optical-2}
\end{figure}

The spectrograph (Figure \ref{fig:optical-2}) includes a 4 mirror collimator followed by a beam splitter reflecting (1-2.1~$\mu$m) and transmitting (2.1-4~$\mu$m) the beam to a short and long-wavelength channel, respectively. Similar beam splitters (with cut-off at 2.4 micron) have been adopted e.g. by NIRCam on JWST. 
We take advantage of the relatively modest resolving power $R=1000$ to make use of prisms instead of gratings as dispersive element. Prisms provide a series of advantages over gratings:\\
\noindent (1) They have a much better transmission especially at the extreme wavelengths of the spectra.\\
\noindent (2) They can have a wavelength range larger than an octave between the extreme wavelengths. In this preliminary design the short wavelength camera has a wavelength range from 1$\mu$m to 2.1$\mu$m, with the possibility of further increasing the total wavelength range up to 0.9$\mu$m to 5$\mu$m in a future trade-off study.\\
\noindent (3) The resolution is more uniform as a function of wavelength. \\
\noindent (4) There is no need for a sorting order filter since there are no orders.\\
\noindent (5) There are no same-bandwidth parasite orders, especially zero and 2, cross-contaminating spectra aligned with each other. These would be unusually strong with our bandwidths where the longest wavelength is the double of the shortest. \\

Each prism directs the beam toward its camera which is made of only one mirror, an unusually low number. This is obtained by making the beam on the prism non-telecentric and correcting the resulting aberrations with the collimator and camera.

For target acquisition it is necessary to directly image the field; as originally proposed for SPACE/Euclid, direct imaging can be achieved through an achromatic grism where grating and prism dispersions cancel each other, with excellent image quality. ATLAS has 2 cameras, but only one is needed for target acquisition. Since the ATLAS short wavelength camera has a wavelength range only slightly different from Euclid, we envision a slightly modified design of the Euclid achromatic grism.

Our preliminary design already has excellent image quality in the spectrograph. The Gaussian Equivalent Full Width at Half Maximum (GEFWHM) is about 1/2 of a DMD mirror image. The image quality of the fore-optics on the DMD is 1/2 of a DMD mirror, satisfactory at this point of the preliminary design.

To attain the performances described above we make use of aspheric surfaces, some of them quite complex but  still well inside the limits of modern manufacturing.

\begin{figure}
\centering
\includegraphics[width=0.8\columnwidth,clip,trim={0 6cm 0 0}]{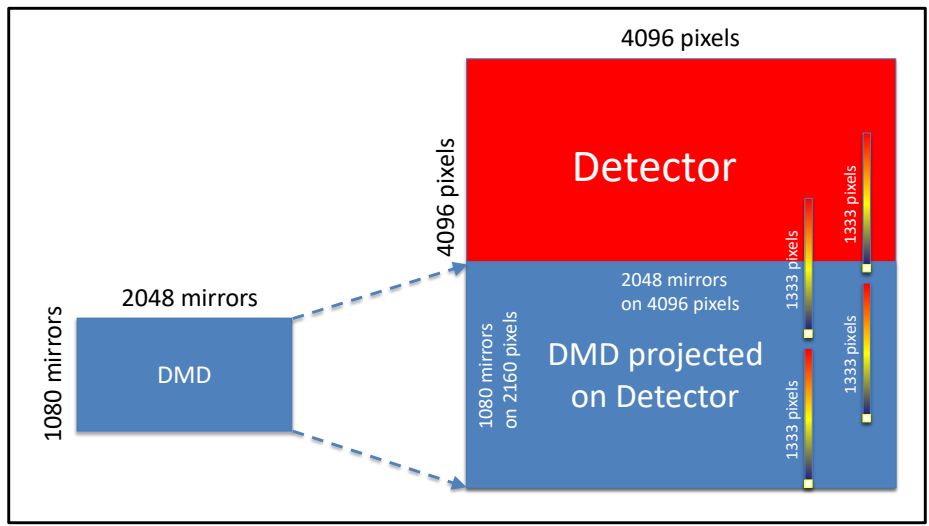}
\caption{Layout of the focal planes for ATLAS Probe: 
left) focal plane projected on the DMD, right) focal plane project on the IR detector.
The DMD is reimaged onto the lower part of the detector, and the detector's upper part is left "dark" to capture the full spectrum of each source, regardless of its position in the reimaged field. The moderate lengths of the $R\simeq 1000$ spectra (1333 pixels for 2 pixels, or 1667 pixels for 2.5 pixels per spectral resolution element sampling) allows for stacking multiple spectra on the same column, increasing source multiplexing. In order for the full spectrum to be captured by the IR detector for any target position in the field, the maximum spectrum length is 1936 pixels, corresponding to about 2.9 pixels per resolution element. 
}
\label{fig:DMD}
\end{figure}

Table \ref{table:optical-parameters} lists the main parameters of our system. Figure \ref{fig:DMD} shows the focal plane layout, with the DMD reimaged onto
the lower part of the detector and the detector's upper part left "dark" to capture the full spectrum of each
source, regardless of its position in the reimaged field.

\begin{table}[ht]
\centering 
\begin{tabular}{|ll|}
\hline
TELESCOPE 	&\\ \hline
Type	& modified Ritchey-Chr\'{e}tien \\\hline
Primary: diameter \& focal ratio	& 150cm; f/1.6 \\\hline
Primary: central obscuration	& 19\% diameter (3.7\% area)\\\hline
              Secondary: diameter &	29 cm \\\hline
              Telescope focal ratio	& f/11.2 \\\hline
PYRAMID MIRROR	& 4 rectangular faces \\\hline
Size	& 4 $\times$ 13.6 cm $\times$ 7.4 cm \\\hline
Field of view 	& 4 $\times$ 25.6$^\prime$ $\times$ 13.5$^\prime$ \\\hline
FORE OPTICS	& \\\hline
f/\# (off axis)	& f/2.3 $\times$ f/2.5  \\\hline
Scale on DMD	& 0.75$^{\prime\prime}$ $\times$ 0.75$^{\prime\prime}$/micromirror (slit size)  \\\hline
COLLIMATOR	& \\\hline
Elements	& 4 mirrors (+ 1 dichroic) \\\hline
DISPERSING ELEMENTS	&  \\\hline
Wavelength ranges	& 1-2.1$\mu$m (NIR); 2.1-4$\mu$m (MIR) \\\hline
Resolving power	& R $\sim$ 1000 \\\hline
Type	& prism  \\\hline
CAMERAS	& \\\hline
Elements              &                  1 mirror (each camera) \\\hline
Sampling on detector	& 0.38$^{\prime\prime}\times$0.39$^{\prime\prime}$ (pixel scale)  \\\hline
\end{tabular}
\caption{ATLAS main optical parameters.}
\label{table:optical-parameters}
\end{table}

\subsection{Detectors and ASICs} 
\label{sec:detector}

Our baseline detector is the Teledyne H4RG-10, the same type currently under development for WFIRST. The long-wavelength cutoff of our spectroscopic channels (2.1$\mu$m and 4$\mu$m) match the standard $\sim\,$2.35$\mu$m and $\sim\,5.37\mu$m cutoff of WFIRST and JWST devices respectively. Fine-tuning the Hg vs. Cd stoichiometric ratio can further reduce dark current, allowing warmer operating temperatures. H4RG arrays tested by WFIRST have typical QE>90\% in the 0.8-2.35$\mu$m range, readout noise $\sim$15e in Double Correlated Mode and mean dark current $<0.01$e/s/pixel. JWST devices have similar performance. The architecture of the multiplexer allows multiple non-destructive reads of each pixel during a single exposure (``sampling up the ramp''); this enables mitigation against cosmic rays and reduction in read noise: typical readout noise with 16 samples drops to about 5e. Teledyne has developed SIDECAR ASICs (Application Specific Integrated Circuit) to manage all aspects of FPA operation and output digitization in cold environment. By keeping analog signal paths as short as possible to reduce noise and output capacitance loading, ASICS improve power consumption, speed, weight and performance. A Teledyne ASIC device is currently driving HST/ACS and several Teledyne ASICs will soon fly on 3 out of 4 JWST instruments; we will adopt such devices for ATLAS. 

\subsection{DMDs and Controller} 
\label{sec:DMD}

The DMD is the core of our system. We base our design on the 2k CINEMA model of 2048$\times$1080 micro-mirrors, 13.7 $\mu$m on a side. DMDs have been successfully used on ground-based spectrographs like RITMOS \citep{Meyer04} and IRMOS \citep{MacKenty06}. A new DMD-based spectrograph, SAMOS, is under construction of the SOAR telescope in Chile \citep{Robberto16}. NASA has funded a Strategic Astrophysics Technology (SAT) program (PI Ninkov, ATLAS Probe core team member) to raise the TRL level of DMDs to TRL5-6 before the 2020 decadal survey. DMDs have successfully passed proton and heavy-ion irradiation testing \citep{Fourspring13,Travinsky16}. Following NASA General Environmental Verification Standard (GEVS), the team performed random vibration, sine burst, and mechanical shock testing of manufacturer-sealed DMDs \citep{Vorobiev16} as well as of devices re-windowed for better UV and IR capabilities \citep{Quijada16}. These GEVS tests suggest that DMDs are robust and insensitive to the potential vibroacoustic environments experienced during launch. Low-temperature testing of DMDs were also performed, the main concern being micro-mirror stiction. Tests at RIT have shown that temperatures as low as 130 K do not affect the performance of DMDs. More recent data obtained at JHU at ~80K confirm these results. A general overview of the results obtained by these test campaigns has been recently presented \citep{Travinsky17}.

These findings place DMDs between TRL levels 5 and 6 (note that level 6 requires testing against the specific environment of a mission, thus is only achievable after a mission has been identified, e.g. ATLAS at L2). DMDs are normally controlled by commercially available boards based on the DLP Discovery 4100 chipset from Texas Instruments. The chipset is not designed to operate in a cryogenic environment or in space. ATLAS core team members at JHU have recently developed custom electronics to operate Cinema DMDs at cryogenic temperatures. For ATLAS, we intend to produce a version of this system based on rad-hard components, suitable for operations at L2.
    
Micro Shutter Arrays (MSA) may represent an alternative to DMDs. However, their limited size implies that one would need 142 devices similar to those on JWST/NIRSpec to obtain the same number of selectable slits achieved with 4 DMDs. Also, the filling factor $\lesssim70$\% of their clear apertures greatly complicates operations, since a large fraction of the visible targets ends up partially occulted. If future developments will lead to the construction of MSAs with $\sim$ 30 times more apertures, clear-aperture ratio larger than 90\%, and cosmetic quality comparable to those of DMDs, we will evaluate their use in the ATLAS Probe design.
    
\section{Mission Architecture and Cost Estimate}
\label{sec:mission-design-cost}

We assume a 5 year ATLAS mission in a halo L2 orbit similar to JWST, enabling long observations in a very stable thermal and radiation environment; the narrow Sun-Spacecraft-Earth angle facilitates passive cooling to the $\sim$50K operating temperature needed for the long wavelength channel detectors. Absolute pointing requirements are relaxed due to the versatility of the DMDs, and pointing stability of 0.1$^{\prime\prime}$ over the 1000s exposure time is appropriate for the 0.39$^{\prime\prime}$ pixels. These requirements will be addressed in the ATLAS Probe design study.

The temperature of the instrument should be low enough to make it immune to changes related to the spacecraft attitude. Note that for a survey telescope the attitude can be more easily maintained within optimal range than for a general observatory (like e.g. HST or JWST) where the pointing and orientation are driven by the science programs, complicating optimal scheduling.

The ATLAS Probe total mission cost (including reserves and the launch vehicle) is estimated to fit within the cost envelope of a NASA probe-class space mission.
This estimate was developed by JPL using Spitzer IRAC and IRS analogies for the instrument, the Smart Telescope Model
(2000) Cost Estimating Relationship for the telescope assembly, and assumed an RSDO spacecraft bus can be used
with minimal modifications. A NICM approach was also used for the instrument with consistent results. A standard wrap
rate is used for the remaining WBS items: project management, system engineering, mission assurance, science, science
data system, MOS, GDS, operations, and I\&T. The launch vehicle cost was based on LSP catalog 2015 range of costs
that would accommodate a LEO, L2, or Earth trailing orbit with a mass range of 1575-5250 kg. 30\% reserves are
included for the observatory and 15\% reserves for operations.

\section{Summary}
\label{sec:summary}

ATLAS Probe is a compelling mission concept for a NASA probe-class mission. It is a follow-up space mission to WFIRST; it multiplexes the scientific return of WFIRST by obtaining $R=1000$ slit spectra of $\sim$ 70\% of all galaxies imaged by the WFIRST High Latitude Survey at $z>0.5$. Enabled by the mature DMD technology that allows a spectroscopic multiplex factor greater than 5,000, ATLAS Probe will lead to ground-breaking science over the entire range of astrophysics: from galaxy evolution to the dark Universe, from objects in the outer Solar System to the dusty regions of the Milky Way.

In future work, we will mature the ATLAS Probe mission design, and further quantify the science return from ATLAS Probe in its diverse science areas. In addition, we will engage the astronomical community in exploring feasible guest observing programs for ATLAS Probe.

\acknowledgments
{\bf Acknowledgments:} We are grateful to Lee Armus, Dominic Benford, Peter Capak, Jeff Kruk, Ian Smail, John Stauffer, Mike Werner, and Rien Weygaert for invaluable discussions.
We acknowledge Falk Baumgarten for help with the computation of the EZmocks, Charles Shapiro for providing information on the detectors, and Alex Merson for advice on cosmic web visualization options.
This work was supported in part by NASA grant 15-WFIRST15-0008 Cosmology with the High Latitude Survey WFIRST Science Investigation Team.
Andrea Cimatti acknowledges the grants ASI n.I/023/12/0, PRIN MIUR 2015 and ASI n.2018-23-HH.0.
Wesley Fraser acknowledges funding from Science and Technology Facilities Council grant ST/P0003094/1.
Christopher Hirata is supported by the US Department of Energy, NASA, the National Science Foundation, and the Simons Foundation.
Jeffrey Newman acknowledges support from DOE grant DOE DE-SC0007914.
Zoran Ninkov acknowledges NASA's Strategic Astrophysics Technology (SAT) Program (grant number NNX14AI62G) in support of the development of DMD technology for use in space.
James Bartlett, Olivier Dor\'{e}, Peter Eisenhardt, Michael Ressler, and Jason Rhodes are supported by JPL, which is run by Caltech under a contract from NASA.

\appendix

\section{Flow-down of the Galaxy Evolution Science Objective to Survey Requirements} \label{sec:error}

\begin{figure}
\centering
\includegraphics[width=0.7 \columnwidth,clip,angle=-0]{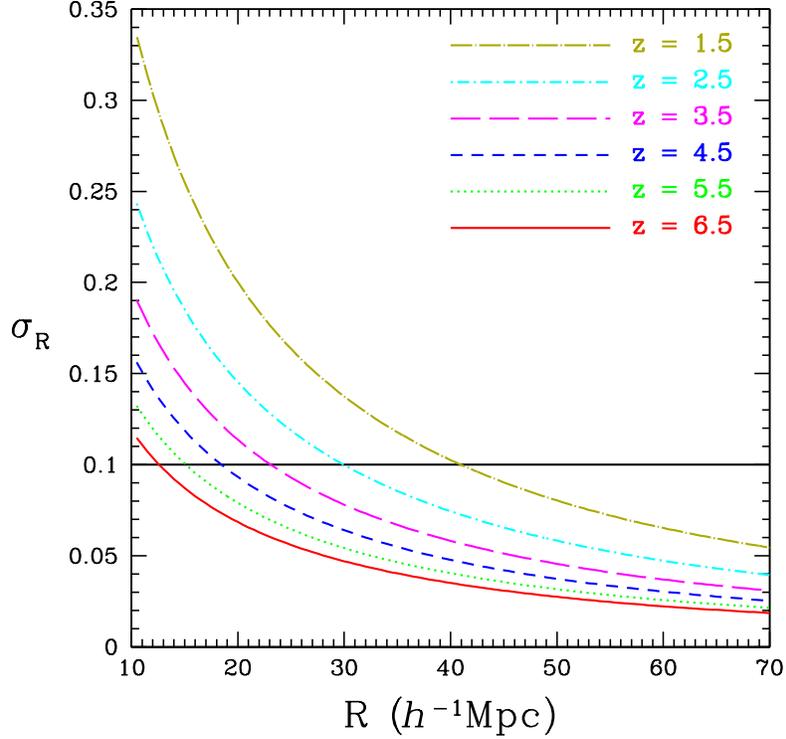}
\vspace{-1.7in}
\caption{Square root of the variance of linear matter density fluctuations in spheres of radius $R$ 
 (in units of $h^{-1}$Mpc), $\sigma_R$, as a function of $R$ at several redshift values in the range explored by ATLAS Probe. 
}
\label{fig:sigR}
\end{figure}

In order to map the cosmic web on the scales relevant for galaxy evolution, we require that ATLAS Probe trace the relation between galaxies and dark matter on those scales with less than 10\% shot noise (Science Objective 1). 
The precision with which we trace that relation can be quantified by the measurement errors on the bias factor between galaxies and matter, $b(r,z)$, on comoving scale $r$ at redshift $z$. 
At a given $z$, the relevant scales are those over which $b(r,z)$ varies with scale $r$. On sufficiently large scales, $b(r,z)$ converges to a constant (the linear bias regime). 
The onset of linearity can be quantified by selecting a low threshold in $\sigma_R$, the square root of the variance of linear matter density fluctuations in spheres of radius $R$. Note that $\sigma_R(R=8\,h^{-1}$Mpc) is the same as the familiar $\sigma_8$ parameter.
Fig.\ref{fig:sigR} shows $\sigma_R$ as a function of $R$ at several redshift values in the range explored by ATLAS Probe.

For the very linear regime define by $\sigma_R < 0.1$, bias is constant. ATLAS Medium and Wide Surveys probe the scales from a few Mpcs (above the galaxy cluster scale) to the onset of linearity as defined by $\sigma_R=0.0683$ (chosen to correspond to $R=100\,h^{-1}$Mpc at $z=0$). This corresponds to $R=57.5, 43.3, 34.3, 28, 23.5, 20\,h^{-1}$Mpc at $z= 1.5, 2.5, 3.5, 4.5, 5.5, 6.5$ respectively.  

The measurement uncertainty on $b(r,z)$ can be estimated using the measurement errors on the galaxy 2-point correlation function (2PCF), $\xi(r,z)$, since $\xi(r,z) = \left[b^2(r,z) + \frac{2}{3} b(r,z)f_g(z) + \frac{1}{5} f_g^2(z)\right]\xi_m(r,z)$, where the matter correlation function $\xi_m(r,z)$ and growth rate $f_g(z)$ can be predicted based on cosmological measurements. Note that $\xi(r,z)$ is the observed galaxy 2PCF in redshift space but $\xi_m(r,z)$ is matter CF in real space. In practice, we will measure $b(r,z)$ by measuring the galaxy 3-point correlation function (3PCF) and the galaxy 2PCF. The uncertainties on the galaxy 2PCF and 3PCF both scale with the effective volume sampled by a survey. For simplicity, we use the errors on the square root of the galaxy 2PCF as an indicator for the bias measurement uncertainty here. In future work, we will measure the galaxy 3PCF and 2PCF jointly from mocks, and derive the measurement errors on $b(r,z)$ in a more realistic manner.

\begin{figure}
\centering
\includegraphics[width=0.5 \columnwidth,clip,angle=-0]{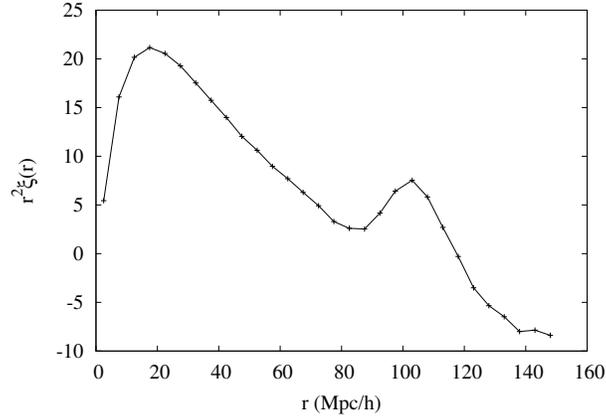}
\caption{
The mean of the correlation functions from 100 mock catalogs (bin size is 5 $h^{-1}$Mpc). 
}
\label{fig:mean_cf}
\end{figure}

\begin{figure}
\centering
\includegraphics[width=0.5 \columnwidth,clip,angle=-0]{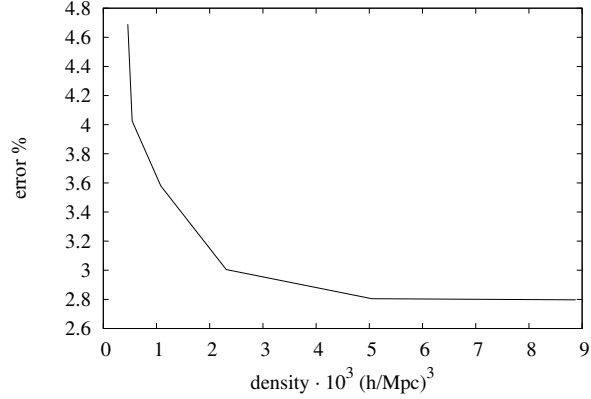}
\caption{
The standard deviation of the correlation function of separation $20 < r < 25$ $h^{-1}$Mpc depending on the number density. The volume is $750^3$ (Mpc/h)$^3$.
}
\label{fig:err_vs_density}
\end{figure}

We estimate the percentage error of the correlation function based on mock galaxy catalogs, by taking into account the number density and the volume of the sample, assuming a bias of $b(z)=1.5+0.4(z-1.5)$ \citep{Spergel15}.\footnote{This bias assumption is conservative for ATLAS Wide, and under-estimates the bias significantly for ATLAS Medium. This means that our error forecasts of the correlation function measurements are very conservative.}
To determine the dependency between the uncertainty and number density of galaxies, we run 100 Zel'dovich-approximation simulations with the size of (750 $h^{-1}$Mpc)$^3$ and the number of the dark matter particles is $288^3$. We populate galaxies using the code EZmock \citep{Chuang:2014vfa} based on these 100 boxes. Fig \ref{fig:mean_cf} shows the mean of the correlation functions (bin size is 5 $h^{-1}$Mpc). By diluting the mock galaxy sample, we compute multiple sets of 100 correlation functions for different number densities. These correlation functions have the same mean values but different uncertainties. Fig \ref{fig:err_vs_density} shows the standard deviation of the separation $20 < r < 25$ $h^{-1}$Mpc depending on the number density. We can then estimate the standard deviations for different number densities by interpolation/extrapolation. In addition, we assume the uncertainty is proportional to $V^{-1/2}$, so that we rescale the uncertainty by a factor $(750^3 /V)^{-1/2}$, where $V$ is the survey volume. 

\begin{figure}
\centering
\includegraphics[width=0.49\columnwidth,clip]{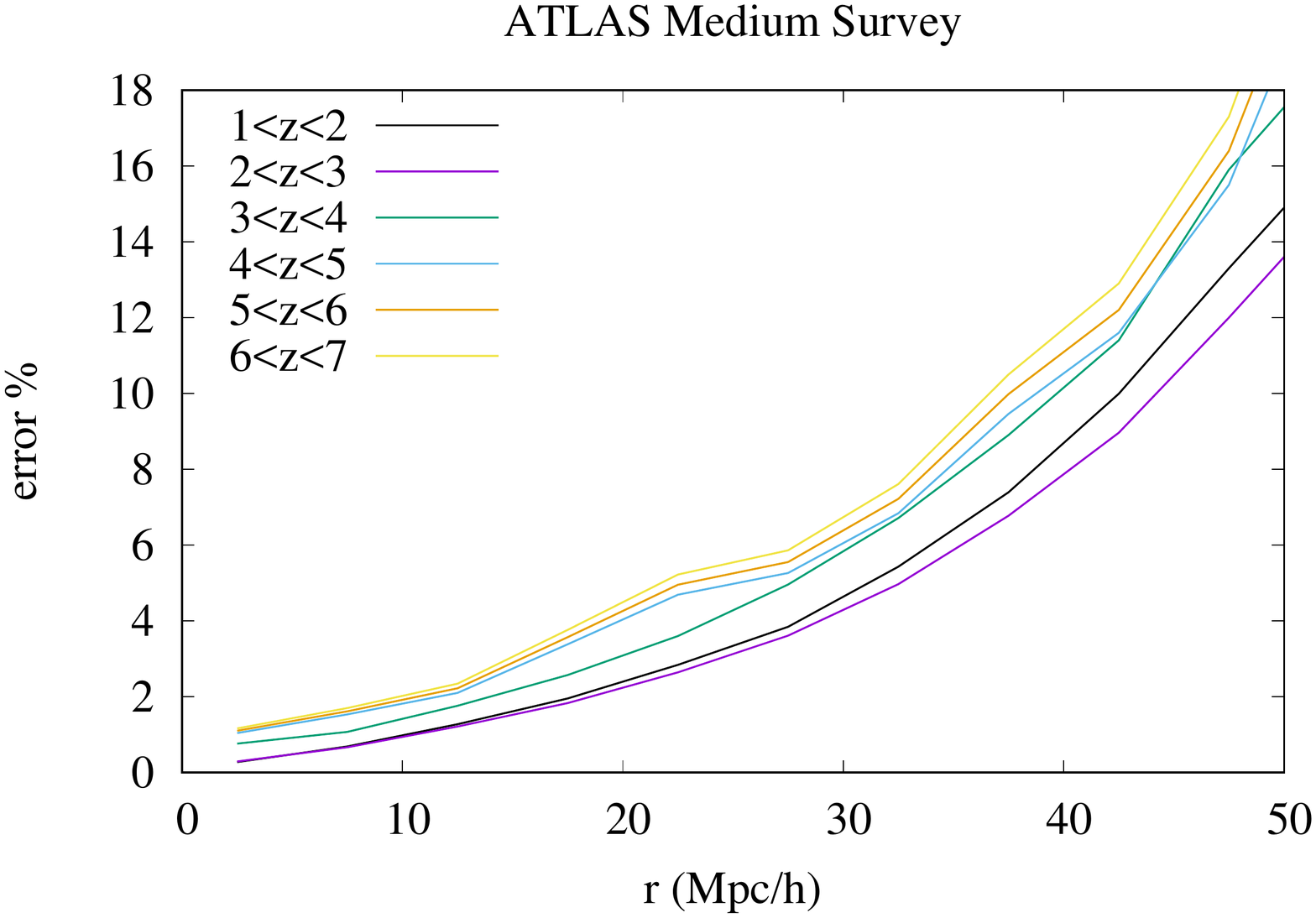}
\includegraphics[width=0.49\columnwidth,clip]{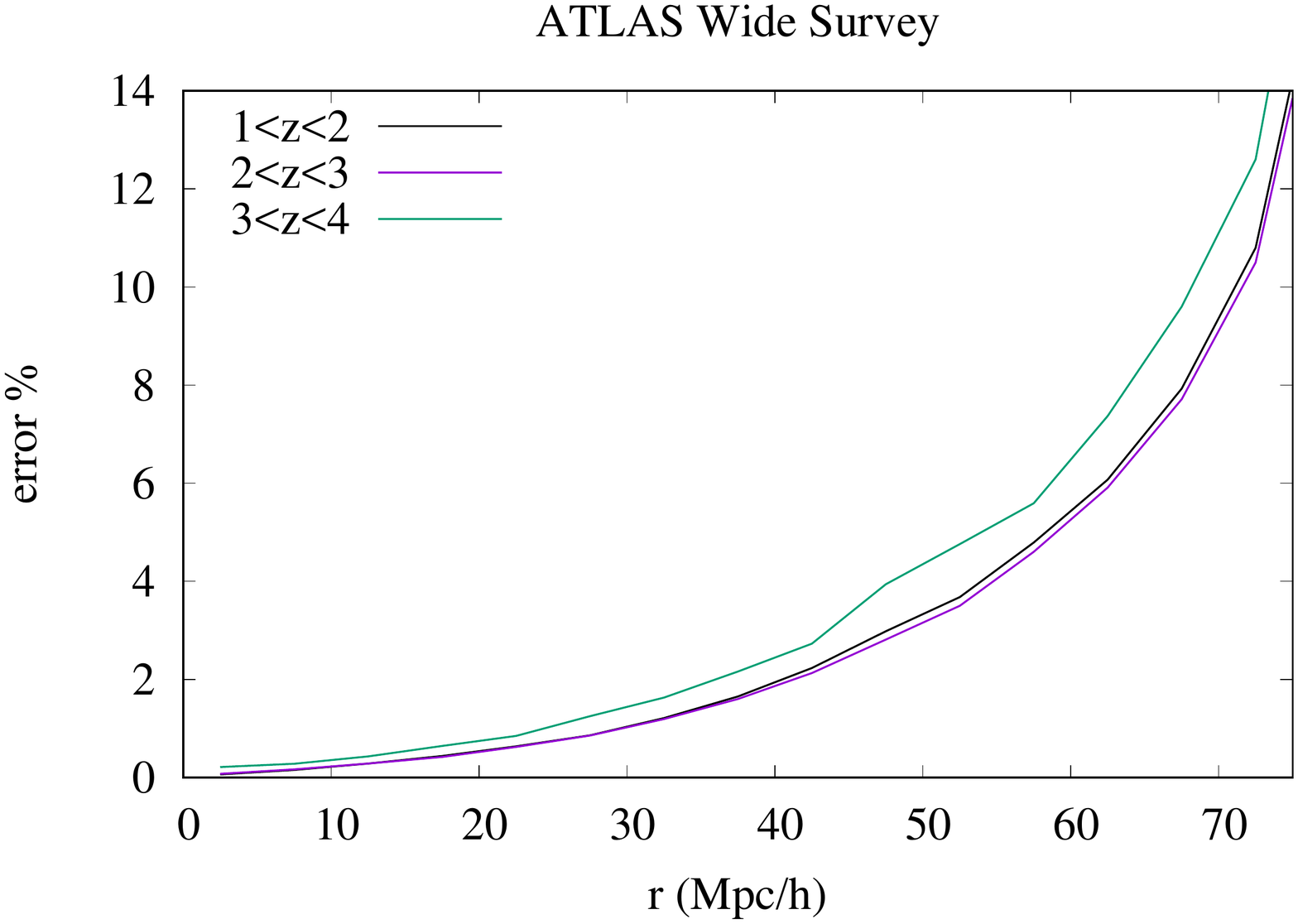}
\caption{
The predicted uncertainties in the measured galaxy correlation functions for ATLAS Medium and ATLAS Wide surveys, with galaxy number densities given in Table \ref{table:h24p5} and Table \ref{table:h23}.
}
\label{fig:h24p5_h23}
\end{figure}

Table \ref{table:h24p5} lists the number densities and volumes for the galaxy sample of the 100 deg$^2$ ATLAS Medium Survey survey at various redshifts.
Table \ref{table:h23} lists the number densities and volumes for the galaxy sample of the 2000 deg$^2$ ATLAS Wide Survey at various redshifts.
Fig.\ref{fig:h24p5_h23} shows the predicted uncertainties corresponding to Table \ref{table:h24p5} and Table \ref{table:h23}. 

Comparing the scales for the onset of linearity in Fig.\ref{fig:sigR} with the measurement errors in Fig.\ref{fig:h24p5_h23}, we see that ATLAS Probe Science Objective 1 flows down to the requirement for ATLAS Medium and Wide Surveys. Note that ATLAS Wide (with the continuum depth of H<23) measures redshifts for 70\% of galaxies in the WFIRST WL sample at $z>0.5$ using emission lines, even though it is shallower than the WFIRST WL sample by $\sim$ 1.7 mag in the continuum. ATLAS Deep provides a validation data set for ATLAS Medium and Wide Surveys; it provides denser sampling of lower mass galaxies of all types at all redshifts, and unique 3D mapping of structure at $5 < z < 7$ with [OIII] line emitters. 

\begin{table}
 \begin{center}
  \begin{tabular}{c c c c} 
     \hline
$	z_{min}	$&$	z_{max}	$&	volume $(Mpc/h)^3$	&	density $(h/Mpc)^3$	\\
$	1	$&$	2	$&$	4.09\times10^8	$&$	1.98\times10^{-2}	$\\
$	2	$&$	3	$&$	4.80\times10^8	$&$	4.81\times10^{-3}	$\\
$	3	$&$	4	$&$	4.62\times10^8	$&$	8.48\times10^{-4}	$\\
$	4	$&$	5	$&$	4.21\times10^8	$&$	1.50\times10^{-4}	$\\
$	5	$&$	6	$&$	3.79\times10^8	$&$	7.23\times10^{-5}	$\\
$	6	$&$	7	$&$	3.40\times10^8	$&$	8.05\times10^{-5}	$\\
     \hline
 \end{tabular}
 \end{center}
\caption{
The number densities and volumes for the galaxy sample of the 100 deg$^2$ ATLAS Medium Survey at various redshifts.
}
\label{table:h24p5}
  \end{table}

\begin{table}
 \begin{center}
  \begin{tabular}{c c c c} 
     \hline
$	z_{min}	$&$	z_{max}	$&	volume $(Mpc/h)^3$	&	density $(h/Mpc)^3$	\\
$	1	$&$	2	$&$	8.17\times10^9	$&$	1.09\times10^{-2}	$\\
$	2	$&$	3	$&$	9.60\times10^{9}	$&$	2.83\times10^{-3}	$\\
$	3	$&$	4	$&$	9.24\times10^{9}	$&$	6.03\times10^{-4}	$\\
     \hline
 \end{tabular}
 \end{center}
\caption{
The number densities and volumes for the galaxy sample of the 2000 deg$^2$ ATLAS Wide Survey at various redshifts.
}
\label{table:h23}
  \end{table}
  
\section{Cosmic web visualization}

The spatial distribution of H$\alpha$ emitters shown in Figs.\ref{fig:cosmic-web1}, \ref{fig:cosmic-web2}, \ref{fig:atlas_cosmic-web}, and \ref{fig:atlas_wide} is obtained by making use of the GALFORM semi-analytical model of galaxy formation \citep{lacey16}. The sizes of the filled circles in Figs.\ref{fig:cosmic-web1}, \ref{fig:cosmic-web2}, and \ref{fig:atlas_wide} are determined by the H mag of the galaxies, with the bigger circles representing brighter galaxies. 

Fig.~\ref{fig:cosmic-web1} compares the spatial distribution of H$\alpha$ emitters predicted with the observed flux limit of ATLAS Medium (AB $< 24.5$, $f_\text{line} > 1.2\times 10^{-18} {\rm erg/s/cm^2}$) against that expected with WFIRST ($f_\text{line} > 10^{-16} {\rm erg/s/cm^2}$), at $z=2$, as tracers (red) of the cosmic web of dark matter (green).
Fig.~\ref{fig:cosmic-web2} shows how the cosmic web of dark matter (green) at $z=4$ is traced by the galaxies (red) from ATLAS Medium (AB $< 24.5$, left) and ATLAS Deep (AB $< 25.5$, right) surveys.
Fig.\ref{fig:atlas_cosmic-web} compares three different redshift surveys (slit, slitless, and photometric) of the same galaxy distribution.
Fig.~\ref{fig:atlas_wide} shows instead what is obtained with ATLAS Wide, with an observed limiting line flux of $5\times 10^{-18} {\rm erg/s/cm^2}$.
Figs.~\ref{fig:cosmic-web1} and \ref{fig:cosmic-web2} show slices of $50~h^{-1}{\rm Mpc}$ and $100 ~h^{-1}{\rm Mpc}$ on a side, respectively,  and $10~h^{-1}{\rm Mpc}$ depth at $z=2$. The figures display the dark matter particle distribution in green, with brighter areas showing a higher concentration of dark matter. Fig.~\ref{fig:atlas_cosmic-web} shows the angular distribution of all galaxies in a slice of $10~h^{-1}{\rm Mpc}$ depth but over the full simulation box.

The GALFORM model predicts the observed properties of galaxies embedded in a large cosmological $N$-body simulation. The haloes and their merger histories are extracted from the {\it Millennium-WMAP7} dark matter-only $N$-body simulation, with halo mass resolution of $\sim 10^{10}~h^{-1} M_{\odot}$ and a box size of $500~h^{-1}{\rm Mpc}$.

The model computes the growth and evolution 
of galaxies by following the evolution of gas, stars and metal components throughout the merger histories of dark matter haloes. Different variants of GALFORM have been shown to predict the abundance 
of H$\alpha$ and {\rm [O \sc{ii}]} emitters in reasonable agreement with observational measurements \citep{orsi10, lagos14, gonzalez17}. The calculation of the H$\alpha$ flux of galaxies is performed 
by first integrating the spectral energy distribution (SED) of each galaxy blue-wards of the Lyman limit to obtain the total production of hydrogen ionizing photons. 
All these photons are assumed to be absorbed by the neutral interstellar medium (ISM), making the escape fraction of Lyman continuum photons $f_{\rm esc} = 0$. 
The intrinsic H$\alpha$ line flux is computed assuming case B recombination and is directly proportional to the number of Lyman continuum photons \citep[see also ][]{orsi14}. 
The observed H$\alpha$ luminosity incorporates the attenuation by dust, which corresponds to that of the continuum at the wavelength of H$\alpha$ \citep[see ][for details of the dust attenuation model]{lacey16}.

\end{document}